\documentclass[a4paper,11pt]{article}
\usepackage{jheppub}
\usepackage{slashed}
\usepackage{soul}
\usepackage{mathtools}
\usepackage{bbold}
\usepackage[mathscr]{euscript}
\usepackage{todonotes}
\usepackage{bm}
\usepackage[normalem]{ulem}

\newcommand{\rL}{{\mathrm{L}}}
\newcommand{\rA}{{\mathrm{A}}}
\newcommand{\rR}{{\mathrm{R}}}

\newcommand{\rd}{{\mathrm{d}}}
\newcommand{\ri}{{\mathrm{i}}}

\newcommand{\CM}{{\mathrm{CM}}}
\newcommand{\CR}{{\mathrm{CR}}}

\newcommand{\PL}{\mathbb{P}_{\rm L}}
\newcommand{\PR}{\mathbb{P}_{\rm R}}

\newcommand{\blue}[1]{{\color{black}#1}}

\newcommand{\imag}{\mathfrak{Im}}
\newcommand{\sgn}{{\rm sgn}}

\newcommand \be{\begin{equation}}
\newcommand \ee{\end{equation}}
\newcommand \bea{\begin{eqnarray}}
\newcommand \eea{\end{eqnarray}}
\renewcommand{\k}{{\bf k}}
\newcommand{\p}{{\bf p}}
\newcommand{\BE}{f_{\rm B}}
\newcommand{\FD}{f_{\rm F}}
\newcommand{\Li}{\textrm{Li}}

\title{CP violation at finite temperature}

\author[a]{K\'aroly Seller,}
\author[b]{Zsolt Sz{\'e}p}
\author[a,b]{and Zolt\'an Tr{\'o}cs\'anyi}

\affiliation[a]{Institute for Theoretical Physics, ELTE E{\"o}tv{\"o}s Lor\'and University,\\
P\'azm\'any P{\'e}ter s{\'e}t\'any 1/A, H-1117 Budapest, Hungary}
\affiliation[b]{HUN-REN-ELTE Theoretical Physics Research Group,\\
P\'azm\'any P{\'e}ter s{\'e}t\'any 1/A, H-1117 Budapest, Hungary}

\emailAdd{karoly.seller@ttk.elte.hu}
\emailAdd{szepzs@achilles.elte.hu}
\emailAdd{zoltan.trocsanyi@cern.ch}

\abstract{
We present a comprehensive study of the finite temperature CP-asymmetry factor needed in the semi-classical treatment of leptogenesis originating from Majorana fermion decays into a lepton and a scalar particle.
The imaginary part of the relevant one-loop integrals are evaluated using both the real time and the imaginary time formalisms of thermal quantum field theory.
In the former we consider the retarded-advanced approach as well as the original thermal cutting method developed by Kobes and Semenoff.
Specific care is directed towards showing the consistency between the various approaches.
We show that the final physical result of the calculation is linear in the statistical factors and is consistent with what is obtained in the Kadanoff-Baym approach.
We also present analytic expressions for the full CP-asymmetry factor in the form of well-behaved triple-integrals, and provide numerical benchmark predictions in a specific particle physics model.
}

\begin{document}
\allowdisplaybreaks
\maketitle

\section{Introduction}

Heavy Majorana neutrino decay has an important role in the CP-violating asymmetry factor that determines the lepton number asymmetry in the early Universe \cite{Fukugita:1986hr}.
The original vacuum calculation was done at energy scales where the Higgs scalar and the leptons are massless \cite{Liu:1993tg}. 
That result was later extended to include the effect of the one-loop neutrino self-energy that can be the dominant source of the asymmetry in case of nearly mass-degenerate neutrinos \cite{Pilaftsis:1997jf}. 
The contribution of both the self-energy and the vertex diagram to the CP-asymmetry was computed at finite temperature in ref.~\cite{Covi:1997dr}, ignoring the thermal mass of the scalar and the leptons.
These latter were included in the influential article ref.~\cite{Giudice:2003jh} that attempted a complete treatment of thermal leptogenesis in the standard model (SM) and its minimal supersymmetric extension based on Boltzmann equations for the particle numbers involved. 

The purpose of the present paper is to further improve the computation of ref.~\cite{Giudice:2003jh} in two aspects. 
On the one hand the computation in ref.~\cite{Giudice:2003jh} neglected two out of the three cuts that appear in the vertex diagram. 
In that paper, the omission of these cuts was justified by pointing out that they have exponentially suppressed contributions in $m_{N}/T\gg 1$, $m_N$ being the mass of the decaying heavy neutrino. 
However, the additional contribution of these cuts may be relevant for low-scale leptogenesis (i.e.,~when $m_{N}\sim T$). 
On the other hand, the expressions for the physically relevant thermal self-energy and vertex function in refs.~\cite{Covi:1997dr,Giudice:2003jh} are quadratic in statistical factors, instead of being linear as one would expect at one-loop level, hence they have to be corrected.

\blue{The origin of this problem was already elucidated within a scalar toy model. 
First it was observed~\cite{Garny:2009qn,Garny:2009rv} that the CP-asymmetry parameter is indeed linear in the statistical factors when derived from the Kadanoff-Baym equations of the non-equilibrium quantum field theory by applying a gradient expansion and a quasiparticle approximation.  
Then, it was demonstrated that the use of the \emph{causal $n$-point function}, introduced by Kobes, leads in the conventional equilibrium approach to a CP-asymmetry parameter linear in the statistical factors, in agreement with the requirement of the first-principle derivation mentioned above~\cite{Garny:2010nj}.
Concerning the vertex function, the problem with the statistical factors in ref.~\cite{Giudice:2003jh} is due to} the incorrect identification of the imaginary part of the vertex function component $\Gamma_{111}$ as the physical quantity.
In fact the causal vertex function related to the physical decay process is a non-trivial combination of independent components of $\Gamma_{abc}$ (with vertex types $a,b,c=1,2)$ obtained in the real time formalism (RTF) of thermal field theory \cite{Kobes:1990kr,Aurenche:1991hi}.
We shall revisit these aspects related to the calculation of the CP-asymmetry factor by working in a phenomenological model involving right-handed neutrinos, a scalar doublet and a left-handed lepton doublet.

Our present purpose is motivated by our interest in low-scale leptogenesis occurring in the superweak extension of the SM (SWSM) \cite{Trocsanyi:2018bkm}, where the effects of the issues mentioned are unclear a priori. 
We shall find that these effects are not negligible. 
Thus,  we shall present the calculation using both the RTF (following ref.~\cite{Giudice:2003jh}) and the imaginary time formalism (ITF), and find agreement between the compared predictions of the two approaches.
The ITF was formerly not employed in this context. 
Due to the intricacies of the RTF and because analytic continuation of results obtained in the ITF might be disconcerting for some practitioners in the field, we devote a few appendices to review the relevant literature on the formalisms and to present the details of the calculation.

The article is structured as follows.
We begin in section \ref{sec:CPviolation} with an overview of the definition of the CP-asymmetry factor both at zero and at finite temperature.
At zero temperature the analytic structure of the result is presented for non-zero masses. 
At finite temperature cutting rules are introduced to evaluate the imaginary part of the one-loop amplitudes and the thermal rate of CP violation is finally given in an integral form.
In section \ref{sec:CalculationCPatFiniteT} we focus purely on the finite temperature part of the CP-asymmetry factor.
After a short introduction to the self-energy diagram at finite temperature, we detail the calculation of the vertex diagram involving all three cuts, and derive the form of the physically relevant vertex function.
We finish this section by evaluating the vertex diagram in the ITF and compare results with those obtained from the RTF.
In section~\ref{sec:results} the formulae for the thermal CP-asymmetry factor are given and numerical results are shown in a particular particle physics model. Finally, a summary is given in section~\ref{sec:summary}.
Identification and evaluation of the causal vertex function that is the physically relevant quantity contributing to the CP-asymmetry factor constitute the main results of this article.
Additionally, there are 8 appendices detailing specific calculations or methods used in the main text.

Throughout the paper we use natural units ($c=\hbar=k_{\rm B} = 1$) appropriate in finite temperature field theory.
Additionally, in all applicable cases we use the Peskin-Schroeder convention for the Minkowski metric and the Feynman rules \cite{Peskin:1995ev}.

\section{CP violating asymmetry}
\label{sec:CPviolation}

The interaction of the right-handed neutrinos ($N$) with the scalar ($\phi$) and the left-handed lepton doublets ($L$) is given by the Lagrangian
\be
\mathcal{L}_N=-\bar L Y \tilde \phi N - \bar N Y^\dagger \tilde \phi^\dagger L
= -\epsilon_{ab} \bar L^a_\alpha Y_{\alpha i} \phi^{*,b} N^i - \epsilon_{ab}^{\rm T}  \bar N^i \phi^a (Y^\dagger)_{i\alpha} L^b_\alpha\,,  
\ee
where $\tilde \phi=\ri \sigma_2 \phi^*$, $Y$ is the complex Yukawa matrix, $a,b=1,2$ denotes components of the doublet, $\epsilon_{ab}=-\epsilon_{ba}$ is a totally antisymmetric matrix with $\epsilon_{12}=1$, while $\alpha$ is the family index of the SM, and $i$ goes over the number of neutrinos involved.
The masses of the particles involved are denoted $m_{N_i}$ for the right-handed neutrinos, $m_\phi$ for the scalars, and $m_L$ for the leptons.
We do not differentiate between the masses of the degrees of freedom within the scalar and lepton doublets as we are only interested in leptogenesis scenarios occurring before the electroweak symmetry breaking, where they are equal.

The complex Yukawa matrix leads to a difference between the probabilities of processes creating leptons and anti-leptons. 
This difference leads to a non-vanishing value of the number density of leptons minus anti-leptons $n_{\Delta L} = n_L-n_{\bar L}$.
In the semi-classical approach to leptogenesis based on the Boltzmann equation, the evolution of the comoving number density associated to the lepton asymmetry, $\mathscr{Y}_{\Delta L}=n_{\Delta L}/s$ where $s$ is the entropy density, is determined by the differential equation \cite{Giudice:2003jh,Frossard:2012pc} (see also e.g.~section 11.3 of ref.~\cite{Xing:2011zza}):
\begin{equation}
    \label{eq:Boltzmann-equation}
    s H z \frac{\rd \mathscr{Y}_{\Delta L}}{\rd z} = \gamma_{\rm D}\Big[\epsilon_i\Big(\frac{\mathscr{Y}_{N_i}}{\mathscr{Y}_{N_i}^{\rm eq}}-1\Big)-\frac{\mathscr{Y}_{\Delta L}}{\mathscr{Y}_L^{\rm eq}}\Big] - \frac{\mathscr{Y}_{\Delta L}}{\mathscr{Y}_L^{\rm eq}}\gamma_{\rm scatterings}\,.
\end{equation}
Here $H$ is the Hubble rate, $z\propto 1/T$ is the inverse temperature up to an arbitrary constant factor, and $\mathscr{Y}_{N_i,L}^{\rm eq}$ are equilibrium comoving number densities of the right-handed neutrinos and leptons respectively.
Moreover, $\gamma_{\rm D}$ is the (tree-level) thermal decay rate of $N_i\to L+\phi$, $\gamma_{\rm scatterings}$ collect the thermal scattering rates, and most importantly for the present work, $\epsilon_i$ is the measure of CP violation in the $N$-decay.
From eq.~\eqref{eq:Boltzmann-equation} it follows that if initially the Universe is fully symmetric, i.e. $\mathscr{Y}_{\Delta L}(z_{\rm init.})=0$, then unless $\epsilon_i\neq 0$ the asymmetry will remain vanishing.
This is ensured by the appearance of the term $\gamma_{\rm D}\mathscr{Y}_{\Delta L}/\mathscr{Y}_L^{\rm eq}$ which is due to a subtraction scheme\blue{, called real intermediate state subtraction (RIS),} used in $|\Delta L|=2$ scattering processes to avoid double counting.

\blue{The RIS procedure was first implemented in refs.~\cite{Kolb:1979qa,Kolb:1979ui} based on a physical argument invoking CPT invariance and the unitarity of the scattering matrix (see in particular sec.~2.3.1 of the former) in order to avoid generation of spurious CP asymmetries in thermal equilibrium, that would have contradicted one of Sakharov's conditions for baryogenesis.
This structure of the Boltzmann equations introduced by Kolb and Wolfram was later obtained in refs.~\cite{Garny:2009rv,Garny:2009qn,Beneke:2010wd} from Kadanoff-Baym equations via a first principle derivation.
In the non-equilibrium framework the RIS subtraction is not done by hand, rather it arises naturally as a result of the consistency of the formalism, as discussed in ref.~\cite{Beneke:2010wd}.
We point to ref.~\cite{Ala-Mattinen:2023rbm} for a recent review of the RIS scheme, while
for a comparison between numerical solutions of the Kadanoff-Baym and Boltzmann equations, see ref.~\cite{Anisimov:2010dk}.
In the remainder of this section we shall formulate the measure of CP-violation $\epsilon_i$ first at zero temperature then at finite temperature.}

\subsection{Zero temperature case}
The CP-violating asymmetry arising from the decay of neutrino through the direct process $N_i\to L_\alpha^a + \phi^b$ and its CP conjugate, $N_i\to \bar L_\alpha^a + \bar \phi^b$, is defined as
\be
\label{Eq:epsT0}
\epsilon_i = \frac{\displaystyle\sum_{a,b,\alpha}\Big[\Gamma(N_i\to L_\alpha^a + \phi^b) - \Gamma(N_i\to \bar L_\alpha^a + \bar \phi^b)\Big]}{\displaystyle\sum_{a,b,\alpha}\Big[\Gamma(N_i\to L_\alpha^a + \phi^b) + \Gamma(N_i\to \bar L_\alpha^a + \bar \phi^b)\Big]}\,.
\ee
The tree-level decay rate, which we denote with a superscript $(0)$, is given by
\begin{equation}
\label{Eq:decayT0_def}
\Gamma^{(0)}(N_i\to L_\alpha^a + \phi^b) = \frac{1}{2m_{N_i}}\int\!\rd\Pi_\phi\rd\Pi_L\,(2\pi)^4\delta(P_N-P_\phi-P_L)\big\langle\big|\mathcal{M}_{\alpha i}^{ab\,(0)}\big|^2\big\rangle,
\end{equation}
where $\rd\Pi_x = \rd^3p_x/((2\pi)^3\, 2E_x)$ is the Lorentz-invariant phase space measure and $\langle\,.\,\rangle$ denotes summation over spin indices. 
We suppressed the dimension of the $\delta$ distribution that will be kept implicit throughout the paper.
An elementary calculation shows that the decay rates of the direct and CP-conjugates processes are equal at lowest order in perturbation theory.
Therefore, $\epsilon_i$ vanishes at tree-level, and to obtain prediction for $\epsilon_i$, the decay rates have to be computed at one-loop level, where the interference between tree- and one-loop decay amplitudes gives a non-vanishing result. 
Neglecting terms of $\mathcal{O}(Y^6)$, we have
\bea
\label{Eq:epsT0_LO}
\epsilon_i(T=0) = \frac{\displaystyle \frac{1}{2m_{N_i}} \int\rd\Pi_\phi \rd\Pi_L~(2\pi)^4\delta(P_N-P_\phi-P_L)\,\epsilon_{\mathcal{M}_i}
\big|M_{i,+}^{(0)}\big|^2
}{\displaystyle\sum_{a,b,\alpha} \Big[\Gamma^{(0)}(N_i\to L_\alpha^a + \phi^b) + \Gamma^{(0)}(N_i\to \bar L_\alpha^a + \bar \phi^b)\Big]}\,,~~
\eea
where the amplitude-level CP-asymmetry factor $\epsilon_{\mathcal{M}_i}$ is defined using the tree- and one-loop amplitudes of the decay process as
\be
\label{Eq:eps_M}
\epsilon_{\mathcal{M}_i} = \frac{\big|M_{i,-}^{[1]}\big|^2}{\big|M_{i,+}^{(0)}\big|^2},\qquad
\big|M_{i,\pm}^{[n]}\big|^2=\sum_{a,b,\alpha}\left[\big\langle\big|{\mathcal{M}}^{ab\,[n]}_{\alpha i}\big|^2\big\rangle \pm \big\langle\big|\overline {\mathcal{M}}^{ab\,[n]}_{\alpha i}\big|^2\big\rangle\right].
\ee
In our notation the superscript $[1]$ means that the squared matrix element is computed at one-loop in perturbation theory, i.e., $M^{[1]}=M^{(0)}+M^{(1)}$.
Introducing $\mathcal{K}=Y^\dagger Y$ with the property $\mathcal{K}^\dagger=\mathcal{K}$, one has
$\big|M_{i,+}^{(0)}\big|^2=8 \mathcal{K}_{ii} P_L\, \cdot P_N$ (no summation over $i$).

\begin{figure}[t]
    \centering
    \includegraphics[width=0.95\linewidth]{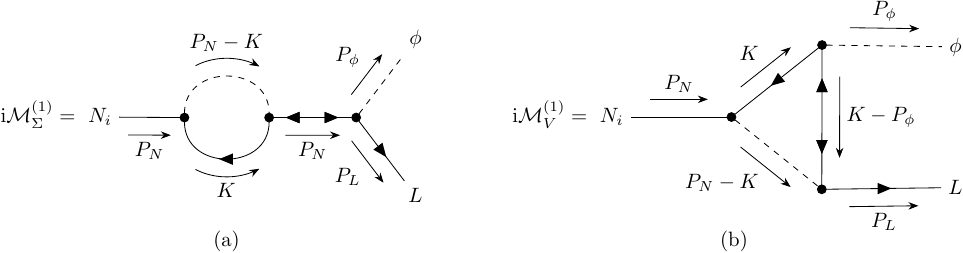}
    \caption{Self-energy and vertex contributions to the one-loop decay amplitude. The internal line with two arrows represent the intermediate Majorana neutrino.}
    \label{fig:1loop_decays}
\end{figure}

The one-loop level Feynman diagrams contributing to the decay amplitude are shown in figure~\ref{fig:1loop_decays}. 
A straightforward calculation, following ref.~\cite{Xing:2011zza} and using the formulae in appendix~\ref{sec:app:vacuum_amplitudes}, gives at lowest order in the coupling:
\be
\label{Eq:M2mbM2}
\begin{aligned}[b]
\big|M_{i,-}^{[1]}\big|^2
=-16 \sum_{j\ne i} \bigg\{&
m_{N_i} m_{N_j} \imag \big[(\mathcal{K}_{ij})^2\big]
\bigg[\imag\,\mathcal{I}^V_{L\phi N_j}(P_N,P_L)
\\ & + P_N\cdot P_L\bigg(1+\frac{m_L^2-m_\phi^2}{P_N^2}\bigg)\,
\imag\bigg(
\frac{\mathcal{B}(P_N;m_L,m_\phi)}{P_N^2-m^2_{N_j}+\ri\varepsilon}
\bigg)
\bigg]
\bigg\}\,,
\end{aligned}
\ee
where $\big|M_{i,-}^{[1]}\big|^2$ was introduced in eq.~\eqref{Eq:eps_M} and the integrals appearing above are defined as
\bea
\label{Eq:int_IV}
\mathcal{I}^V_{L\phi N_j}(P_N,P_L) &=& -\int_K K\cdot P_L \, D(K-P_N+P_L,m_{N_j}) D(K-P_N,m_\phi) D(K,m_L)\,,\quad\\
\label{Eq:bubble}
\mathcal{B}(P;m_1,m_2) &=& -\ri\,\int_K D(K,m_1) D(P-K,m_2)\,.
\eea
Here $D(K,m)=\ri/(K^2-m^2+\ri\,\varepsilon)$ is the scalar Feynman propagator and $\int_K=\int\!\frac{\rd^4 K}{(2\pi)^4}$ with $K=(k_0, {\k})$ being the four-momentum in the Minkowski metric.
In eq.~\eqref{Eq:int_IV} we suppressed the masses in the argument of the vertex integral in favor of ordered indices to shorten the notation (cf.~eq.~\eqref{Eq:SV_def}).
In writing eq.~\eqref{Eq:M2mbM2}, we dropped a term $\propto P^2_N \imag (\mathcal{K}_{ij} \mathcal{K}_{ji})$ from the contribution of the self-energy, since it vanishes due to the property $\mathcal{K}_{ij}={\mathcal{K}}^*_{ji}$. 
Also, we could leave out the term $j=i$ from the sum, as $\mathcal{K}_{ii}$ is real.   

Eq.~\eqref{Eq:M2mbM2} represents the generalization to finite scalar and lepton masses of the vacuum vertex and self-energy contributions summarized in chapter~11.3 of ref.~\cite{Xing:2011zza}.
For a recent vacuum calculation of the vertex contribution to the thermal
CP-asymmetry factor in the massive case see appendix~C of ref.~\cite{Bhattacharya:2024ohh}.
Using $P_\phi=P_N-P_L$ and the identity
\[
K\cdot P_L = \frac{1}{2}\left[(K-P_\phi)^2 - m_{N_j}^2 - \big((K-P_N)^2 - m_\phi^2\big) + m_{N_j}^2 + P_N^2 - P_\phi^2 \right],
\]
the integral $\mathcal{I}^V_{L\phi N_j}$ defined in eq.~\eqref{Eq:int_IV} can be written in terms of basic one-loop scalar integrals encountered in the usual Passarino-Veltman reduction (see e.g.~ref.~\cite{Ellis:2007qk})
\bea
\mathcal{I}^V_{L\phi N_j}(P_N,P_L) &=& \frac{1}{2}(m_{N_j}^2-m_\phi^2+P_N^2-P_\phi^2) \mathcal{C}_{L\phi N_j}(P_N,P_L) 
\nonumber \\
&&+\frac{1}{2} \Big[\mathcal{B}(P_N;m_L,m_\phi) - \mathcal{B}(P_\phi;m_L,m_{N_j})\Big],
\label{Eq:V_reduced}
\eea
where the scalar triangle integral $\mathcal{C}_{L\phi N_j}(P_N,P_L)$ has the same form as the integral in eq.~\eqref{Eq:int_IV}, but without the factor $K\cdot P_L $ in the integrand (see eq.~\eqref{Eq:SV_def} for its definition).

We mention that for on-shell external momenta and depending on the relation between the masses, the bubble integral eq.~\eqref{Eq:bubble} can also contribute to the imaginary part of the triangle graph besides the contribution from the scalar triangle integral. 
Namely, when $m^2_\phi>(m_L+m_{N_j})^2$ then $\imag\,\mathcal{B}(m_\phi;m_L,m_{N_j})$ contributes to $\imag\,\mathcal{I}^V_{L\phi N_j}$ obtained by cutting its two fermionic lines, while $\imag\,\mathcal{B}(m_{N_i};m_L,m_\phi)$ contributes to $\imag\,\mathcal{I}^V_{L\phi N_j}$ obtained by cutting the scalar and lepton lines whenever $m_{N_i}^2>(m_\phi+m_L)^2$ is satisfied. 

At finite temperature Lorentz invariance is broken, thus eq.~\eqref{Eq:V_reduced} does not hold, that is, simply using the finite temperature bubble and scalar triangle integrals in the formula is not sufficient (see also eq.~\eqref{Eq:intSigmaKPL} for the self-energy contribution). 
For this reason in section~\ref{sec:CalculationCPatFiniteT} we compute the amplitude-level CP-asymmetry factor differently.

\subsection{Finite temperature case}
\label{sec:CPviolatingasymmetry_FiniteTemperatureCase}

At finite temperature the CP-asymmetry factor is defined by replacing the vacuum decay rates in eq.~\eqref{Eq:epsT0} with the thermal interaction rates
\be
\gamma_{N_i\to L_\alpha^a + \phi^b} =\! \int\! \rd\Pi_N \rd\Pi_\phi \rd\Pi_L\,(2\pi)^4\delta(P_N-P_\phi-P_L)\big\langle\big|\mathcal{M}_{\alpha i}^{ab}\big|^2\big\rangle\FD(E_N)[1+\BE(E_\phi)][1-\FD(E_L)].
\ee
Here $\mathcal{M}^{ab}_{\alpha i}$ is computed using finite temperature Feynman rules, which only matters from one-loop level, and $f_{\rm B/F}(E)$ denotes the Bose-Einstein or Fermi-Dirac statistical factors at vanishing chemical potential,
\begin{equation}
    \label{eq:fBEandfFDdefinitions}
    f_{{\rm B}/{\rm F}}(E) = \frac{1}{e^{\beta E}\mp 1}\,,
\end{equation}
where $\beta=T^{-1}$ is the inverse temperature.
The finite temperature counterpart of eq.~\eqref{Eq:epsT0_LO} is given at leading order in the coupling by
\be
\label{Eq:epsT_LO}
\epsilon_i=\frac{\displaystyle \int \rd\Pi_N \rd\Pi_\phi \rd\Pi_L\,(2\pi)^4\delta(P_N-P_\phi-P_L)\,\epsilon_{\mathcal{M}_i} \big|M_{i,+}^{(0)}\big|^2 \FD(E_N)[1+\BE(E_\phi)][1-\FD(E_L)]}{\displaystyle\sum_{a,b,\alpha} \Big[\gamma^{(0)}_{N_i\to L_\alpha^a + \phi^b} + \gamma^{(0)}_{N_i\to \bar L_\alpha^a + \bar \phi^b}\Big]}.
\ee

At finite temperature, especially in cosmological applications, the calculation is usually performed in the rest frame of the plasma, called the \emph{cosmic rest frame} (CR).
Moreover, it is also conventional to add thermal corrections to the vacuum masses and use the so-called \emph{thermal masses} \blue{in the Boltzmann equations and thus in the expression of the CP-asymmetry parameter.
When the quantum Boltzmann equations are derived from the quantum kinetic Kadanoff-Baym equations as e.g. in ref.~\cite{Frossard:2012pc}, the thermal mass appears as a result of using the quasiparticle approximation to the spectral function (see eqs.~(57) and (65) in ref.~\cite{Frossard:2012pc} for leptons and scalars, respectively).
App.~\ref{sec:app:thermal-mass} explains how the introduction of the thermal mass can be achieved in the present case.
Following ref.~\cite{Espinosa:1992kf}, $\overline{m}^2$ will denote the thermal mass squared,
except in cases where it is not essential to specify the type of mass used.

In the context of leptogenesis, the inclusion of the thermal masses provides an easy way to understand how changes in occupation numbers due to interactions impact particle numbers~\cite{Davidson:1994gn}.}
Their role in CP-violation was investigated in ref.~\cite{Frossard:2012pc}, where it was shown that while the decay and washout reaction densities are enhanced by thermal effects, the inclusion of the thermal masses reduces this enhancement and even turns it into suppression at high temperature.
Further relativistic effects on lepton-number conserving and violating rates due to more complicated dispersion relations were investigated in ref.~\cite{Garbrecht:2019zaa}.  In this reference the hard thermal loop resummation scheme was considered for the leptons, which included in their spectral function the effect of $2\leftrightarrow 2$ $t$-channel scattering processes mediated by virtual leptons.

In the CR frame the decaying $N_i$ has energy $E_N\geq \overline{m}_{N_i}$ and we choose the direction of the momentum as $\p_N\parallel \hat{x}$.
When one calculates the thermal decay rate one has to perform an appropriate boost from the center-of-mass (CM) to the CR frame,
\begin{equation}
\label{eq:CRenergyDefinitions}
\begin{gathered}
    E^\CR_L
    =\gamma E_L^\CM+v \gamma p_{x}^\CM \equiv E_L^\CR(E_N,\theta)\,,\\
    E^\CR_\phi=\gamma E_\phi^\CM-v \gamma p^\CM_{x}  \equiv E^\CR_\phi(E_N,\theta)\,,\\
    E_{L/\phi}^\CM = \frac{\overline{m}_{L/\phi}^2-\overline{m}_{\phi/L}^2+\overline{m}_{N_i}^2}{2\overline{m}_{N_i}}\,, \qquad p_{x}^\CM = \sqrt{\big(E_L^\CM\big)^2-\overline{m}_L^2}\,\cos\theta\,,
\end{gathered}
\end{equation}
where $v=p_N/E_N$ is the velocity of the decaying particle, $\gamma=E_N/\overline{m}_{N_i}$ is the Lorentz factor, and $\theta$ is the decay angle in the CM frame.

Since $E_{L/\phi}^\CR$ depends only on the energy of the decaying particle and the decay angle $\theta$, in eq.~\eqref{Eq:epsT_LO} one can use the identity
\begin{equation}
    \FD(E_N)[1+\BE(E_\phi)][1-\FD(E_L)]=\FD(E_N) \BE(E_\phi) \FD(E_L) e^{\beta E_N}\,,
\end{equation}
with the definition of the tree-level vacuum decay rate in eq.~\eqref{Eq:decayT0_def} to perform the final state integrals (with the exception of the integral over the scattering angle $\theta$) and write
\begin{equation}
\begin{split}
    \label{eq:thermalAvgEpsilonDenominator}
    \gamma^{(0)}_{N_i\to L_\alpha^a + \phi^b} = & \gamma^{(0)}_{N_i\to \bar L_\alpha^a + \bar \phi^b} = \frac{\overline{m}_{N_i}}{4\pi^2}\Gamma^{(0)}(N_i\to L_\alpha^a + \phi^b) \\
    &\times\int_{\overline{m}_{N_i}}^\infty\rd E\sqrt{E^2-\overline{m}_{N_i}^2}\FD(E)e^{\beta E}\int_{-1}^1\!\rd\cos\theta \,\BE\big(E_\phi^\CR\big)\FD\big(E_L^\CR\big)\,.
\end{split}
\end{equation} 
The phase space measure for the neutrino was replaced as $\rd\Pi_N=\sqrt{E^2-\overline{m}^2_{N_i}} \rd E/(2\pi)^2$.
Note that the angular integral can be evaluated analytically, see eq.~\eqref{eq:integral_fBfF_costheta}.

What remains to be done in order to have $\epsilon_i(T)$ is to perform the initial and final state integrals in eq.~\eqref{Eq:epsT_LO}.
When computing the final state integrals we use the relation $\displaystyle \big|M_{i,+}^{(0)}\big|^2=2\sum_{a,b,\alpha}\langle|\mathcal{M}^{ab\,(0)}_{\alpha i}|^2\rangle$ to find the vacuum decay rate and then the numerator of $\epsilon_i$ is
\begin{equation}
\label{eq:thermalAvgEpsilonNumerator}
\begin{split}    
    \sum_{a,b,\alpha}&\Big[\gamma_{N_i\to L_\alpha^a+\phi^b}^{(1)}-\gamma_{N_i\to \bar L_\alpha^a+ \bar \phi^b}^{(1)}\Big]
    =\frac{\overline{m}_{N_i}}{2\pi^2}  \sum_{a,b,\alpha}\Gamma^{(0)}(N_i\to L_\alpha^a + \phi^b) \\
    &\times\int_{\overline{m}_{N_i}}^\infty\!\rd E\,\sqrt{E^2-\overline{m}_{N_i}^2}\FD(E)e^{\beta E}\int_{-1}^1\!\rd\cos\theta\,\epsilon_{\mathcal{M}_i}(E,\cos\theta)\BE\big(E_\phi^\CR\big)\FD\big(E_L^\CR\big).
\end{split}
\end{equation}
We see that the vacuum decay rate cancels in the expression of $\epsilon_i(T)$ in eq.~\eqref{Eq:epsT_LO}, which becomes the thermal average of $\epsilon_{\mathcal{M}_i}$ over the particle distributions. 
The expression of $\epsilon_{\mathcal{M}_i}$ can be read off from eqs.~\eqref{Eq:eps_M} and \eqref{Eq:M2mbM2}. 
An alternative way to obtain $\epsilon_{\mathcal{M}_i}$ used in ref.~\cite{Giudice:2003jh} is presented in the next subsection.

\subsection{Calculating the interference using cutting rules}

Let the relevant amplitude corresponding to the CP-violating decay process (obtained through standard Feynman rules) be given by $\ri\mathcal{M}$.
The CP-conjugate amplitude $\overline{\mathcal{M}}$ has the same structure as $\mathcal{M}$ but with all couplings complex conjugated.
Neglecting indices, a schematic representation of the amplitude is
\begin{equation}
    \mathcal{M}=\sum_{\ell} g_\ell \mathcal{I}^{(\ell)} \qquad\to\qquad \overline{\mathcal{M}}=\sum_{\ell} g_\ell^*\mathcal{I}^{(\ell)}\,.
\end{equation}
Here $g_\ell = \mathcal{O}(g^{2\ell+1})$ is the overall coupling that appears at $\ell$-loop and $\mathcal{I}^{(\ell)}$ is the $\ell$-loop amplitude stripped of all couplings.
In this notation the amplitude-level CP-asymmetry factor introduced in eq.~\eqref{Eq:eps_M} at leading order in perturbation theory reads as
\begin{equation}
    \label{eq:epsilonA}
    \epsilon_{\mathcal{M}} =  -2\frac{\imag(g_0^* g_1)}{|g_0|^2}\sum_{\rm spins}\frac{\imag\left(\mathcal{I}^{(0)*}\mathcal{I}^{(1)}\right)}{|\mathcal{I}^{(0)}|^{2}}\,.
\end{equation}

The imaginary part of amplitudes can be decomposed using the finite temperature cutting rules \cite{Kobes:1985kc,Kobes:1986za,Gelis:1997zv}, see appendix~\ref{sec:app:FTcuttingRules} for our notations and specific rules.
The cutting rules presented apply to the $n$-point Green's function $G_n$.
However, from field theory we know that the relation between the two is simply $G_n\equiv\ri\mathcal{M}$.
Using the one-loop imaginary part formula in eq.~\eqref{eq:nPoint1loopImag}, one finds for the imaginary part of the one-loop amplitude:
\begin{equation}
    \label{eq:ImagAmplitude}
    -2\imag \,\ri G_n^{(1)}\equiv 2\imag\mathcal{M}^{(1)}=\sum_{\rm circlings}G_n^{(1)}\,,
\end{equation}
where the right-hand side is obtained through cutting rules, and the sum refers to all possible circlings of vertices where at least one circled and one uncircled vertex exists.

For the sterile neutrino decay into a scalar field and a lepton the tree-level amplitude is purely given as the spinor product $\mathcal{I}^{(0)} = -\Bar{u}_L\PR u_{N_i}$.
We can also write the one-loop amplitude as 
\begin{equation}
\label{eq:strippedAmplitudeDef}
    \mathcal{I}^{(1)}\big(\{P_i\}\big) = \int_K\mathcal{S}^{(1)}\big(\{P_i\},K\big)\,\widetilde{\mathcal{M}}^{(1)}\big(\{P_i\},K\big)\,,    
\end{equation}
with the spinor chain contained in $\mathcal{S}^{(1)}$ and the remaining part involving the propagator denominators in $\widetilde{\mathcal{M}}^{(1)}$.
We call $\widetilde{\mathcal{M}}$ the \emph{stripped amplitude} as it is $\mathcal{M}$ with couplings, spinors, and the loop momentum integration stripped off\footnote{The factors of $-\ri$ coming from the Feynman rules for the vertex are taken into account in $\widetilde{\mathcal{M}}$.}.
Here we used a notation for the arguments that differentiates the independent external momenta given in parentheses as $\{P_1,P_2,\dots\}$ from the loop momentum $K$.
The spinor chains of $\mathcal{I}^{(0)*}$ and $\mathcal{S}^{(1)}$ are contracted using the spin summation and we find \cite{Gluza:1991wj}
\begin{equation}
\label{eq:DiracTrace}
\begin{split}
    \mathcal{T}(K)   
    &= \sum_{\rm spins}\left(-\Bar{u}_L \PR u_{N_i}\right)^*\big[\Bar{u}_L \PR(\slashed{Q}+\overline{m}_{N_j})\PR[(\slashed{K}+\overline{m}_L)]\PL \Bar{u}_{N_i}\big] \\
    &= -2\overline{m}_{N_i}\overline{m}_{N_j}(K\cdot P_L)\,,
\end{split}
\end{equation}
where $Q$ is momentum of the intermediate neutrino that differs between the self-energy ($Q=P_N$) and the vertex diagram ($Q=K-P_\phi$).
Using $\PR\slashed{Q}\PR = 0$ we find that the momentum of the neutrino does not affect the final result and the spinor traces are equal for either case.
Since $\mathcal{T}(K)$ is a real number, it can be taken out of the imaginary part of the interference in eq.~\eqref{eq:epsilonA}.
Using eq.~\eqref{eq:ImagAmplitude} finally one has
\begin{equation}
    \label{eq:epsilonAFinalDef}
    \imag\bigg(\sum_{\rm spins}\mathcal{I}^{(0)*}\mathcal{I}^{(1)}\bigg) = \int_K\mathcal{T}(K)\imag\,\widetilde{\mathcal{M}}^{(1)} = \frac{1}{2}\int_K\mathcal{T}(K)\sum_{\rm circlings}\ri\widetilde{\mathcal{M}}^{(1)}\,.
\end{equation}
The calculation of the CP-asymmetry factor is thus reduced to the calculation of finite temperature cuts involving the one-loop self-energy and vertex diagrams.
Putting everything together, with $\langle|\mathcal{I}^{(0)}|^2\rangle=2P_N\cdot P_L$ the amplitude-level CP-asymmetry factor is
\begin{align}
\label{eq:CPasymmetryFactorImagAmplitudeDef}
    \epsilon_{\mathcal{M}} 
    &= 2 G\frac{\overline{m}_{N_i}\overline{m}_{N_j}}{P_N\cdot P_L}\int_K(K\cdot P_L)\imag\widetilde{\mathcal{M}}^{(1)}\,.
\end{align}
where the overall coupling is ($g_0=Y_{\alpha i}$ and $g_1=Y_{\alpha j} \mathcal{K}_{ij}$):
\begin{equation}
\label{eq:couplingCorrespondence}
    G = \frac{\imag(g_0^*g_1)}{|g_0^2|} = \frac{\imag\big[(\mathcal{K}_{ij})^2\big]}{\mathcal{K}_{ii}}\,.
\end{equation}
We present the calculation of the cuts in the next section.

\section{Calculation of the CP-asymmetry factor at finite temperature}
\label{sec:CalculationCPatFiniteT}

At one loop there are two non-vanishing Feynman diagram contributions to the CP asymmetry produced by the decay of a Majorana neutrino $N$ decaying into a scalar field $\phi$ and a lepton $L$: (i) the neutrino self-energy diagram and (ii) the vertex diagram, shown in figure~\ref{fig:1loop_decays}.
Both diagrams had been considered in a thermal setting in ref.~\cite{Giudice:2003jh}.
In the following two subsections we focus separately on the self-energy diagram and then the full vertex contribution.
The former is discussed for completeness, while the latter involves the calculation of all cut diagrams and their physical contributions that have been absent in the literature that uses equilibrium formulation.

\subsection{CP violation in the self-energy diagram}

In the one-loop neutrino decay, shown as graph (a) in figure~\ref{fig:1loop_decays}, the imaginary part of the amplitude originates from the sterile neutrino self-energy.
The one-loop self-energy is given by the following bubble diagram:
\begin{equation}
\label{eq:SigmaDefBubble}
\begin{aligned}
    \includegraphics{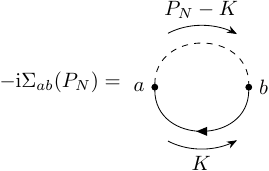}
\end{aligned}
\end{equation}
where $a,b=1,2$ are the closed-time-path (CTP) indices of the RTF in the Kobes-Semenoff (KS) formalism \cite{Kobes:1985kc,Kobes:1986za} and we suppressed the dependence on the masses of the particles in the loop.

The thermal self-energy is a well-known function \cite{Weldon:1983jn,Kobes:1985kc,Kobes:1990kr} and here we only present a short derivation using the RTF and cutting rules (see also in chapter 3.4 of ref.~\cite{Bellac:2011kqa}).
Application of the Feynman rules set up in appendix~\ref{sec:app:FTcuttingRules} leads to a total of 8 distinct circled diagrams (or 4 cut diagrams).
These diagrams are not all independent.
We define the only topologically possible cut, i.e.,~where both propagators in the loop are on-shell, as shown in eq.~\eqref{eq:SelfEnergyABcutDef}.
Note that the self-energy, as defined in eq.~\eqref{eq:SigmaDefBubble}, is equal to an imaginary unit times the two-point function: $\Sigma_{ab} = \ri G^{ab}_{2}$, i.e.,~it is the function that appears in eq.~\eqref{eq:ImagAmplitude}.
It follows that the imaginary part of the self-energy is (see also section 5. in ref.~\cite{Kobes:1985kc}):
\begin{equation}
\label{eq:SelfEnergyABcutDef}
\begin{aligned}
    \includegraphics{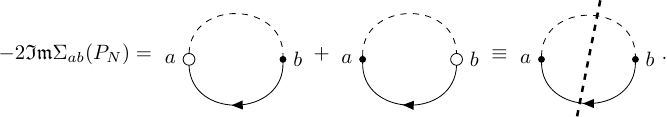}
\end{aligned}
\end{equation}
In eq.~\eqref{eq:epsilonAFinalDef} we found that the spinor structure in the amplitude can be taken into account by a multiplicative factor of $\mathcal{T}(K)$ and for the CP-asymmetry factor in eq.~\eqref{eq:epsilonAFinalDef} we only need the stripped amplitude as defined in eq.~\eqref{eq:strippedAmplitudeDef}.
It is then convenient to define the stripped fermion propagator $\Tilde{S}_{ab}(P,m)$ through
\begin{equation}
    \label{eq:StrippedFermionPropagator}
    S_{ab}(P,m) = (\slashed{P}+m)\Tilde{S}_{ab}(P,m)\,.
\end{equation}
Using momenta as shown in eq.~\eqref{eq:SigmaDefBubble} and the Feynman rules given in appendix~\ref{sec:app:FTcuttingRules}, one finds that the imaginary parts of the components of the stripped thermal self-energy $\widetilde{\Sigma}_{ab}(\{P_N\},K)$ are related as:
\begin{equation}
\begin{split}
    \label{eq:ImagSigmaABrelations}
    \imag\widetilde{\Sigma}_{11}\big(\{P_N\},K\big) &= \imag\widetilde{\Sigma}_{22} \big(\{P_N\},K\big) \\
    =-\frac{1}{2}\Big[\Tilde{S}_-(&K,\overline{m}_L)G_-(P_N-K,\overline{m}_\phi) + \Tilde{S}_+(K,\overline{m}_L) G_+(P_N-K,\overline{m}_\phi)\Big] \\
    \imag\widetilde{\Sigma}_{12}(\{P_N\},K) &= -\imag\widetilde{\Sigma}_{21}(\{P_N\},K) = \Tilde{S}_{12}(K,\overline{m}_\phi)G_{12}(P_N-K,\overline{m}_\phi)\,.
\end{split}
\end{equation}
Using the relations between the off-diagonal components and the positive or negative frequency propagators given in eqs.~\eqref{eq:boson_propagator_offdiag_relation}--\eqref{eq:fermion_propagator_offdiag_relation}, we find that $\imag\Sigma_{11}$ and $\imag\Sigma_{12}$ are not independent,
\begin{equation}
    \label{eq:relation_sigma_11_12}
    \imag\Sigma_{11}(P_N) = \sinh\left(\frac{\beta p_N^0}{2}\right)\imag\Sigma_{12}(P_N)\,.
\end{equation}
As indicated in this equation, the relation holds also for the full self-energy components, not only for the stripped ones.
The entire self-energy matrix has one independent component which is conventionally taken to be $\Sigma_{11}$.

The self-energy function that appears in the Feynman propagator is a specific linear combination of the self-energy components $\Sigma_{ab}$ as explained in appendix~\ref{sec:app:RAformalism}.
In the retarded-advanced (RA) formalism, one uses the retarded and advanced propagators instead of $G_{ab}$.
These two sets of propagators are related to each other via ``rotation" matrices denoted as $\mathbf{U}^{\rm (B/F)}$ and $\mathbf{V}^{\rm (B/F)}$ (for definitions, see appendix~\ref{sec:app:RAformalism}).
The formula for the causal self-energy is defined as $\Sigma(P)\equiv \Sigma_{\rR;\rR}(P;P)$ \cite{Kobes:1985kc}, where the notation on the right-hand side means an incoming and an outgoing retarded propagator (i.e.~flowing in the same temporal direction), both with momentum $P$.
By definition, one then finds \cite{Kobes:1985kc,Kobes:1990kr,Aurenche:1991hi}:
\begin{equation}
\begin{gathered}
    \label{eq:CausalSigmaDef}
    \Sigma(P_N)\equiv\Sigma_{\rR;\rR}(P_N;P_N) = \sum_{a,b=1}^2\mathbf{V}_{\rR a}^{\rm (F)}(P_N) \Sigma_{ab}(P_N)\mathbf{U}^{\rm (F)}_{b\rR}(P_N)
    \\
    \imag \Sigma(P_N) = \coth\left(\frac{\beta p_N^0}{2}\right)\imag \Sigma_{11}(P_N)\,.
\end{gathered}
\end{equation}
As we will see later, the hyperbolic function plays the crucial role of removing terms in $\imag\Sigma_{11}$ that are quadratic in the statistical factors.
Note that eq.~\eqref{eq:CausalSigmaDef} is a general formula for the thermal self-energy of fermions, for bosons one simply has to change the hyperbolic function $\coth\to\tanh$, as usual in thermal field theory.
We also mention that eq.~\eqref{eq:relation_sigma_11_12} follows in the RA formalism from the vanishing of the self-energy component for purely incoming or outgoing lines of the same type, for example for two incoming particles of the retarded type $\Sigma_{\rR\rR}(P_N,-P_N)=0$.

By direct substitution of the thermal propagators one finds for the imaginary part of the stripped self-energy: 
\begin{equation}
\label{eq:ImagSigmaTildeIntegralRepresentation}
\begin{split}
    \imag\widetilde{\Sigma}\big(\{P_N\},K\big) = -2\pi^2{\rm sgn}(k^0)&{\rm sgn}(p_N^0-k^0)\delta(K^2-\overline{m}_L^2)\delta\big((P_N-K)^2-\overline{m}_\phi^2\big)\\ &\times\big(1+\BE(p_N^0-k^0)-\FD(k^0)\big)\,.
\end{split}
\end{equation}
The resulting formula includes two Dirac-delta distributions corresponding to the two on-shell propagators in the loop, it is also at most linear in the statistical factors, and does not vanish in the limit $T\to 0$.

The self-energy contribution to the thermal CP-asymmetry factor defined in eq.~\eqref{eq:CPasymmetryFactorImagAmplitudeDef} is given as\footnote{Note that the fermion self-energy is defined with a minus sign compared to the amplitude, which causes the sign difference between eq.~\eqref{eq:CPasymmetryFactorImagAmplitudeDef} and eq.~\eqref{eq:EpsilonSigmaFinalFormula}.}
\begin{equation}
    \label{eq:EpsilonSigmaFinalFormula}
    \epsilon_{\Sigma_i}(P_{N}) = -4G\sum_{j\neq i}\frac{\overline{m}_{N_i}\overline{m}_{N_j}}{P_{N}\cdot P_L}\int_K \imag\widetilde{\Sigma}\big(\{P_{N}\},K\big) \frac{K\cdot P_L}{P_{N}^2-\overline{m}_{N_j}^2}\,.
\end{equation}
There is an extra factor of 2 because the loop in eq.~\eqref{eq:SigmaDefBubble} can contain either pair of the SU(2)$_\rL$ doublets of $L^a$ and $\phi^b$.
We used the notation introduced in eq.~\eqref{eq:strippedAmplitudeDef} and below for the arguments of the stripped self-energy function. 
Note that the propagator of the $N_j$ Majorana neutrino is resonant when $P_{N}^2\equiv \overline{m}_{N_i}^2\simeq \overline{m}_{N_j}^2$ leading to an amplification of the CP-asymmetry factor \cite{Pilaftsis:1997jf}.

To compare with the vacuum result of eq.~\eqref{Eq:M2mbM2} we take the limit $T\to 0$ of $\epsilon_{\Sigma_i}(P_{N})$.
The statistical factors vanish for $T\to 0$, and since in this limit $\coth(\beta p^0_{N}/2)\to 1$, we also find $\imag\Sigma_{11}(T=0)=\imag\Sigma(T=0)$.
For $p_N^0>0$ one simply has from eq.~\eqref{eq:ImagSigmaTildeIntegralRepresentation}
\begin{equation}
     \lim_{T\to 0}\imag\widetilde{\Sigma}\big(\{P_N\},K\big) =  -\frac{(2\pi)^2}{2}\theta(k^0)\theta(p_N^0-k^0)\delta(K^2-m_L^2)\delta\big((P_N-K)^2-m_\phi^2\big)\,,
\end{equation}
which is exactly what would be obtained from the application of the $T=0$ cutting rules (see, e.g.~chapter 24.1.2 in ref.~\cite{Schwartz:2014sze}).
Additionally,
\begin{equation}
    \lim_{T\to 0}\int_K\imag\widetilde{\Sigma}\big(\{P_N\},K\big) = \imag\mathcal{B}(P_N;m_L,m_\phi)\,,
\end{equation}
where the bubble integral was given in eq.~\eqref{Eq:bubble}.
It is then simple to check that eq.~\eqref{eq:EpsilonSigmaFinalFormula} can be cast into a form that appears in eq.~\eqref{Eq:M2mbM2}.

At finite temperature the calculation presented in appendix~\ref{app:self-energy_and_first_cut_vertex} for the self-energy gives at $P_N^2>(\overline m_\phi+\overline m_L)^2$:
\begin{align}
&\int_K\!\big(K\cdot P_L\big) \imag\widetilde{\Sigma}\big(\{P_{N}\},K\big) =
 \frac{P_N\cdot P_L}{2}\bigg(1+\frac{\overline{m}_L^2-\overline{m}_\phi^2}{P_N^2}\bigg)\,
 \imag\,\mathcal{B}^T(P_N;\overline{m}_L,\overline{m}_\phi) \nonumber \\
 &\quad + \frac{T}{16\pi p_N} \bigg(E_N \frac{\p_N\cdot \p_L}{p_N} - E_L\,p_N\bigg)\Bigg\{
   \frac{\lambda^{\frac{1}{2}}(\overline{m}^2_\phi,\overline{m}^2_L,P_N^2)}{2 P_N^2}
   \ln\frac{\big(1+e^{-\beta\omega^{(1)}_+}\big)\big(1+e^{-\beta\omega^{(1)}_-}\big)}{\big(1-e^{-\beta E^{(1)}_+}\big)\big(1-e^{-\beta E^{(1)}_-}\big)}\nonumber \\
  &\qquad + \frac{T}{p_N}
 \Big[\Li_2\Big(e^{-\beta E^{(1)}_+}\Big)-\Li_2\Big(e^{-\beta E^{(1)}_-}\Big)-\Li_2\Big(-e^{-\beta\omega^{(1)}_+}\Big)+\Li_2\Big(-e^{-\beta\omega^{(1)}_-}\Big)\Big]\Bigg\},
 \label{Eq:intSigmaKPL}
\end{align}
where $p_N=|\p_N|$, $E^{(1)}_\pm=E_N-\omega^{(1)}_\mp$, $\lambda(x,y,z)$ is the K{\"a}ll{\'e}n function given in eq.~\eqref{Eq:Kallen_fv}, and $\Li_2(x)$ is the dilogarithm (see eq.~\eqref{eq:polylog}).
In the given range of $P_N$ the expression of the imaginary part of the bubble integral is 
\begin{equation}
\imag\,\mathcal{B}^T(P_N;\overline m_L,\overline m_\phi) =-\frac{\lambda^{\frac{1}{2}}(\overline m^2_\phi,\overline m^2_L,P_N^2)}{16\pi P_N^2}
-\frac{T}{16\pi p_N} \ln\frac{\big(1+e^{-\beta\omega^{(1)}_+}\big)\big(1-e^{-\beta E^{(1)}_+}\big)}{\big(1+e^{-\beta\omega^{(1)}_-}\big)\big(1-e^{-\beta E^{(1)}_-}\big)}.
\end{equation}

Eq. \eqref{Eq:intSigmaKPL} shows explicitly that in addition to the thermal correction to the bubble integral, one has additional thermal contributions to the vacuum expression of $\epsilon_{\Sigma_i}$ due to the breaking of the Lorentz symmetry occurring at finite temperature. 

The expression in eq.~\eqref{Eq:intSigmaKPL} can be written as $-P_L^\mu L_\mu/(32\pi)$, with the components of $L_\mu$ given in eqs.~(46) and (47) of ref.~\cite{Biondini:2017rpb} (see also eq.~(D13) in ref.~\cite{Frossard:2012pc}). 
This shows that the self-energy contribution to the thermal CP-asymmetry factor calculated from eq.~\eqref{eq:EpsilonSigmaFinalFormula} agrees with the \blue{first term in eq.~(45) of ref.~\cite{Biondini:2017rpb}.
This also shows that the way we treat the thermal mass (see app.~\ref{sec:app:thermal-mass}) corresponds to the quasiparticle approximation used in ref.~\cite{Frossard:2012pc}.}

\subsection{CP violation in the vertex correction \label{sec:vertex}} 

In this subsection we focus on the vertex contribution to the CP-asymmetry factor in the sterile neutrino decay.
Our goal here is to show that the imaginary part of the physical vertex diagram that contributes to $\epsilon_\mathcal{M}$ in eq.~\eqref{eq:CPasymmetryFactorImagAmplitudeDef} is given as a non-trivial combination of the thermal vertex components $\imag\Gamma_{abc}$ in the KS formalism.
Namely, with momenta assigned as shown in graph (b) of figure~\ref{fig:1loop_decays},
\begin{equation}
    \label{eq:ImGammaC_sec_beginning_formula}
    -\imag\Gamma_{\rm C}
    = \coth\left(\frac{\beta p_N^0}{2}\right)\imag\Gamma_{111}^{\sf (cut\,1)} + \tanh\left(\frac{\beta p_\phi^0}{2}\right)\imag\Gamma_{111}^{\sf (cut\,2)} + \coth\left(\frac{\beta p_L^0}{2}\right)\imag\Gamma_{111}^{\sf (cut\,3)}.
\end{equation}
Here $\Gamma_{\rm C}$ is the \emph{causal vertex function} introduced in ref.~\cite{Kobes:1990kr} in the context of the $\phi^3$ scalar theory.
Furthermore, $\Gamma_{111}^{\sf (cut\,i)}$ corresponds to the three cuts of the vertex diagram with only type 1 CTP indices, as explained below.
This symmetric structure of the causal vertex function is our new reformulation of known results that used distinct components of the vertex function instead of the aforementioned cuts of one specific component.
We shall see that using $\imag\Gamma_{111}^{\sf (cut\,i)}$ as the independent degrees of freedom of the vertex function leads to a result that is formally simpler than previously used relations.
Note that the right-hand side of eq.~\eqref{eq:ImGammaC_sec_beginning_formula} is the finite temperature generalization of the imaginary part of the vacuum scalar triangle integral originally introduced in ref.~\cite{tHooft:1978jhc} that is summarized in appendix~\ref{sec:app:vacuumscalarintegral}.

First we compute the thermal vertex function in the KS formalism, in particular relations between the vertex components will be derived and the formulae for the above mentioned three cuts are introduced. Next, the thermal vertex function will be discussed in the RA formalism and $\Gamma_{\rm C}$ will be defined as the $\Gamma_{\rm ARR}$ component of the vertex function.
Finally, the two formalisms are compared and the causal vertex function is expressed in the KS formalism as shown above.

\subsubsection{Vertex function in the KS formalism}
\label{sec:KS_vertex}

The diagram for the one-loop vertex correction for the CP violating neutrino decay into a scalar and a lepton is given as graph (b) in figure~\ref{fig:1loop_decays}.
Based on this diagram, we define the one-loop thermal vertex function as 
\begin{equation}
\begin{aligned}
    \includegraphics{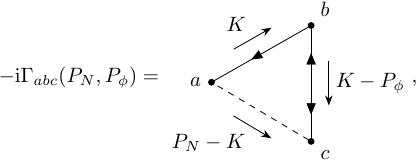}
\end{aligned}
\end{equation}
where $a,b,c=1,2$ are CTP indices.
We use this assignment of the CTP indices and momenta in the rest of this chapter, unless we explicitly specify differently.

\begin{figure}[t]
    \centering
    \includegraphics[width=0.9\linewidth]{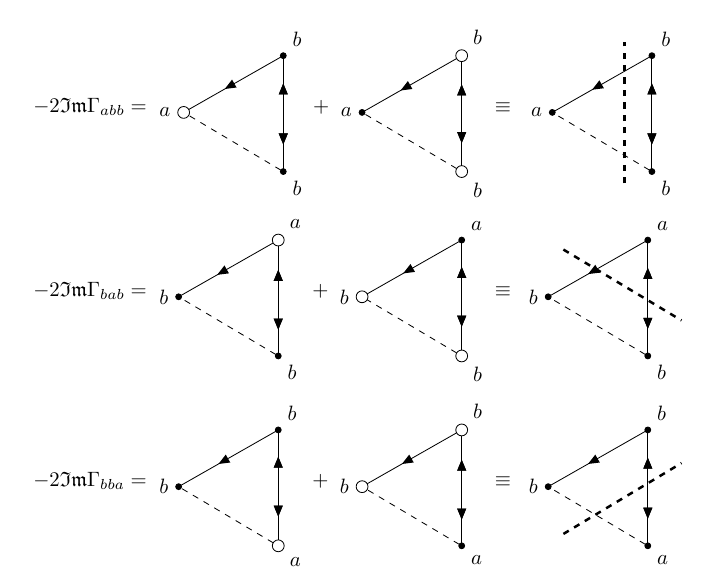}
    \caption{Imaginary parts of the off-diagonal components of the vertex function in the KS formalism with CTP indices $a,b=1,2$ and $a\neq b$.
    Each component can be represented by a single cut diagram as indicated.}
    \label{fig:GammaBBAcuts}
\end{figure}

First we discuss the off-diagonal components of the vertex function, i.e.,~$\Gamma_{[abb]}$ where $a\neq b$ and $[abb]$ denotes any permutation of the indices.
Using the cutting rules given in appendix~\ref{sec:app:FTcuttingRules} it is simple to show that for any propagators and permutation of the indices
\begin{equation}
\label{eq:cancellation}
\begin{aligned}
    \includegraphics{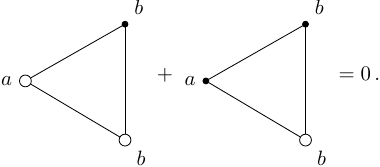}
\end{aligned}
\end{equation}
Specifically, we used that the propagator $G_{ab}$ (or $S_{ab}$) is independent of the circling for $a\neq b$ and that circling reverses the sign of the vertex.
It follows from eq.~\eqref{eq:cancellation} that 4 out of the 6 possible circled diagrams cancel in the expressions for $\imag\Gamma_{[abb]}$, therefore one cut diagram may be assigned to the remaining two circled diagrams for each instance of $\Gamma_{[abb]}$, as shown in figure~\ref{fig:GammaBBAcuts}.

The diagrams in figure~\ref{fig:GammaBBAcuts} are evaluated using the finite temperature cutting rules of appendix~\ref{sec:app:FTcuttingRules}.
We focus on the stripped amplitudes $\Tilde{\Gamma}_{abc}(\{P_N,P_\phi\},K)$ and use the propagator relation $S_{11}-S_{22}=2\ri\imag S_{11}$ with a similar one for bosons.
Using the symmetry relations between the off-diagonal components of the thermal propagators (see eqs.~\eqref{eq:boson_propagator_11_12} and \eqref{eq:fermion_propagator_11_12}), pairs of components corresponding to each line in figure~\ref{fig:GammaBBAcuts} are equal in absolute value.
For example, for the vertex type ${abb}$ in the first line of figure~\ref{fig:GammaBBAcuts} we have:
\begin{equation}
    \label{eq:G122_and_G211}
    \imag\Tilde{\Gamma}_{122} = \Tilde{S}_{12}(K,\overline{m}_L)G_{12}(P_N-K,\overline{m}_\phi)
    \imag\Tilde{S}_{11}(K-P_\phi,\overline{m}_{N_j}) = -\imag\Tilde{\Gamma}_{211}\,.
\end{equation}
Similar relations are found for the second and third lines of figure~\ref{fig:GammaBBAcuts}.
In summary, the off-diagonal components of the vertex function are not independent and we have the following three relations for the full vertex function (not only for the stripped one):
\begin{equation}
    \label{eq:ABB_imaginary_part_relations}
    \imag\Gamma_{112} = -\imag\Gamma_{221}\,,\quad \imag\Gamma_{121} = \imag\Gamma_{212}\,,\quad \imag\Gamma_{211} = -\imag\Gamma_{122}\,.
\end{equation}
As a quick rule of thumb, the imaginary parts shown above are equal if the CTP index that appears once corresponds to a bosonic leg, and there is a minus sign if the corresponding leg is fermionic.

Next, we continue with the diagonal components of the thermal vertex function, $\Gamma_{111}$ and $\Gamma_{222}$.
Here we do not have a cancellation such as that shown in eq.~\eqref{eq:cancellation} for the off-diagonal components, and we need to calculate all 6 circled diagrams, corresponding to 3 cut diagrams for each diagonal component.
There exists a pairwise equality between specific circled diagrams: if one simultaneously flips all CTP indices and circlings then the resulting diagrams have the same amplitude.
As an example, the following two diagrams are equal irrespective of the specific propagators used,
\begin{equation}
\begin{aligned}
    \includegraphics{Figures/inline_vertex_abb_cancellation.pdf}
\end{aligned}
\end{equation}
Exploiting this relation, after a similar calculation as that for the off-diagonal components, one finds that 
\begin{equation}
    \label{eq:AAA_imaginary_part_relations}
    \imag\Gamma_{111}=\imag\Gamma_{222}\,.
\end{equation}

\begin{figure}[t]
    \centering
    \includegraphics[width=0.85\linewidth]{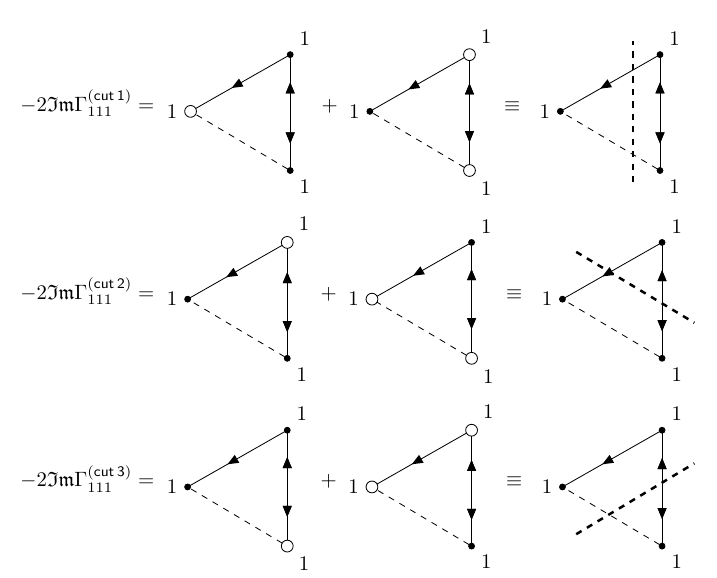}
    \caption{Imaginary part of the 111 component of the vertex function. The six circled diagrams are grouped into 3 pairs, each corresponding to a cut diagram as indicated. The full imaginary part of $\Gamma_{111}$ is the sum of the three cut contributions.}
    \label{fig:Gamma111cuts}
\end{figure}

The imaginary part of the diagonal components of the vertex function, in particular that of $\Gamma_{111}$, is separated into three terms corresponding to three cuts,
\begin{equation}
    \label{eq:111cutsDefinition}
    \imag\Gamma_{111}(P_N,P_\phi) = \sum_{i=1}^{3}\imag\Gamma_{111}^{\sf (cut\,i)}(P_N,P_\phi)\,.
\end{equation}
Each cut diagram is indexed with a superscript as $\imag\Gamma_{111}^{\sf (cut\,i)}$ and they are related to the circled diagrams as indicated in figure~\ref{fig:Gamma111cuts}.
At zero temperature, if the neutrino in the loop is heavier than the decaying one, then only $\imag\Gamma_{111}^{\sf (cut\,1)}$ (first line in figure~\ref{fig:Gamma111cuts}) contributes and the other two cuts vanish (see eq.~\eqref{eq:ImCmassrelation}).
However, at finite temperature the energy flow in the cut propagators is not fixed, and therefore all diagrams give non-zero contribution \cite{Garbrecht:2010sz}.
In models with very heavy sterile neutrinos the contribution of the second and third cuts are exponentially small for temperatures $T\ll m_{N_j}$ due to a Boltzmann-suppression factor $\exp(-\beta m_{N_j})\ll 1$ that appears through the cut neutrino propagator \cite{Giudice:2003jh}.
However, we shall see that in general these diagrams are also relevant for the measure of the CP-asymmetry and their contribution should not be neglected.

An important detail about the $\imag\Gamma_{111}^{\sf (cut\,i)}$ diagrams is that the separate cuts in general are complex while the sum of the three cuts is of course real.
The cancellation of the imaginary parts is equivalent to the cancellation of terms that involve the product of three Dirac-delta distributions (i.e., terms where all three intermediate states are on-shell which may be possible for specific masses).
In order to be able to talk about the imaginary part of the separate cuts of the 111 component of the vertex diagram, it is beneficial to make the replacement
\begin{equation}
    \label{eq:G11prescriptionForReal}
    G_{11}(P,\overline{m}) \to \ri\,\imag G_{11}(P,\overline{m})\quad\text{ and }\quad S_{11}(P,\overline{m})\to \ri\,\imag S_{11}(P,\overline{m})\,,
\end{equation}
which leaves the sum of the diagrams unchanged such that each cut diagram becomes real as well.
This argumentation is valid to the vertex function.
In essence, each circled diagram will involve the product of two Dirac-delta distributions (corresponding to two cut propagators) and an explicitly off-shell third propagator.
From here on, we assume that eq.~\eqref{eq:G11prescriptionForReal} has been utilized and $\imag\Gamma_{111}^{\sf (cut\,i)}$ are real functions.

Each cut of $\imag\Gamma_{111}$ (see figure~\ref{fig:Gamma111cuts}) is non-trivially related to one cut diagram of $\imag\Gamma_{[abb]}$ with similar topology (see figure~\ref{fig:GammaBBAcuts}).
To see this, recall that the propagators $G_\pm$ and $S_\pm$ are related to $G_{12/21}$ and $S_{12/21}$ via eqs.~\eqref{eq:boson_propagator_offdiag_relation}--\eqref{eq:fermion_propagator_offdiag_relation}.
For example, for the first line in figure~\ref{fig:Gamma111cuts}: 
\begin{equation}
\begin{aligned}
    \imag\Tilde{\Gamma}_{111}^{\sf (cut\,1)}&= -\frac{1}{2}\imag\Tilde{S}_{11}(K-P_\phi,\overline{m}_{N_j})
    \\
    &\qquad\times\left[\Tilde{S}_-(K,\overline{m}_L)G_-(P_N-K,\overline{m}_\phi) + \Tilde{S}_+(K,\overline{m}_L)G_+(P_N-K,\overline{m}_\phi)\right] \\
    &= \imag\Tilde{S}_{11}(K-P_\phi,\overline{m}_{N_j}) \Tilde{S}_{12}(K,\overline{m}_L)G_{12}(P_N-K,\overline{m}_\phi)\sinh\left(\frac{\beta p_N^0}{2}\right)\,.
\end{aligned}
\end{equation}
We see the same propagator product appearing in $\imag\Tilde{\Gamma}_{122}$ in eq.~\eqref{eq:G122_and_G211}.
After similar calculations for the other cuts, one finds the following relations between the imaginary parts of the diagonal and the off-diagonal components of the thermal vertex function:
\begin{subequations}
\label{eq:imag_relations}
\begin{align}
    \imag\Gamma_{111}^{\sf (cut\,1)} &= -\imag\Gamma_{211}\sinh\left(\frac{\beta p_N^0}{2}\right)\,, \\
    \imag\Gamma_{111}^{\sf (cut\,2)} &= -\imag\Gamma_{121}\cosh\left(\frac{\beta p_\phi^0}{2}\right)\,, \\
    \imag\Gamma_{111}^{\sf (cut\,3)} &= +\imag\Gamma_{112}\sinh\left(\frac{\beta p_L^0}{2}\right)\,.
\end{align}
\end{subequations}
As indicated, these relations hold for the full vertex function, not only for the stripped ones.
For later convenience we rewrite eq.~\eqref{eq:imag_relations} as an explicit constraint equation between the components of the thermal vertex function.
Using eq.~\eqref{eq:111cutsDefinition} one finds
\begin{equation}
    \label{eq:MainTextVertexConstraint}
    \imag\Gamma_{111} + \imag\Gamma_{211}\sinh\left(\frac{\beta p_N^0}{2}\right) + \imag\Gamma_{121}\cosh\left(\frac{\beta p_\phi^0}{2}\right) - \imag\Gamma_{112}\sinh\left(\frac{\beta p_L^0}{2}\right)=0\,.
\end{equation}

Eqs.~\eqref{eq:ABB_imaginary_part_relations}--\eqref{eq:AAA_imaginary_part_relations} with eq.~\eqref{eq:MainTextVertexConstraint} show that the thermal vertex function only has three independent components out of the 8 total, and it is convenient to choose these as the three cuts $\imag\Gamma_{111}^{\sf (cut\,i)}$.
A similar relation to eq.~\eqref{eq:MainTextVertexConstraint} is found in ref.~\cite{Kobes:1990kr} for the vertex function in the $\phi^3$ scalar theory.
We find the same relation in appendix~\ref{sec:app:RAformalism} as a general consequence of the RA formalism (see, eq.~\eqref{eq:GammaRRRconstraint}), as such it provides a good consistency check between the two methods, as here we obtained this equation in a purely diagrammatic way with no reference to the RA formalism.

In summary, we choose the three independent degrees of freedom of the thermal vertex function in the KS formalism as the (real part of the) three cuts of the 111 component:
\begin{subequations}
\begin{align}
    \label{eq:ImGamma111cut1}
    \imag\Tilde{\Gamma}_{111}^{\sf (cut\,1)} &= \imag\Tilde{S}_{11}(K-P_\phi,\overline{m}_{N_j}) \Tilde{S}_{12}(K,\overline{m}_L)G_{12}(P_N-K,\overline{m}_\phi)\sinh\left(\frac{\beta p_N^0}{2}\right)\,,
    \\
    \label{eq:ImGamma111cut2}
    \imag\Tilde{\Gamma}_{111}^{\sf (cut\,2)} &= \imag G_{11}(P_N-K,\overline{m}_\phi)\Tilde{S}_{12}(K,\overline{m}_L)\Tilde{S}_{12}(K-P_\phi,\overline{m}_{N_j})\cosh\left(\frac{\beta p_\phi^0}{2}\right)\,,
    \\
    \label{eq:ImGamma111cut3}
    \imag\Tilde{\Gamma}_{111}^{\sf (cut\,3)} &= \imag\Tilde{S}_{11}(K,\overline{m}_L)G_{12}(P_N-K,\overline{m}_\phi)\Tilde{S}_{12}(K-P_\phi,\overline{m}_{N_j})\sinh\left(\frac{\beta p_L^0}{2}\right)\,.
\end{align}
\end{subequations}
With the matrix of the thermal vertex function now fully defined, we turn to find the combination of the components that appears for physical processes and thus influence the CP-asymmetry factor.
This is most easily derived using the RA formalism.

\subsubsection{Vertex function in the RA formalism}
\label{sec:RA_vertex}

In ref.~\cite{Kobes:1990kr} it was argued in the context of scalar field theory that the physically relevant, so-called \emph{causal vertex function} ($\Gamma_{\rm C}$) is given as a combination of various vertex function components $\Gamma_{abc}$.
It was later found in ref.~\cite{Aurenche:1991hi} that this combination is equivalent to a specific component of the vertex function in the RA formalism.
In the following we start with briefly introducing the relevant details of the RA formalism and refer the reader to appendix~\ref{sec:app:RAformalism} or ref.~\cite{Aurenche:1991hi} for more details.
Our goal is to derive the one-loop physical vertex function for the $N\to \phi L$ decay and to connect this with the causal vertex function defined by Kobes in ref.~\cite{Kobes:1990kr}. 

The tree level vertex function in the RA formalism is given in a general form in eq.~\eqref{eq:general_stripped_tree_vertex_RA}.
We are particularly interested in the $N\phi L$ vertex that has 2 fermionic and 1 bosonic external leg.
After direct substitution, the components of the vertex function are found explicitly:
\begin{equation}
    \label{eq:FBF_relations}
    \begin{aligned}
        &\Tilde{\gamma}_{\rm RRR}^{\rm FBF}(P,Q) = \Tilde{\gamma}_{\rm AAA}^{\rm FBF}(P,Q) = 0\,, \\
        &\Tilde{\gamma}_{\rm RRA}^{\rm FBF}(P,Q) = \Tilde{\gamma}_{\rm RAR}^{\rm FBF}(P,Q) = \Tilde{\gamma}_{\rm ARR}^{\rm FBF}(P,Q) = 1\,, \\
        &\Tilde{\gamma}_{\rm RAA}^{\rm FBF}(P,Q) = \BE(p^0+q^0)+\FD(q^0) \,, \\
        &\Tilde{\gamma}_{\rm ARA}^{\rm FBF}(P,Q) = -\left[1-\FD(p^0)-\FD(q^0)\right]\,, \\
        &\Tilde{\gamma}_{\rm AAR}^{\rm FBF}(P,Q) = \BE(p^0+q^0)+\FD(p^0)\,.
    \end{aligned} 
\end{equation}
Notice that $\Tilde{\gamma}_{\rm AAR}^{\rm FBF}(P,Q)=\Tilde{\gamma}_{\rm RAA}^{\rm FBF}(Q,P)$ due to cyclicity, irrespective of whether the exchanged fermionic external legs are of the same type.
The vanishing of the RRR and AAA type vertices is a consequence of causality and momentum conservation.
In the rest of this section we will use only this type of vertices between 2 fermions and 1 boson, so for simplicity we will neglect to write the F and B superscripts.

We move on to consider the one-loop $N\phi L$ vertex correction to the heavy neutrino decay.
We assign the momenta and RA indices as
\begin{equation}
\label{eq:RAvertexDiagram}
\begin{aligned}
    \includegraphics{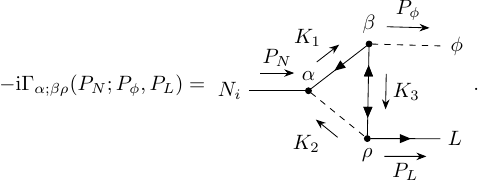}
\end{aligned}
\end{equation}
We use the notation where a semicolon separates the incoming and the outgoing external momenta.
Due to momentum conservation $P_\phi=P_N-P_L$ and the momenta in the loop are given as $K_2=K_1-P_N$ and $K_3=K_1-P_N+P_L$.

When all external momenta are directed as in eq.~\eqref{eq:RAvertexDiagram}, each vertex point has both incoming and outgoing momenta, whereas the vertices in eq.~\eqref{eq:FBF_relations} were defined for purely incoming momenta.
The inversion of a momentum direction also changes the corresponding R/A index \cite{Aurenche:1991hi}, for example
\begin{equation}
    \label{eq:semicolonNotation}
    \Tilde{\gamma}_{\alpha\beta;\rho}(P_1,P_2;P_3) \equiv \Tilde{\gamma}_{\alpha\beta\Bar{\rho}}(P_1,P_2,-P_3)\,,\text{ where }P_1+P_2=P_3\,.
\end{equation}
Here $\Bar{\rho}$ indicates the ``conjugation" of the index $\rho$, meaning the exchange R$\leftrightarrow$A.
The absence of the semicolon means by definition that all momenta are incoming.

In ref.~\cite{Aurenche:1991hi} the authors defined the physical amplitude corresponding to a process $\sum_i\psi_{i}(P_i)\to \sum_j\psi_j(P_j)$ for any number and type of particles as $\mathcal{M}_{\rR\rR\dots ;\rR\rR\dots }(\{P_i\};\{P_j\})$.
It follows that for the decay of the sterile neutrino as depicted in eq.~\eqref{eq:RAvertexDiagram} one requires the component $\Gamma_{\rR;\rR\rR}(P_N;P_\phi,P_L)$.
In order to write this amplitude, we use the vertices as defined in eq.~\eqref{eq:FBF_relations} and the tree-level R/A propagators 
\begin{equation}
    \label{eq:RApropagator}
    D_{\rR/\rA}(K,m) = \frac{\ri}{K^2-m^2\pm \ri \epsilon k^0} = \ri\mathcal{P}\frac{1}{K^2-m^2} \pm {\rm sgn}(k^0)\pi\delta(K^2-m^2)\,.
\end{equation}
Here the second equality is known as the Plemelj formula and $\mathcal{P}$ denotes the principal value.
The RA propagators are same for bosons and fermions in the sense that for fermions we mean the stripped version.
Contrary to the KS formalism, in the RA scheme all temperature dependence is contained in the vertices and the propagators are simply the vacuum ones.
Using eq.~\eqref{eq:semicolonNotation} to take into account the specific momentum flows in the vertices we find
\begin{equation}
\label{eq:IntegralDDDggg}
\begin{aligned}
    \int_{K_1}\!\Tilde{\Gamma}_{\rR;\rR\rR}\big(&\{P_N,P_L\},K_1\big)
    = -\sum_{\alpha,\beta,\rho=\rR,\rA}\int_{K_1}\!D_{\alpha}(K_1,\overline{m}_L)D_{\beta}(K_2,\overline{m}_\phi)D_{\rho}(K_3,\overline{m}_{N_j}) \\
    &\times
    \Tilde{\gamma}_{\rR\beta\Bar{\alpha}}(P_N,-K_1)
    \Tilde{\gamma}_{\Bar{\rho}\rA\alpha}(-K_3,K_1)
    \Tilde{\gamma}_{\rho\Bar{\beta}\rA}(K_3,-P_L)
    \,.
\end{aligned}
\end{equation}
After evaluating the sum there are in total 8 terms out of which 4 vanish due to $\Tilde{\gamma}_{\rR\rR\rR}=\Tilde{\gamma}_{\rA\rA\rA}=0$.
We can further simplify the expression by using the relations for the products of statistical factors given in eqs.~\eqref{eq:statistical_factor_products}--\eqref{eq:BEFDperFD__momentum_inversion}.
For example for $\alpha=\beta=\rho=\rR$ one has the vertex function product
\begin{equation}
\label{eq:gamma3_example}
\begin{aligned}
    \Tilde{\gamma}_{\rR\rR\rA}(P_N,-K_1)&\Tilde{\gamma}_{\rA\rA\rR}(-K_3,K_1)\Tilde{\gamma}_{\rR\rA\rA}(K_3,-P_L) \\ 
    &= \big[\BE(-k_3^0+k_1^0)+\FD(-k_3^0)\big]\big[\BE(k_3^0-p_L^0)+\FD(-p_L^0)\big] \\
    &= \big[\BE(p_N^0-p_L^0)+\FD(-p_L^0)\big]\big[1+\BE(k_2^0)-\FD(p_N^0)\big]\,.
\end{aligned}
\end{equation}
We separated the vertex function into a product of two terms, one that only depends on the external momenta and another that depends on the loop momentum as well.
Upon integration we use $\int_{K_1}\!D_\alpha(K_1,\overline{m}_L)D_\alpha(K_2,\overline{m}_\phi)D_\alpha(K_3,\overline{m}_{N_j})=0$ for $\alpha=\rR/\rA$ to cancel some terms in eq.~\eqref{eq:IntegralDDDggg}.
After similar calculations for the other terms finally one has
\begin{equation}
\label{eq:GammaRpRR_DDD_form}
\begin{split}
    \int_{K_1}\!\Tilde{\Gamma}_{\rR;\rR\rR}\!\big(\{&P_N,P_L\},K_1\big) = 
    \big[\BE(p_N^0-p_L^0)+\FD(-p_L^0)\big]\int\frac{\rd^4 K_1}{(2\pi)^4}\Big\{ \\   
    &\FD(k_1^0) D_\rA(K_2,\overline{m}_\phi) D_\rA(K_3,\overline{m}_{N_j}) \left[D_\rR(K_1,\overline{m}_L)-D_\rA(K_1,\overline{m}_L)\right] \\
    -&\BE(k_2^0) D_\rR(K_1,\overline{m}_L) D_\rR(K_3,\overline{m}_{N_j}) \left[D_\rR(K_2,\overline{m}_\phi) - D_\rA(K_2,\overline{m}_\phi)\right] \\
    +& \FD(k_3^0) D_\rR(K_1,\overline{m}_L) D_\rA(K_2,\overline{m}_\phi) \left[D_\rR(K_3,\overline{m}_{N_j}) - D_\rA(K_3,\overline{m}_{N_j})\right]
    \Big\}\,.
\end{split}
\end{equation}
The prefactor of the integral is the tree-level vertex function
\begin{equation}
    \Tilde{\gamma}_{\rR;\rR\rR}(P_N;P_N-P_L,P_L) = \Tilde{\gamma}_{\rR\rA\rA}(P_N,-P_L) = \BE(p_N^0-p_L^0)+\FD(-p_L^0)\,.
\end{equation}
This is a general result, one can always factorize the corresponding tree-level vertex function in the expressions of one-loop diagrams \cite{Aurenche:1991hi}.
Due to this prefactor being independent of the loop momentum and being equal to the tree-level vertex function, it cancels when calculating the amplitude-level CP-asymmetry factor, see eq.~\eqref{Eq:eps_M}.
In the following, we shall take this into account explicitly and consider the vertex correction as
\begin{equation}
    \label{eq:VRpRR_definition}
    \Tilde{V}_{\rR;\rR\rR}\big(\{P_N,P_L\},K_1\big) = \frac{\Tilde{\Gamma}_{\rR;\rR\rR}\big(\{P_N,P_L\},K_1\big)}{\Tilde{\gamma}_{\rR;\rR\rR}(P_N;P_N-P_L,P_L)}\,.
\end{equation}
The one-loop vertex correction is thus linear in the statistical factors, cf.~eq.~\eqref{eq:GammaRpRR_DDD_form}, as generally required by thermal field theory.

The vertex correction $\Tilde{V}_{\rR;\rR\rR}$ is the sum of three contributions that can be decomposed into the sum of the three cuts that we have seen in section \ref{sec:KS_vertex}.
Using the second equality in eq.~\eqref{eq:RApropagator}, the difference of the retarded and advanced propagators gives the on-shell part of the propagator,
\begin{equation}
    D_\rR(P,m)-D_\rA(P,m) = {\rm sgn}(p^0) 2\pi \delta(P^2-m^2)\,.
\end{equation}
The difference is real, thus in order to find the imaginary part of the vertex function we need to determine the imaginary part of the remaining product of two R/A propagators in each line of eq.~\eqref{eq:GammaRpRR_DDD_form}.
With $\alpha_i=\rR/\rA$ we have:
\begin{equation}
    \imag\big[ D_{\alpha_1}(P_1,m_1)D_{\alpha_2}(P_2,m_2) \big] = \sum_{\overset{i,j=1,2}{i\neq j}}(-1)^{\delta_{\alpha_i \rA}}{\rm sgn}(p_i^0)\pi\delta(P_i^2-m_i^2)\mathcal{P}\frac{1}{P_j^2-m_j^2}\,.
\end{equation}
Finally, after direct substitution and some organization of terms such that each line collects terms with the same propagator structure one finds
\begin{equation}
\label{eq:ImVRpRR}
\begin{split}
    \int_{K_1}\!\imag&\Tilde{V}_{\rR;\rR\rR}\big(\{P_N,P_L\},K_1\big)=-2\pi^2\int\!\frac{\rd^4K_1}{(2\pi)^4}\bigg\{ 
    \\
    &{\rm sgn}(k_1^0){\rm sgn}(k_2^0)\mathcal{P}\frac{1}{K_3^2-\overline{m}_{N_j}^2}\delta(K_1^2-\overline{m}_L^2)\delta(K_2^2-\overline{m}_\phi^2)\big[\FD(k_1^0)+\BE(k_2^0)\big] \\
    +&{\rm sgn}(k_1^0){\rm sgn}(k_3^0)\mathcal{P}\frac{1}{K_2^2-\overline{m}_\phi^2}\delta(K_1^2-\overline{m}_L^2)\delta(K_3^2-\overline{m}_{N_j}^2)\big[\FD(k_1^0)-\FD(k_3^0)\big] \\
    +&{\rm sgn}(k_2^0){\rm sgn}(k_3^0)\mathcal{P}\frac{1}{K_1^2-\overline{m}_L^2}\delta(K_2^2-\overline{m}_\phi^2)\delta(K_3^2-\overline{m}_{N_j}^2)\big[\FD(k_3^0)+\BE(k_2^0)\big]\bigg\}\,.
\end{split}
\end{equation}
Each line above corresponds to one specific cut of the one-loop vertex diagram as each line involves two propagators put on-shell by the $\delta$ distributions, and an off-shell propagator corresponding to the uncut propagator.
In the numbering convention introduced for the cuts in figure~\ref{fig:Gamma111cuts} and with $K_i$ defined as below eq.~\eqref{eq:RAvertexDiagram}, the first line of eq.~\eqref{eq:ImVRpRR} corresponds to cut 1, the second line is cut 2, and the third line is cut 3.

In ref.~\cite{Kobes:1990kr} the author introduced the causal vertex function as the one-loop amplitude at finite temperature corresponding to the $2\to 1$ inverse decay process.
For incoming momenta $P_{1,2}$ (and outgoing momentum $P_1+P_2$) their definition could be expressed in terms of the RA formalism as \cite{Kobes:1990kr,Aurenche:1991hi}
\begin{equation}
    \Gamma_{\rm C}(P_1,P_2)\equiv \Gamma_{\rR\rR;\rR}(P_1,P_2;P_1+P_2)\,.
\end{equation}
While this definition was introduced in the context of a $\phi^3$ scalar field theory, we are free to extend it to any theory involving other types of fields as well.
As the corresponding tree-level vertex is trivial $\Tilde{\gamma}_{\rR\rR;\rR}=1$ in any theory, we see that $\Gamma_{\rm C}$ immediately gives the one-loop vertex correction $V_{\rR\rR;\rR}$ to the coupling, where the omission of the tilde is understood as defined in and below eq.~\eqref{eq:strippedAmplitudeDef}.
In fact, this is the reason why the $\Gamma_{\rR\rR;\rR}$ component was considered in the first place.
Naturally, a similarly defined causal vertex function in the context of the $N_i\phi L$ interaction is expected to be related to the one-loop vertex correction that we derived in eq.~\eqref{eq:ImVRpRR}.

In our assignment of the momenta shown in eq.~\eqref{eq:RAvertexDiagram} it is more convenient to rewrite the causal vertex function using eq.~\eqref{eq:semicolonNotation} and reorganizing the indices as
\begin{equation}
    \Gamma_{\rm C}(P_N,P_L) = V_{\rA;\rA\rA}(P_N;P_N-P_L,P_L)=V^*_{\rR;\rR\rR}(P_N,P_N-P_L,P_L)\,.
\end{equation}
The last equality is a general consequence of the RA formalism upon switching all indices as $\rR\leftrightarrow\rA$ \cite{Aurenche:1991hi}.
More concretely, it follows from the retarded and advanced propagators being complex conjugates of each other, see eq.~\eqref{eq:RApropagator}.
The imaginary part of the vertex correction in eq.~\eqref{eq:ImVRpRR} is then related to the causal vertex function as
\begin{equation}
    \label{eq:VRpRRandGammaC}
    \imag V_{\rR;\rR\rR}(P_N;P_N-P_L,P_L)=-\imag\Gamma_{\rm C}(P_N,P_L)\,.
\end{equation}

\subsubsection{Connecting the KS and RA formalisms}

In the previous subsection we introduced the imaginary part of the vertex correction $\Tilde{V}_{\rR;\rR\rR}\big(\{P_N,P_L\},K_1\big)$ as the physically relevant quantity for the CP-asymmetry factor.
In this subsection, we show how the same result is obtained from the KS formalism using the connection to the RA formalism given by the rotation matrices $\mathbf{U}^{\rm (B/F)}$ and $\mathbf{V}^{\rm (B/F)}$ introduced in more detail in appendix~\ref{sec:app:RAformalism}.
For the decay $N\to \phi+L$,
\begin{equation}
    \label{eq:GammaRpRR_def_rotation}
    \Gamma_{\rR;\rR\rR}(P_N;P_N-P_L,P_L)=\sum_{a,b,c=1,2}\mathbf{V}^{\rm (F)}_{\rR a}(P_N)\Gamma_{abc}(P_N,P_L)\mathbf{U}^{\rm (B)}_{b\rR}(P_N-P_L)\mathbf{U}^{\rm (F)}_{c\rR}(P_L)\,.
\end{equation}
We take the imaginary part of both sides and exploit the relations between the imaginary part of the  components of the KS vertex function, see eqs.~\eqref{eq:ABB_imaginary_part_relations} and \eqref{eq:AAA_imaginary_part_relations}.
After the evaluation of the sums for the CTP indices, we factor out the tree-level vertex function $\gamma_{\rR;\rR\rR}$ as in eq.~\eqref{eq:VRpRR_definition} and obtain the one-loop vertex correction
\begin{equation}
    \label{eq:imVRpRR_no_cut_form}
    \imag V_{\rR;\rR\rR} \! = \! \coth\Big(\frac{\beta p_N^0}{2}\Big)\!\left[
    \imag\Gamma_{111}+\imag\Gamma_{121}\frac{\cosh(\beta p_L^0/2)}{\cosh(\beta p_N^0/2)}+\imag\Gamma_{112}\frac{\sinh(\beta p_\phi^0/2)}{\cosh(\beta p_N^0/2)}
    \right].
\end{equation}
The same formula holds in scalar theory with $\cosh(x)\leftrightarrow\sinh(x)$ interchange for the hyperbolic functions with fermionic momentum in the argument \cite{Kobes:1990kr}.

The convenient degrees of freedom to use in the vertex function are the separate cuts of the 111 component, as mentioned in section~\ref{sec:KS_vertex}.
We use the relations we derived between the off-diagonal components of the KS vertex function and the separate cuts of the 111 component, given in eq.~\eqref{eq:imag_relations}, and the hyperbolic function relations of eq.~\eqref{eq:hyperbolic_magic} to find 
\begin{equation}
    \label{eq:imagVRpRR_final_result}
    \imag V_{\rR;\rR\rR} = \coth\Big(\frac{\beta p_N^0}{2}\Big)\imag\Gamma_{111}^{\sf (cut\,1)} + \tanh\Big(\frac{\beta p_\phi^0}{2}\Big)\imag\Gamma_{111}^{\sf (cut\,2)} + \coth\Big(\frac{\beta p_L^0}{2}\Big)\imag\Gamma_{111}^{\sf (cut\,3)}\,.
\end{equation}
As explained in eq.~\eqref{eq:VRpRRandGammaC} this result is equal to the causal vertex function up to a sign, and with this information we arrive at the formula for $\Gamma_{\rm C}$ that we presented at the beginning of this section.
Comparing eq.~\eqref{eq:imagVRpRR_final_result} to eq.~\eqref{eq:imVRpRR_no_cut_form} we can appreciate the simplicity of the final formula that we derived based on the cuts of the 111 component.
As before, the vertex involving 3 bosonic external legs is simply obtained by replacing $\coth\to\tanh$ in eq.~\eqref{eq:imagVRpRR_final_result}.

In contrast with $\Gamma_{abc}$, the imaginary part of the causal vertex function is linear in the statistical factors at one loop, as expected.
To see this, we substitute the definitions of $\imag\tilde\Gamma_{111}^{\sf (cut\,i)}$ in eqs.~\eqref{eq:ImGamma111cut1}--\eqref{eq:ImGamma111cut3} into $\imag \tilde V_{\rR;\rR\rR}$ given in eq.~\eqref{eq:imagVRpRR_final_result}.
First, note that since $\Tilde{S}_{12}(K)\propto \FD(|k^0|)$ and $G_{12}(K)\propto \BE(|k^0|)$, formally one has $\imag\Tilde\Gamma_{111}^{\sf (cut\,i)}\sim f_{\rm B/F}^2$.
However, as mentioned before, the hyperbolic functions in eq.~\eqref{eq:imagVRpRR_final_result} play the role of ``removing" these products of statistical factors, as we will see now explicitly.
As an example, the first cut diagram evaluates to
\begin{align}
    \nonumber
    \int_K\! \imag&\Tilde{\Gamma}_{111}^{\sf (cut\,1)}
    = -(2\pi)^2\int_K\mathcal{P}\frac{1}{(K-P_\phi)^2-\overline{m}_{N_j}^2}\sgn(k^0)\delta(K^2-\overline{m}_L^2)\delta\big((P_N-K)^2-\overline{m}_\phi^2\big) \\
    &\times \FD(|k^0|)\BE(|p_N^0-k^0|) \exp\left(\frac{\beta |k^0|}{2}\right)\exp\left(\frac{\beta |p_N^0-k^0|}{2}\right)\sinh\left(\frac{\beta p_N^0}{2}\right)\,.
\end{align}
We can expand the absolute value in the arguments of the statistical factors and the exponential functions using eq.~\eqref{eq:useful_f_exp_relations}.
We eventually find that the term proportional to the first cut $\imag\Gamma_{111}^{\sf (cut\,1)}$ in $\imag V_{\rR;\rR\rR}$ simplifies to
\begin{align}
    \nonumber
    \coth&\left(\frac{\beta p_N^0}{2}\right)\int_K\!\imag\Tilde{\Gamma}_{111}^{\sf (cut\,1)}\big(\{P_N,P_\phi\},K\big) = \\ 
    \nonumber
    &-2\pi^2\int_K\!\mathcal{P}\frac{1}{(K-P_\phi)^2-\overline{m}_{N_j}^2}\sgn(k^0)\sgn(k^0-p_N^0)\delta(K^2-\overline{m}_L^2)\delta\big((K-P_N)^2-\overline{m}_\phi^2\big) \\
    &\qquad\qquad\qquad\times \Big(\FD(k^0)+\BE(k^0-p_N^0)\Big)\,.
    \label{eq:ImGammaCut1}
\end{align}
This is exactly the first line of the RA vertex function given in eq.~\eqref{eq:ImVRpRR}.
Furthermore, the expression for the first cut of the vertex diagram is also very similar to the self-energy contribution we derived in eq.~\eqref{eq:ImagSigmaTildeIntegralRepresentation}.

Similarly, for the contributions of the other two cuts, see eqs.~\eqref{eq:ImGamma111cut2}--\eqref{eq:ImGamma111cut3}, one finds:
\begin{align}
    \nonumber
    \tanh&\left(\frac{\beta p_\phi^0}{2}\right)\int_K\!\imag\Tilde{\Gamma}_{111}^{\sf (cut\,2)}\big(\{P_N,P_\phi\},K\big) = \\ 
    \nonumber
    &-2\pi^2\int_K\!\mathcal{P}\frac{1}{(P_N-K)^2-\overline{m}_{\phi}^2}\sgn(k^0)\sgn(k^0-p_\phi^0)\delta(K^2-m_L^2)\delta\big((K-P_\phi)^2-\overline{m}_{N_j}^2\big) \\
    &\qquad\qquad\qquad\times \Big(\FD(k^0)-\FD(k^0-p_\phi^0)\Big)\,,
    \label{eq:ImGammaCut2}
\end{align}
and
\begin{align}
    \nonumber
    &\coth\left(\frac{\beta p_L^0}{2}\right)\int_K\!\imag\Tilde{\Gamma}_{111}^{\sf (cut\,3)}\big(\{P_N,P_\phi\},K\big) = \\ 
    \nonumber
    &-2\pi^2\int_K\!\mathcal{P}\frac{1}{K^2-\overline{m}_L^2}\sgn(k^0-p_N^0)\sgn(k^0-p_\phi^0)\delta\big((K-P_N)^2-\overline{m}_\phi^2\big)\delta\big((K-P_\phi)^2-\overline{m}_{N_j}^2\big) \\
    &\qquad\qquad\qquad\times \Big(\BE(k^0-p_N^0)+\FD(k^0-p_\phi^0)\Big)\,.
    \label{eq:ImGammaCut3}
\end{align}
We see that these are all linear in the statistical factors, and they all display the cutting structure as expected: two propagators are on shell, as indicated by the $\delta$-distributions, while the third one is explicitly off-shell, as indicated by the principal value.

In summary, the vertex contribution to the CP-asymmetry factor given in eq.~\eqref{eq:CPasymmetryFactorImagAmplitudeDef} is properly defined with the imaginary part of the $\Tilde{V}_{\rR;\rR\rR}$ component of the vertex function in the RA formalism.
To close this section, we give the final formula for the vertex contribution:
\begin{equation}
    \label{eq:epsilonVi}
    \epsilon_{V_i}(P_N) = -2G\sum_{j\neq i}\frac{\overline{m}_{N_i}\overline{m}_{N_j}}{P_{N}\cdot P_L}\int_K\!(K\cdot P_L)\imag \Tilde{V}_{\rR;\rR\rR}\big(\{P_N,P_\phi\},K\big)\,.
\end{equation}
Here the integrand is a sum of three terms. 
The evaluation of the corresponding integrals can be found in appendix~\ref{sec:EvaluationOfIntegrals}.

\subsection{The scalar triangle function in the imaginary time formalism}

Although it is customary to write the imaginary part of the triangle graph using cutting rules, we show bellow how to obtain it in the ITF.
To compare with the stripped vertex function given in eq.~\eqref{eq:ImVRpRR}, it is sufficient to focus on the scalar triangle function.

\begin{figure}[t]
    \centerline{\includegraphics[width=0.26\textwidth]{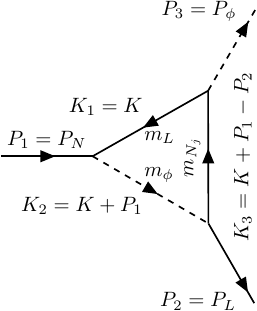}}
    \caption{
    \label{fig:scalartriangle}
    Particle and momentum assignment in the scalar triangle function.}
\end{figure}

At finite temperature the scalar triangle function, represented diagrammatically in figure~\ref{fig:scalartriangle}, is given as
\be
\label{Eq:SV_T}
\mathcal{C}^T_{L\phi N_j}(\bar P_N,\bar P_L) = T\sum_m\int\frac{\rd^3 k}{(2\pi)^3} \Delta(\bar K+\bar P_N-\bar P_L,m_{N_j}) \Delta(\bar K+\bar P_N,m_\phi) \Delta(\bar K,m_L)\,.
\ee
This expression is obtained from the vacuum expression of eq.~\eqref{Eq:SV_def} through the substitutions
\bea
\int_K \to \ri T\int\!\frac{\rd^3 k}{(2\pi)^3} \sum_{m=-\infty}^\infty
\quad\textrm{and}\quad
D(K,m_L)\to -\ri \Delta(\bar K,m_L)\,.
\eea
In doing so we also replaced the Minkowskian four momenta with Euclidean ones (indicated with a bar on the momenta), 
\begin{equation*}
    \bar K = (\ri \nu_m, \k),\quad\bar P_N = (\ri \nu_n, \p_N),\quad\bar P_L = (\ri \nu_l, \p_L),\quad{\rm and}\quad\bar P_\phi=\bar P_N-\bar P_L=(\ri\omega_h,\p_\phi)\,,
\end{equation*}
where we introduced the Matsubara frequencies $\nu_i$ ($i = l,\,m,\,n$) for fermions and $\omega_h$ for the boson, which take discrete values $\nu_i = (2i+1) \pi T$ and $\omega_h=2h\pi T$ with $i,\,h\in\mathbb{Z}$. They satisfy the usual energy conservation relation $\omega_h=\nu_n-\nu_l$.
As we decoupled the spinor structure, the tree level vacuum propagators for bosons and fermions have the same form, yet they differ at finite temperature in their Matsubara frequencies.

The spectral representation of the Euclidean propagators is formally the same for fermions and bosons with the specific Matsubara frequency determining the particle nature.
For example for the lepton we have
\be
\label{Eq:spectral_rep}
\Delta(\bar K,m_L) =\int_{-\infty}^\infty \frac{d k^0}{2\pi}\frac{\rho_L(K)}{k^0-\ri \nu_m}, \quad
\rho_L(K) = 2\pi\,\textrm{sgn}(k^0)\delta(K^2-m^2_L)\,.
\ee
One can then write the thermal scalar triangle function in the form
\be
\label{Eq:SV_T_sr}
\begin{aligned}
\mathcal{C}^T_{L\phi N_j}(\bar P_N, \bar P_L)&=\int\frac{\rd^3 k}{(2\pi)^3}\int_{-\infty}^\infty \frac{\rd k_1^0}{2\pi}\int_{-\infty}^\infty \frac{\rd k_2^0}{2\pi}\int_{-\infty}^\infty \frac{\rd k_3^0}{2\pi}
\\
&\bigg[\rho_L(k_1^0,\k)\rho_\phi(k_2^0,\k+\p_N)\rho_{N_j}(k_3^0,\k+\p_\phi)~T\sum_{m\in\mathbb{Z}} \prod_{j=1}^3\frac{1}{k_j^0-\ri \omega_j}\bigg]\,,
\end{aligned}
\ee
where $\omega_1=\nu_m$, $\omega_2=\nu_m+\nu_n$, and $\omega_3=\nu_m+\nu_n-\nu_l=\nu_m+\omega_h$. 

The Matsubara sum can be computed using standard methods (see e.g.~section~4.2.3 in ref.~\cite{Bellac:2011kqa}). 
We use Gaudin's method summarized in  appendix~\ref{sec:app:Gaudin}, which applied to the present case leads to the following identity:
\begin{equation}
\begin{aligned}
\label{Eq:Matsu_sum_res}
T&\sum_{m\in\mathbb{Z}}\prod_{j=1}^3\frac{1}{k_j^0-\ri \omega_j}=
-\frac{\FD(k_1^0)}{\big(k_2^0 - (k_1^0+\ri \nu_n)\big)\big(k_3^0 - (k_1^0+\ri \omega_h)\big)}
\\ 
&+ \frac{\BE(k_2^0)}{\big(k_1^0 - (k_2^0 - \ri \nu_n)\big)\big(k_3^0 - (k_2^0 - \ri \nu_l)\big)}
-\frac{\FD(k_3^0)}{\big(k_1^0 - (k_3^0 -\ri \omega_h)\big)\big(k_2^0 - (k_3^0 + \ri \nu_l)\big)}\,.
\end{aligned}
\end{equation}

Plugging eq.~\eqref{Eq:Matsu_sum_res} in eq.~\eqref{Eq:SV_T_sr} one can use the spectral representation given in eq.~\eqref{Eq:spectral_rep} to perform in each term two integrals over the energies (real frequencies) that do not appear as argument of the statistical factor.
The remaining energy integral can be combined with the integral over $\k$ after utilizing the finite shifts\footnote{The integrals are finite, hence finite shifts are allowed.} of the integration variable $\k\to\k-\p_N$ and $\k\to\k-\p_N+\p_L =\k-\p_\phi$ in the second and third terms of the sum eq.~\eqref{Eq:Matsu_sum_res}. 
Thus we obtain
\be
\begin{aligned}
  \mathcal{C}^T_{L\phi N_j}(\bar P_N, \bar P_L)=\int_K
  \Big[
    &-\FD(k^0)\, \rho_L(K) \Delta(K+\bar P_N,m_\phi) \Delta(K+\bar P_N-\bar P_L,m_{N_j})\\
    &+\BE(k^0)\, \rho_\phi(K) \Delta(K-\bar P_N,m_L) \Delta(K-\bar P_L,m_{N_j})\\
    &-\FD(k^0)\, \rho_{N_j}(K) \Delta(K-\bar P_N+\bar P_L,m_L) \Delta(K+\bar P_L,m_\phi)
    \Big]\,.
\end{aligned}
\ee

Now, one does the analytic continuation back to real frequencies, for which we use $\ri \nu_n\to p_N^0 + 2 \ri \varepsilon$, $\ri \nu_l\to p_L^0 + \ri \varepsilon$ and $\ri \omega_h \to p_\phi^0 + \ri \varepsilon$, in accordance with the energy conservation of the decay process, which is $\nu_n = \nu_l + \omega_h$ in Euclidean space. 
Using the retarded and advanced propagators defined in eq.~\eqref{eq:RApropagator}, one can write the scalar triangle integral at finite temperature and for real momenta in the form
\be
\begin{aligned}
  \!\!\!\mathcal{C}^T_{L\phi N_j}(P_N, P_L)=\int_K
  \Big[
    &\FD(k^0)\,\rho_L(K) D_\rR(K+P_N,m_\phi) D_\rR(K+P_N-P_L,m_{N_j})\\
    -&\BE(k^0)\,\rho_\phi(K) D_\rA(K-P_N,m_L) D_\rA(K-P_L,m_{N_j})\\
    +&\FD(k^0)\,\rho_{N_j}(K) D_\rA(K-P_N+P_L,m_L) D_\rR(K+P_L,m_\phi)
    \Big]\,.
\end{aligned} 
\ee
To obtain our final form for the scalar triangle integral, we shift the momenta 
in the second and third terms and use the abbreviations $K_1=K$, $K_2=K+P_N$, and $K_3=K+P_N-P_L$ (see figure \ref{fig:scalartriangle}):
\begin{equation}
\begin{aligned}
  \mathcal{C}^T_{L\phi N_j}(P_N, P_L)=\int_{K_1}
  \Big[
    &\FD(k_1^0)\, \rho_L(K_1) D_\rR(K_2,m_\phi) D_\rR(K_3,m_{N_j})\\
    -&\BE(k_2^0)\, \rho_\phi(K_2) D_\rA(K_3,m_{N_j}) D_\rA(K_1,m_L)\\
    +&\FD(k_3^0)\, \rho_{N_j}(K_3) D_\rA(K_1,m_L) D_\rR(K_2,m_\phi)
    \Big]\,.
\end{aligned} 
\end{equation}

As a last step, we use the Plemelj formula introduced in the second equality of eq.~\eqref{eq:RApropagator}.
Thus, using thermal masses ($m_X\to \overline{m}_X$) we obtain for the imaginary part of $\mathcal{C}_{L\phi N_j}^T$ the expression given in eq.~\eqref{eq:ImVRpRR}, that is
\be
\imag\,\mathcal{C}^T_{L\phi N_j}(P_N, P_L) =  -\int_{K}\imag\,\Tilde{V}_{\rR;\rR\rR}\big(\{P_N,P_L\},K\big)\,,
\ee
which completes the derivation of the imaginary part of the scalar triangle integral in the imaginary time formalism, relating its expression to the one obtained in the real time formalism.

\section{Results and discussion}
\label{sec:results}

In this section we present our final results for the CP-asymmetry factor for the mass hierarchy $\overline{m}_{N_j}\gtrsim \overline{m}_{N_i}>\overline{m}_\phi+ \overline{m}_L$.
First, we show the final formulae for contributions of the self-energy and the three cuts of the vertex function to the thermally averaged CP-asymmetry factor.
We relegated most details of the calculation into appendices \ref{sec:app:ThermalVertexCuts} and \ref{sec:EvaluationOfIntegrals}.
Afterwards, we show numerical results for the thermal CP-asymmetry in the context of a specific particle physics model with right-handed neutrinos, and compare them to various approximations that may be found in the literature.
We end this section with a discussion on the relevance of using the full expression of the CP-asymmetry and its possible consequences in leptogenesis.

\subsection{Thermally averaged CP-asymmetry factors}


The CP-asymmetry factor at finite temperature, given in eq.~\eqref{Eq:epsT_LO}, is the sum of the thermal averages of the amplitude-level CP-asymmetries that we computed in section~\ref{sec:CalculationCPatFiniteT}.
In eq.~\eqref{eq:thermalAvgEpsilonNumerator} we see that apart from the loop-integrals present in $\epsilon_{\mathcal{M}_i}$, we also have to integrate over the initial state phase space, that could be reduced to an integral over the initial state energy and the decay angle $\theta$.
The full calculation of the loop integral is presented in detail in appendix~\ref{sec:EvaluationOfIntegrals}, here we go through the steps schematically and present the final formulae that could be numerically integrated to find $\epsilon_i(T)$ in a given model.

The key ingredient in evaluating these integrals is to choose a suitable reference frame and integration variables.
The reference frame is the CR frame, however we still have the freedom to orient the spatial part of the coordinate system in any direction we want.
The natural system to use changes with the various diagrams:
(i) for the self-energy and first cut of the vertex function the natural choice is using $E_N$ and $\p_N$ aligned with the $x$ axis ($\p_N\parallel\hat{x}$), (ii) for the second cut we use $E_\phi$ and $\p_\phi\parallel\hat{x}$, and (iii) for the third cut we use $E_L$ and $\p_L\parallel\hat{x}$.
Naturally, the integration over the initial phase space in eq.~\eqref{eq:thermalAvgEpsilonNumerator} always involves integrating over the energy of the decaying neutrino ($E_N$) that is easily done in case (i), however, we must change variables $E_N\to E_{L,\phi}$ in cases (ii) and (iii).
Such changes to the integration variable is non-trivial due to the relationship between the initial- and final-state energies not being in a one-to-one relationship, as explained in great detail in appendix~\ref{sec:app:ThermalVertexCuts}.

First, we deal with the loop momentum integration in the amplitude-level CP-asymme-try factors.
Recall, that in both the self-energy and the vertex contributions the integrands involve the product of two $\delta$ distributions due to two cut (on-shell) propagators in each corresponding diagram.
These Dirac-deltas could be used to trivially evaluate 2 out of the 4 integrals in
\begin{equation*}
    \int_K = \frac{1}{(2\pi)^4}\int_{-\infty}^\infty\!\rd\omega\int_0^\infty\!\rd k\int_0^{2\pi}\!\rd\varphi'\int_{-1}^1\!\rd\cos\theta'\,.
\end{equation*}
We choose to perform the $k$ and the $\cos\theta'$ integrations with the $\delta$ distributions, putting the spatial momentum on-shell as $k\to \sqrt{\omega^2-\overline{m}^2}$ and turning $\cos\theta'$ into a function of the external and loop energies, $\cos\theta'\to f_{\theta'}(E,\omega)$.
The $k$-integral selects the positive energy solution as $k\geq 0$. However for the $\cos\theta'$-integral the requirement that $f_{\theta'}(E,\omega)\in[-1,1]$ leads to a finite domain for the loop energy $\omega\in[\omega_-,\omega_+]$, where $\omega_\pm(E)$ are functions of only the external energy $E$.
From the remaining two integrals, the one for the polar angle can be evaluated analytically, with the result generally denoted as $\mathcal{I}_{\varphi'}(E,\theta,\omega)$ where $\theta$ is the scattering angle in the CM frame.
The functions $f_{\theta'}$, $\omega_\pm$, and $\mathcal{I}_{\varphi'}$ depend on which cut diagram we specifically compute.

In summary, the amplitude-level CP-asymmetry factor is analytically reduced to a single, finite domain integral over the loop energy $\omega\in[\omega_-,\omega_+]$.
The explicit expressions are given in eq.~\eqref{eq:epsilon_Sigma_final} for the self-energy contribution, and in eqs.~\eqref{eq:epsilon_V_cut1_final}, \eqref{eq:epsilon_V_cut2_final}, and \eqref{eq:epsilon_V_cut3_final} for the first, second, and third cuts of the vertex contribution.

The last step of the calculation is to evaluate the thermal averaging over the initial particle phase space.
The normalization of $\epsilon_i$ is an integral over a product of statistical factors, cf. eq.~\eqref{eq:thermalAvgEpsilonDenominator}, that we denote as
\begin{equation}
    \label{eq:epsilon_Normalization}
    \mathcal{N}^{-1}(T)=\int_{\overline{m}_{N_i}}^\infty\!\rd E_N\,\sqrt{E_N^2-\overline{m}_{N_i}^2}\FD(E_N)e^{E_N/T}\int_{-1}^1\!\rd\cos\theta\,\FD\big(E_L^\CR\big)\BE\big(E_\phi^\CR\big)\,.
\end{equation}
We mention again that the $\cos\theta$ integral is analytic and its general form is given in eq.~\eqref{eq:integral_fBfF_costheta}.
Substituting $\epsilon_{\Sigma_i}$ given in eq.~\eqref{eq:epsilon_Sigma_final} into eq.~\eqref{eq:thermalAvgEpsilonNumerator} we eventually find 
the following parametric representation for the self-energy contribution to the thermal CP-asymmetry:
\begin{equation}
\begin{aligned}
    \epsilon_i^{(\Sigma)}(T)\!=\!\frac{\mathcal{N}G}{4\pi E_L^\CM}\sum_{j\neq i}\frac{\overline{m}_{N_j}}{\overline{m}_{N_i}^2-\overline{m}_{N_j}^2}\!\int_{\overline{m}_{N_i}}^\infty\!\rd E_N \!\int_{\omega_-^{(1)}}^{\omega_+^{(1)}}\!\rd\omega\!
    \int_{-1}^1\!\rd\cos\theta\,\mathscr{F}_{1}(E_N,\theta,\omega)\big(K\cdot P_L\big)\Big|_{\varphi'=\frac{\pi}{2}}\,,
\end{aligned}
\end{equation}
where the integral over $\cos\theta$ is analytic, the scalar product is computed in the CR frame with spatial orientation such that $\p_N\parallel\hat{x}$, and
\begin{equation}
    \mathscr{F}_1(E_N,\theta,\omega) = \FD(E_N)\FD(E_L^\CR)\BE(E_\phi^\CR)e^{E_N/T}\big(1+\BE(E_N-\omega)-\FD(\omega)\big)\,.
\end{equation}
The CR frame energies of the final state particles are functions of $E_N$ and $\cos\theta$ and their definitions may be found in eq.~\eqref{eq:CRenergyDefinitions}.
The remaining integrals over $E_N$ and $\omega$ have to be evaluated numerically.

The contribution of the first cut of the vertex function is the thermal average of the amplitude-level CP-asymmetry given in eq.~\eqref{eq:epsilon_V_cut1_final}.
We find the parametric representation
\begin{equation}
\begin{aligned}
    \epsilon_i^{\sf (cut\,1)}(T) = \frac{\mathcal{N}G}{8\pi E_L^\CM}&\sum_{j\neq i}\overline{m}_{N_j}\int_{\overline{m}_{N_i}}^\infty\!\rd E_N \!\int_{\omega_-^{(1)}}^{\omega_+^{(1)}}\!\rd\omega\!\int_{-1}^1\!\rd\cos\theta\,\mathscr{F}_{1}(E_N,\theta,\omega)\mathcal{I}^{\sf (cut\,1)}_{\varphi'}(E_N,\theta,\omega)\,,
\end{aligned}
\end{equation}
where the result of the $\varphi'$-integral is the function $\mathcal{I}_{\varphi'}^{\sf (cut\,1)}$ given explicitly in eq.~\eqref{eq:Ivarphi1}.
The product of statistical factors $\mathscr{F}_1$ is the same as for the self-energy.
The final result is reduced to a triple integral as the $\cos\theta$ integration is no longer analytic.

To simplify the calculation, for the second and third cuts we use specific coordinate systems where $\p_{\phi,L}\parallel\hat{x}$.
Additionally, we integrate over $E_{L,\phi}$ instead of $E_N$ as explained at the beginning of this section and in appendix~\ref{sec:app:ThermalVertexCuts}.
The final formula for the contribution of the second vertex cut to the amplitude-level CP-asymmetry factor was derived in eq.~\eqref{eq:epsilon_V_cut2_final}.
Substituting into eq.~\eqref{eq:thermalAvgEpsilonNumerator} one finds a parametric representation as a sum of three integrals that emerge due to the variable change $E_N^\pm(E_\phi,\cos\theta)$, and will be expressed using an integral operator to account for the transformation of the integral domain,
\begin{equation}
\begin{aligned}
    \iint_{\mathscr{D}_s}\!\rd E_\phi\,\rd\cos\theta &= \int_0^1\!\rd\cos\theta\int_{E_\phi^{\rm crit}}^\infty\!\rd E_\phi \frac{\partial E_N^+}{\partial E_\phi}\delta_{s,+} + \int_{-1}^0\!\rd\cos\theta\int_{E_\phi^\CM}^\infty\!\rd E_\phi\frac{\partial E_N^-}{\partial E_\phi}\delta_{s,-}
    \\
    &+ \int_0^1\!\rd\cos\theta\int_{E_\phi^\CM}^{E_\phi^{\rm crit}}\!\rd E_\phi \frac{\partial E_N^-}{\partial E_\phi}\delta_{s,-}\,.
\end{aligned}
\label{eq:IntOp}
\end{equation}
Then the the contribution of the second cut can be expressed as
\begin{equation}
\begin{aligned}
    \epsilon_i^{\sf (cut\,2)}(T) = \frac{\mathcal{N}G}{8\pi E_L^\CM}&\sum_{j\neq i}\overline{m}_{N_j}\sum_{s=\pm}\iint_{\mathscr{D}_s}\!\rd E_\phi\rd\cos\theta\,\frac{\sqrt{(E_N^s)^2-\overline{m}_{N_i}^2}}{\sqrt{E_\phi^2-\overline{m}_{\phi}^2}} \\
    &\times \int_{\omega^{(2)}_-}^{\omega^{(2)}_+}\!\rd\omega\,\mathcal{I}_{\varphi'}^{{\sf (cut\,2)},s}(E_\phi,\theta,\omega)\mathscr{F}^{s}_2(E_\phi,\theta,\omega)\,,
\end{aligned}
\end{equation}
where $\mathcal{I}_{\varphi'}^{{\sf (cut\,2)},\pm}$ is given in eq.~\eqref{eq:Ivarphi_cut2}, the inverted energy relations $E_N^\pm(E_\phi,\cos\theta)$ are defined in eq.~\eqref{eq:ENplusminus_solution}.
The integration defined by the operator in eq.~\eqref{eq:IntOp} is equivalent to what is given in eq.~\eqref{eq:HiggsRegionIntegrals}, we only suppressed the arguments here for visual clarity.
The product of the statistical factors in $\epsilon_i^{\sf (cut\,2)}$ is
\begin{equation}
    \mathscr{F}^{\pm}_2(E_\phi,\theta,\omega) = \FD(E_N^\pm)\FD(E_N^\pm-E_\phi)\BE(E_\phi)e^{E_N^\pm/T}\big(\FD(E_\phi-\omega)-\FD(-\omega)\big)\,.
\end{equation}

The contribution due to the third cut of the vertex function to the CP-asymmetry factor is evaluated similarly to the second cut.
The amplitude-level CP-asymmetry was derived in eq.~\eqref{eq:epsilon_V_cut3_final}, and we find
\begin{equation}
\begin{aligned}
    \epsilon_i^{\sf (cut\,3)}(T) = &\mathcal{N}^{-1}(T) \frac{G}{8\pi E_L^\CM}\sum_{j\neq i}\overline{m}_{N_j}\sum_{s=\pm}\iint_{\mathscr{D}_s}\!\rd E_L\rd\cos\theta\,\frac{\sqrt{(E_N^s)^2-\overline{m}_{N_i^2}}}{\sqrt{E_L^2-\overline{m}_{L}^2}} \\
    &\times \int_{\omega^{(3)}_-}^{\omega^{(3)}_+}\!\rd\omega\,\big[(P_N-K)\cdot P_L\big]\mathcal{I}_{\varphi'}^{{\sf (cut\,3)},s}(E_L,\theta,\omega)\mathscr{F}^s_3(E_L,\theta,\omega)\,,
\end{aligned}
\end{equation}
where $\mathcal{I}_{\varphi'}^{\sf (cut\,3),\pm}$ is given in eq.~\eqref{eq:Ivarphi_cut3} and the inverted energy relations are defined in eq.~\eqref{eq:ENplusminus_solution}.
The integration variable change $E_N\to E_L$ is also reflected in the transformation of the integration domain, as detailed in eq.~\eqref{eq:LeptonRegionIntegrals}, and consequently the integral operator that corresponds to eq.~\eqref{eq:IntOp} in the case of the third cut is
\begin{equation}
\begin{aligned}
    \iint_{\mathscr{D}_s}\!\rd E_L\,\rd\cos\theta &=
    \int_{-1}^0\!\rd\cos\theta \int_{E_L^{\rm crit}}^\infty\!\rd E_L \frac{\partial E_N^+}{\partial E_L}\delta_{s,+} + \int_0^1\!\rd\cos\theta\int_{E_L^\CM}^\infty\!\rd E_L\frac{\partial E_N^-}{\partial E_L}\delta_{s,-} \\
    &+ \int_{-1}^0\!\rd\cos\theta\int_{E_L^\CM}^{E_L^{\rm crit}}\!\rd E_L\frac{\partial E_N^-}{\partial E_L}\delta_{s,-}\,.
\end{aligned}
\end{equation}
The product of the statistical factors in $\epsilon^{\sf (cut\,3)}_i$ is
\begin{equation}
    \mathscr{F}^\pm_3(E_L,\theta,\omega) = \FD(E_N^\pm)\FD(E_L)\BE(E_N^\pm-E_L)e^{E_N^\pm/T}\big(\BE(-\omega)+\FD(E_L-\omega)\big)\,.
\end{equation}

In summary, we analytically reduced the expression of the full CP-asymmetry factor to a sum of 4 contributions: one double integral for the self-energy and three triple integrals for the vertex function.
The complete derivation of the relevant computations is presented in the appendices \ref{sec:app:ThermalVertexCuts} and \ref{sec:EvaluationOfIntegrals}.
Finally, the full thermal CP-asymmetry is 
\begin{equation}
    \label{eq:epsilonT_simpledef}
    \epsilon_i(T) = \epsilon_i^{(\Sigma)}(T) + \sum_{{\sf j}=1}^3\epsilon_i^{\sf (cut\, j)}(T)\,,
\end{equation}
for which we present numerical results in the following subsection.

\subsection{CP-asymmetry in the superweak extension of the SM}

The superweak extension of the Standard Model \cite{Trocsanyi:2018bkm} is a phenomenological particle physics model where 3 right-handed neutrinos, a SM-singlet scalar field, and a new U(1)$_z$ force carrier are added to the spectrum of the SM.
The model was devised to be a simple extension to the SM such that it is capable of simultaneously explaining various beyond the SM phenomena, for example dark matter \cite{Iwamoto:2021fup} and vacuum stability \cite{Peli:2022ybi}.

\begin{figure}[t]
    \centering
    \includegraphics[width=\linewidth]{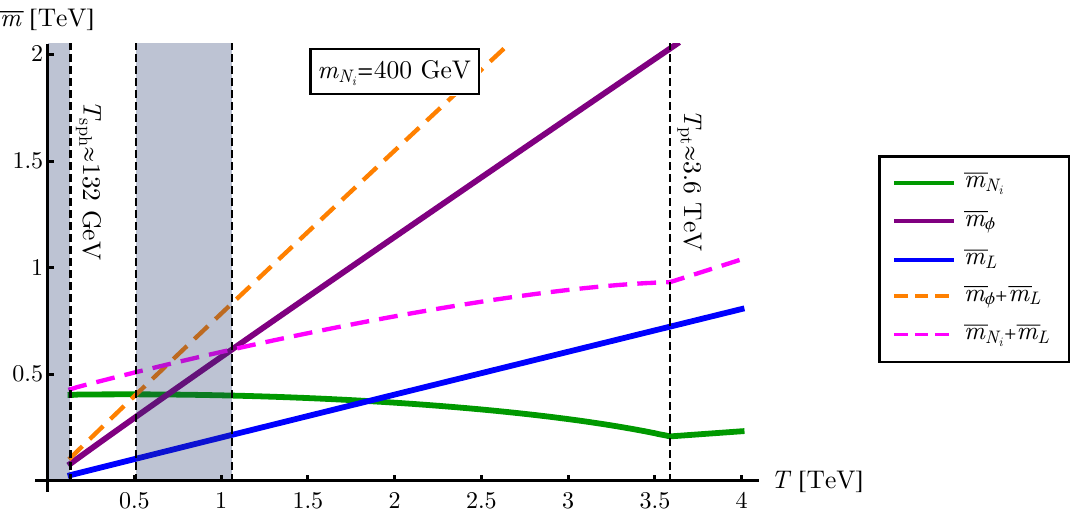}
    \caption{Thermal masses in the SWSM.
    In the shaded region at intermediate temperatures the thermal masses are such that decays involving sterile neutrinos, leptons, and scalar fields are kinematically excluded.
    We indicate a low temperature cutoff at $T_{\rm sph}\simeq 132$~GeV where the sphaleron processes freeze-out \cite{DOnofrio:2014rug}.
    Here we used that the two heavy neutrinos have similar vacuum masses of $m_{N_2}\simeq m_{N_3}=400$~GeV and that the  mass of the singlet scalar is $m_\chi=650~$GeV with the singlet vacuum expectation value being $w_0=10v_0$, where $v_0=246.22$~GeV is the SM Higgs vacuum expectation value.
    With these parameters the phase transition occurs around $T_{\rm pt}\approx 3.6$~TeV, as indicated by the kink in the sterile neutrino mass at this temperature.}
    \label{fig:ThermalMasses}
\end{figure}

For our discussion the relevant detail of the particle physics model is the generation of the Majorana mass term of the sterile neutrinos through a Higgs-mechanism facilitated by the singlet scalar acquiring a non-zero vacuum expectation value and spontaneously breaking the new U(1)$_z$ symmetry.
The corresponding phase transition is not expected to be strongly first order and it had been studied within the context of the SWSM in ref.~\cite{Seller:2023xkn}.
Once the temperature-dependence of the vacuum expectation value of the singlet scalar field is determined, one can write the full thermal mass of the sterile neutrinos as the sum of the vacuum term and thermal corrections, both being non-zero below the symmetry-breaking temperature.
The vacuum term eventually outgrows the purely thermal contribution and the sterile neutrino becomes heavier than the leptons and scalar fields at relatively low temperatures, see figure~\ref{fig:ThermalMasses}, allowing the process $N_i\to \phi+L$ to take place.
We refer the reader to ref.~\cite{Seller:2023xkn} for additional details regarding the particle physics model, in particular for the definitions of the thermal masses, and the computation of the temperature-dependence of the vacuum expectation values.

As indicated in figure~\ref{fig:ThermalMasses} the thermal mass-hierarchy changes with temperature and the $N_i\to\phi+L$ sterile neutrino decay is only available below some temperature $T_1$ where the threshold $\overline{m}_{N_i}(T_1)=\overline{m}_L(T_1)+\overline{m}_\phi(T_1)$ is satisfied.
For the parameters used in figure~\ref{fig:ThermalMasses} this temperature is $T_1\approx 511\,$GeV. 
Exactly at $T=T_1$ the CM momentum vanishes, see eq.~\eqref{eq:CRenergyDefinitions}, and consequently the loop-energy integration boundaries defined in eq.~\eqref{eq:omegapm1} become equal:
\begin{equation}
    \omega_\pm^{(1)}\Big|_{\overline{m}_{N_i}=\overline{m}_L+\overline{m}_\phi} = \frac{E_N \overline{m}_L}{\overline{m}_{N_i}} \quad\rightarrow\quad \int_{\omega_-^{(1)}}^{\omega_+^{(1)}}\rd\omega\, f(\omega) = 0\,.
\end{equation}
As a result, at $T=T_1$ both contributions of the self-energy and of the first cut of the vertex function to the CP-asymmetry factor are zero.
This is easily understood as in both of these cases the CP-asymmetry is proportional to the decay rate (optical theorem) that vanishes due to no available phase-space for the final states.
Importantly, this is not the case for the second and third cuts of the vertex function, see eqs.~\eqref{eq:omegapm2} and \eqref{eq:omegapm3}, and there is no reason to assume that these cuts have vanishing contributions at $T=T_1$.
In fact, we expect that these contribute in a non-trivial way, as these cuts correspond to physical, necessarily thermal processes where particles are absorbed from the plasma \cite{Weldon:1983jn}.
This means that at and close to $T_1$, we expect that contributions from the second and third vertex cuts dominate over those of the first cut and the self-energy.

\begin{figure}[t]
    \centering
    \includegraphics[width=\linewidth]{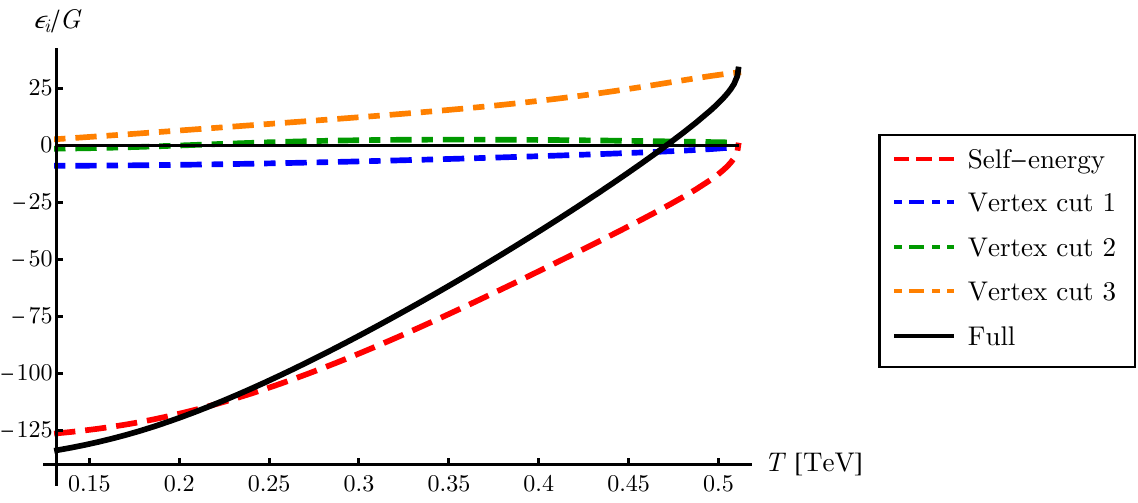}
    \caption{The full CP-asymmetry factor as a function of the temperature normalized with the common constant coupling factor $G$.
    We used two heavy neutrino mass states with a 10\% mass gap.
    The contributions from the self-energy diagram and from the cuts of the vertex function are shown in dashed and dash-dotted lines respectively, whereas their sum is given by the solid black line.}
    \label{fig:CPasymmetryFull}
\end{figure}

In figure~\ref{fig:CPasymmetryFull} we show the CP-asymmetry factor $\epsilon_i$ given in eq.~\eqref{eq:epsilonT_simpledef} and its separate contributions from the self-energy and vertex diagrams.
Using a 10\% mass gap between the two heavy sterile neutrino states (with the mass of the neutrino in the loop being the larger), the self-energy contribution mostly dominates over the vertex contributions due to the resonant behavior explained in and around eq.~\eqref{eq:EpsilonSigmaFinalFormula}.
In the figure we plot a normalized form of the CP-asymmetry factor where we factored out the constant coupling factor $G$, see~ eq.~\eqref{eq:couplingCorrespondence}.
As explained in the previous paragraph, the self-energy and the first cut of the vertex function vanish at $T=T_1$, while the contribution due to the other two cuts are finite (although it is difficult to see this in the figure for the contribution of the second cut).
In fact, the third cut dominates here, providing a large non-zero rate of CP-violation even at temperatures where the usually dominant self-energy contribution is small.
Since the second and third cuts are purely thermal, they vanish when $T\to 0$, this tendency is also visible in the figure. 

An interesting consequence of taking into account all cuts of the vertex function is that at the edge of the kinematically excluded region (see figure~\ref{fig:ThermalMasses}) the CP-asymmetry is not vanishing.
One would expect that the asymmetry function to be continuous for all temperatures, however, here the excluded region seemingly contradicts this expectation.
However, in a thermal plasma the allowed processes must necessarily also include interactions with particle absorption from the plasma, not only $1\to 2$ decays.
Indeed, the reasons for the second and third cuts being purely thermal were directly related to similar arguments.
It is thus expected that when all forms of interaction are taken into account, the CP-asymmetry factor is a continuous and non-zero function of the temperature for any $T$.

\section{Summary and outlook \label{sec:summary}}

In this article, we presented the evaluation of the one-loop CP-asymmetry factor at finite temperature in various approaches to thermal quantum field theory.
In particular, to evaluate the imaginary parts of Feynman diagrams in the real time formalism, we used (i) the finite temperature cutting rules originally introduced by Kobes and Semenoff and (ii) the retarded-advanced formalism, and showed that the two approaches lead to the same result.
We have also shown how this result may be obtained in the imaginary time formalism.

We showed that by considering the physical thermal vertex function, the resulting one-loop CP-asymmetry factor is linear in the statistical factor, as expected on general grounds in thermal field theory and found in studies using the Kadanoff-Baym formalism.
In agreement with previous literature \cite{Garny:2010nj}, we demonstrated that there is {\it no difference} between the thermal CP-asymmetry factors obtained in the various approaches, the perceived discrepancy may only be due to an incorrect identification of the physical quantity.

A new feature of our computations is that we have taken into account all contributions from the three cut diagrams of the vertex diagram, in addition to that of the self-energy diagram, with finite thermal masses for all particles involved. 
Furthermore, we have also derived a new, symmetric formula for the causal vertex function in terms of the cuts of the one-loop vertex diagram with only type-1 external CTP indices in the Kobes-Semenoff approach.
Finally, we derived compact parametric double- and triple-integrals for the thermal CP-asymmetry factors and presented numerical predictions indicating that the contributions due to all cut diagrams are to be considered when the temperature is comparable to the masses of the heaviest particles involved.

We have presented results in the special case when the mass hierarchy was fixed such that the neutrino in the loop was heavier than the decaying one.
For a complete study of the thermal CP-asymmetry factor, and consequently leptogenesis scenarios, different mass hierarchies should also be considered (as done in ref.~\cite{Garbrecht:2010sz}), for example a light internal neutrino in the vertex correction would also introduce contributions due to the previously not considered cuts that are not vanishing even in the vacuum.
Additionally, in case of right-handed neutrinos of comparable (same order of magnitude) but not equal masses the kinematically forbidden region may be closed, i.e. for a given mass hierarchy the forbidden regions of either decay can overlap with the allowed region of the other.
The fermion dispersion relation also affects the kinematically excluded region as already studied in ref.~\cite{Garbrecht:2019zaa} using hard thermal loop expansion technique.

\acknowledgments
This research was supported by the 
Excellence Programme of the Hungarian Ministry of Culture and 
Innovation under contract TKP2021-NKTA-64 and by the National Research, Development and Innovation Fund under contract NKKP-150794.
K.~S.~was partially supported by ÚNKP-23-4 New National Excellence Program of the Ministry for Culture and Innovation from the source of the National Research, Development and Innovation Fund.

\appendix 

\section{Vacuum decay amplitudes}
\label{sec:app:vacuum_amplitudes}

Using the Feynman rules given in ref.~\cite{Gluza:1991wj}, the tree and one-loop contribution to the decay amplitude reads ($u_L(s)\equiv u(P_L,s)$)
\begin{subequations}
\begin{align}
\mathcal{M}^{ab[1]}_{\alpha i} = \epsilon_{ab} \bar u_L(s) 
\mathbb{P}_{\rm R} \big[Y_{\alpha i} + V_{\alpha i}(P_N,P_L) + 
Y_{\alpha j}\, S(P_N,m_{N_j}) \big(-\ri\,\Sigma_{j_i}(P_N)\big) \big] u_N(s'),\\
\overline{\mathcal{M}}^{ab[1]}_{\alpha i} = \epsilon_{ab} \bar u_L(s) 
\mathbb{P}_{\rm L} \big[Y^*_{\alpha i} + \overline{V}_{\alpha i}(P_N,P_L) + 
Y^*_{\alpha j}\, S(P_N,m_{N_j}) \big(-\ri\,\Sigma_{j_i}(P_N)\big) \big] u_N(s'),
\end{align}
\end{subequations}
where $\mathbb{P}_{\rm R/L}=(1\pm\gamma_5)/2$ are the chiral projection operators, $S(K,m)=\ri/(\slashed{K} - m + \ri\varepsilon)$ is the fermion Feynman propagator, $\Sigma_{j_i}(P_N)$ denotes the one-loop self-energy matrix of the Majorana fermions, while $V_{\alpha i}(P_N,P_L)$ and $\overline{V}_{\alpha i}(P_N,P_L)$ are the one-loop vertex-functions for the direct and the CP conjugate decay.
Using relations such as $\mathbb{P}_{\rm L} (\slashed{P} + m) \mathbb{P}_{\rm R}= \slashed{P} \mathbb{P}_{\rm R}$ and $\mathbb{P}_{\rm R} (\slashed{P}_1 + m_1) \mathbb{P}_{\rm R} (\slashed{P}_2 + m_2) \mathbb{P}_{\rm L} = m_1 \mathbb{P}_{\rm R} \slashed{P}_2$, one can write
\begin{subequations}
\bea
\Sigma_{j_i}(P_N) &=& -2\ri\, \big(\mathcal{K}_{ji}\mathbb{P}_{\rm R} + \mathcal{K}_{ij} \mathbb{P}_{\rm L}\big)\int_K\! \slashed{K} D(K,m_L) D(P_N-K,m_\phi),
\\
V_{\alpha i}(P_N,P_L) &=& m_{N_j} Y_{\alpha j} (\mathcal{K}^{\rm T})_{ji} \mathcal{J}(P_N,P_L;m_\phi,m_L,m_{N_j}), \\
\overline{V}_{\alpha i}(P_N,P_L) &=& m_{N_j} Y^*_{\alpha j} \mathcal{K}_{ji} \mathcal{J}(P_N,P_L;m_\phi,m_L,m_{N_j}),
\eea
\end{subequations}
where in the self-energy we have taken into account the two possible fermion orientation in the loop and introduced the integral
\be
\mathcal{J}(P_N,P_L;m_L,m_\phi,m_{N_j}) = -\int_K\! \slashed{K} D(K-P_N+P_L,m_{N_j}) D(K-P_N,m_\phi) D(K,m_L).
\ee

\section{Vacuum scalar triangle integral}
\label{sec:app:vacuumscalarintegral}

The scalar triangle integral was computed at zero temperature in ref.~\cite{tHooft:1978jhc}. Introducing $Q=P-P'$, one has
\bea
\label{Eq:SV_def}
\mathcal{C}_{123}(P,P')&\equiv&\mathcal{C}(P,P';m_1, m_2, m_3) = -\int_K\! D(K+Q,m_3) D(K+P,m_2) D(K,m_1) \nonumber \\
&=&\frac{1}{16\pi^2}\int_0^1\! \rd x \int_0^x\! \rd y \big(a x^2 + b y^2 + c x y + d x + e y + f \big)^{-1}\nonumber \\
&=& \frac{\lambda^{-1/2}(P^2,P'^2,Q^2)}{16\pi^2} \sum_{k=1}^3 \int_0^1\! \rd y \frac{\ln\big(f_k(y)-\ri\,\varepsilon\big) - \ln\big(f_k(y_k)-\ri\,\varepsilon\big)}{y-y_k}.\ \ 
\eea
The second line in eq.~\eqref{Eq:SV_def} is the representation of the integral in terms of Feynman parameters, in which
\begin{gather}
a=P'^2,\quad b=P^2,\quad c=Q^2-P^2-P'^2,\quad d=-P'^2 + m_2^2 - m_3^2,\nonumber\\ \quad e = P'^2 - Q^2 + m_1^2 - m_2^2,\quad f=m_3^2 - \ri\,\varepsilon.
\end{gather}
This particular form, used as a starting point in ref.~\cite{tHooft:1978jhc}, is obtained from the many equivalent choices (see section~5.1.4 of ref.~\cite{Bardin:1999ak}) by assigning $x_i$ to the propagator with mass $m_i$, doing the shift $K\to K-P$ before the Wick rotation, using the Dirac-delta $\delta(x_1+x_2+x_3-1)$ to do the $x_2$ integral, followed by changing the variables $x_3=1-x$ and $x_1=y$.

The third line in eq.~\eqref{Eq:SV_def} is the main result of ref.~\cite{tHooft:1978jhc} concerning the scalar triangle integral. 
It was obtained with ingenious splitting of the Feynman integral and changes of variables in order to relate to the three possible cuts that determine the imaginary part. 
The Källén function appearing in this expression is defined by
\be
\label{Eq:Kallen_fv}
\lambda(x,y,z) = x^2 + y^2 + z^2 -2 x y -2 x z -2 y z, 
\ee
while $f_k(y)$ can be given in terms of the function $f(y;K,m,M)=K^2 y^2 + (-K^2 + m^2 -M^2) y + M^2$, which is familiar from the bubble integral with two different masses\footnote{It is easy to see that for $0\le y\le 1$ one can have $f(y;K,m,M)<0$ only for $K^2>0$.}:
\bea
\label{Eq:fi_y}
f_1(y) = f(y;P,m_1,m_2),\quad f_2(y) = f(y;Q,m_1,m_3),\quad f_3(y) = f(y;P',m_2,m_3).
\eea
The values $y_k$ are defined as
\bea
y_1=y_0+\alpha_+,\quad y_2=\frac{y_0}{1-\alpha_+}, \quad y_3 = -\frac{y_0}{\alpha_+},
\eea
where $y_0=-\frac{d+e\alpha_+}{c+2b\alpha_+}$ with $\alpha_+=(-c+\sqrt{c^2-4ab})/(2b)$ chosen to be the largest solution of $b \alpha^2 + c \alpha + a=0$, case in which $y_0=-(d+e\alpha_+)\lambda^{-1/2}(P^2,P'^2,Q^2)$. 

In our case of interest, i.e., the decay of a Majorana neutrino with mass $m_i$ into a scalar and a lepton, we have the correspondence $m_1=m_L$, $m_2=m_\phi$, $m_3=m_{N_j}$ for the masses, with $m_{N_j}$ being the mass of the intermediate Majorana neutrino, $P=P_N$, $P'=P_L$, $Q=P_N-P_L=P_\phi$ for the four-momenta and the mass-shell conditions $P_N^2=m_{N_i}^2$, $P_\phi^2=m_\phi^2$, $P_L^2=m_L^2$. 
When calculating the decay of the neutrino in the CM frame one has also $P_L\cdot P_N = (m_{N_i}^2 + m_L^2 - m_\phi^2)/2$, and therefore
\begin{subequations}
\begin{align}
    \alpha_+ &= \frac{1}{2m_{N_i}^2}\big(m_{N_i}^2+m_\phi^2-m_L^2+\lambda^{1/2}(m_\phi^2,m_L^2,m_{N_i}^2)\big), \\ 
    y_0 &= \frac{m_{N_j}^2 + (m_\phi^2-m_L^2)(1-2\alpha_+)}{\lambda^{1/2}(m_\phi^2,m_L^2,m_{N_i}^2)}.
\end{align}
\end{subequations}
In the CM frame the functions introduced in eq.~\eqref{Eq:fi_y} can be written for $m^2_{\phi,L}\ne0$ as
\be
f_k(y)=c_k g(y;a_k,b_k),\qquad  g(y;a,b) = y^2 + (a-b-1)y + b,
\ee
with $c_1=m_{N_i}^2$, $c_2=m_\phi^2$, $c_3=m_L^2$, $a_{1,2}=m_L^2/c_{1,2}$, $b_1=m_\phi^2/c_1$, $a_3=m_\phi^2/c_3$, and $b_{2,3}=m_{N_j}^2/c_{2,3}$.

Since $\imag \ln(x-\ri\,\varepsilon)=-(1-\textrm{sgn}(x))\pi/2$, one can have imaginary part if $f_k(y)<0$\footnote{Note that for $m_\phi=m_L=0$, one has $f_2(y)=f_3(y)=m_{N_j}^2(1-y)>0$ and therefore the second and third cuts do not contribute to the imaginary part of the scalar triangle integral.}. 
A bit of algebra shows that this is possible only when $y_\pm^{(k)}\in [0,1]$ ($y_\pm^{(k)}$ are the two solutions of $f_k(y)=0$), condition which is satisfied when $1>\sqrt{a_k}+\sqrt{b_k}$, $k=1,2,3$. 
When $y_k\in[y_-^{(k)},y_+^{(k)}]$ the contributions of the two logarithms in the last expression of eq.~\eqref{Eq:SV_def} cancel each other. 
On the other hand, when $y_k\notin[y_-^{(k)},y_+^{(k)}]$, or equivalently $\big(y_k-y_+^{(k)}\big)\big(y_k-y_-^{(k)}\big)>0$, one has contribution only from the first logarithm, namely,
\bea
\imag \int_0^1\! \rd y \frac{\ln\big(f_k(y)-\ri\,\varepsilon\big)}{y-y_k} = -\pi \int_{y_-^{(k)}}^{y_+^{(k)}}\! \rd y \frac{1}{y-y_k} = -\pi \ln\frac{y_k-y_+^{(k)}}{y_k-y_-^{(k)}},
\eea
provided that $1>\sqrt{a_k}+\sqrt{b_k}$.

Putting everything together, using the kinematics of the decay of the neutrino in the CM frame, the imaginary par of the scalar triangle integral is non-vanishing provided that $\big[m_{N_i}^2 m_{N_j}^2-(m_\phi^2-m_L^2)^2\big] \big[m_{N_i}^2+m_{N_j}^2-2 (m_\phi^2+m_L^2)\big]>0$ and has the expression
\be
\label{eq:ImCmassrelation}
\!\!\!\!\imag\,\mathcal{C}_{L\phi N_j}(P_N,P_L)\Big|_\textrm{CM} = -\frac{1}{16\pi \sqrt{\lambda_i}}
\left\{
\begin{array}{ll}\displaystyle
      \ln\frac{1}{1+\frac{\lambda_i}{m_{N_i}^2m_{N_j}^2-(m_\phi^2-m_L^2)^2}}, & \ m_{N_i} > m_L + m_\phi, \\[4pt]
      \ln\frac{F_2 - \sqrt{\lambda_i \lambda_j}}{F_2 + \sqrt{\lambda_i \lambda_j}}, & \ m_\phi > m_{N_j} + m_L, \\[4pt]
      \ln\frac{F_3 - \sqrt{\lambda_i \lambda_j}}{F_3 + \sqrt{\lambda_i \lambda_j}}, & \ m_L > m_{N_j} + m_\phi, \\
\end{array} 
\right.
\ee
where we used the shorthands $\lambda_{i,j}=\lambda(m_\phi^2,m_l^2,m_{N_{i,j}}^2)$ and
\begin{subequations}
\begin{align}
    F_2 &= m_{N_i}^2 m_{N_j}^2 + (m_{N_i}^2 + m_{N_j}^2 - 3 m_\phi^2 - m_L^2)(m_\phi^2-m_L^2),\\
    F_3 &= m_{N_i}^2 m_{N_j}^2 + (m_{N_i}^2 + m_{N_j}^2 - 3 m_L^2 - m_\phi^2)(m_L^2-m_\phi^2).
\end{align}
\end{subequations}
In the case when $m_{N_i}>m_L+m_\phi$, that is when the decay is non-vanishing at tree-level, one can make the following observations: (i) cut 1 always contributes and for $m_\phi=m_L=0$ it is the only contribution to the imaginary part 
\[
\lim_{m_{\phi,L}\to 0}\imag \mathcal{C}_{L\phi N_j}(P_N,P_L)\Big|_\textrm{CM}=-\frac{1}{16\pi m_{N_i}}\ln\frac{m^2_{N_j}}{m^2_{N_i}+m^2_{N_j}},
\]
(ii) when $m_{N_j}=m_{N_i}$, then one has contribution only from cut 1, (iii) when $m_{N_j}<m_{N_i}$ one can have additional contribution from either cut 2 or cut 3, depending on the masses (cut 2 and cut 3 do not contribute simultaneously).  

\section{Finite temperature cutting rules}
\label{sec:app:FTcuttingRules}

In this appendix we introduce the method for evaluating the imaginary parts of Feynman amplitudes at finite temperature.
We follow the so-called circling method devised by Kobes and Semenoff \cite{Kobes:1985kc,Kobes:1986za} (see also refs.~\cite{Aurenche:1991hi,Gelis:1997zv}) in the context of the RTF of the thermal quantum field theory (see e.g. \cite{Bellac:2011kqa}).
We begin with introducing the RTF propagators for bosons and fermions, then we present the cutting rules at finite temperature, and finally we give the formula for the imaginary part of an arbitrary $n$-point function.

In the RTF the usual time ordering operator along the real time axis is exchanged with a so-called path ordering operator along the Schwinger-Keldysh contour (closed-time-path, CTP) \cite{Schwinger:1960qe,Keldysh:1964ud} that spans from $t=-\infty$ to $t=\infty$ and back with an additional piece in the imaginary direction needed in equilibrium studies.
Due to the doubling of the contour along the real time axis the thermal propagator becomes a $2\times 2$ matrix with each component corresponding to a propagator connecting time coordinates lying on either the ``forward" temporal path (corresponds to real particles) or the ``backward" temporal path (corresponds to ghosts).

Subject to periodic (for bosons) and anti-periodic (for fermions) boundary conditions, called the Kubo–Martin–Schwinger (KMS) conditions, the propagators in momentum space are as follows.
In the so-called {\emph symmetric assignment} (see chapter 3.3 of ref.~\cite{Bellac:2011kqa}) the bosonic thermal propagators are
\begin{equation}
\label{eq:boson_propagator_11_12}
\begin{aligned}
    G_{11}(P,m)&=G_{22}^*(P,m)=\ri\mathcal{P}\left(\frac{1}{P^2 - m^2}\right) + \pi\delta\big(P^2-m^2\big)\big(1+2\BE(|p^0|)\big)\,, \\
    G_{12}(P,m)&=G_{21}(P,m)=2\pi\delta\big(P^2-m^2\big)e^{\beta|p^0|/2}\BE(|p^0|)\,,
\end{aligned}
\end{equation}
while for fermions we express the propagators as $S_{ij}(P,m)=(\slashed{P}+m)\Tilde{S}_{ij}(P,m)$, where
\begin{equation}
\label{eq:fermion_propagator_11_12}
\begin{aligned}
    \nonumber
    \Tilde{S}_{11}(P,m)&=\Tilde{S}_{22}^*(P,m) = \ri\mathcal{P}\left(\frac{1}{P^2-m^2}\right)+\pi\delta\big(P^2-m^2\big)\big(1-2\FD(|p^0|)\big)\,, \\
    \Tilde{S}_{12}(P,m)&=-\Tilde{S}_{21}(P,m) = -\mathrm{sgn}(p^0)\,2\pi\delta\big(P^2-m^2\big)e^{\beta |p^0|/2}\FD(|p^0|)\,.
\end{aligned}
\end{equation}
We denoted the principal value with $\mathcal{P}$. 
In both cases the temperature-dependent part of the propagator is explicitly on-shell and is linear in the statistical factor.
In fact, {\it all} temperature dependence appears through $f_{\rm B/F}$ (this is also generally true in finite temperature field theory).
The thermal parts of these propagators explicitly break Lorentz-invariance, consequently the notation $G_{ij}(P,m)$ is meant as a shorthand to indicate that the propagator depends on the \emph{components} of the 4-momentum.

We also introduce the positive and negative frequency propagators $G_{\pm}(P,m)$ (see e.g. chapter 4.6 in ref.~\cite{Greiner:1996zu}).
These are trivially connected with the off-diagonal elements of the propagator matrix,
\begin{subequations}
\begin{gather}
    \label{eq:boson_propagator_offdiag_relation}
    G_\pm(P,m) = e^{\pm\beta p^0/2}G_{12}(P,m)\,, \\
    \label{eq:fermion_propagator_offdiag_relation}
    S_\pm(P,m) = \mp e^{\pm \beta p^0/2}S_{12}(P,m)\,,
\end{gather}
\end{subequations}
or explicitly:
\begin{subequations}
\begin{gather}
    G_\pm(P,m) = 2\pi\delta(P^2-m^2)\left[\theta(\pm p^0) + \BE(|p^0|)\right]\,, \\
    \Tilde{S}_\pm(P,m) = 2\pi\delta(P^2-m^2)\left[\theta(\pm p^0) - \FD(|p^0|)\right]\,.
\end{gather}
\end{subequations}
Note that the temperature-dependent part of the propagators is not vanishing for any sign of the particle energy thus the direction of the energy flow is not constrained as it is in vacuum.

Each vertex in a Feynman diagram corresponds to a specific space-time point.
However, in thermal field theory these vertices can lie either on the ``forward" or the ``backward" temporal path of the Schwinger-Keldysh contour.
It follows that we can draw $2^{\rm V}$ distinct diagrams, where V is the number of vertices.
The vertices are labeled with a CTP index 1 or 2 that reflect their position on the Schwinger-Keldysh contour.
In addition to the CTP indices, for the evaluation of the imaginary part of the diagram we also introduce \emph{circling} of vertices.
Circled vertices are simply the complex conjugates of the non-circled ones.
The formula for the imaginary part of an $n$-point function is then given by a double sum over all possible circlings of vertices and all possible CTP index assignments (see, e.g.~\cite{Aurenche:1991hi}).
The Feynman rules for evaluating these graphs are given at the end of section 3 in ref.~\cite{Kobes:1985kc}.
Diagrammatically one has the following Feynman rules.
There are 2 types of vertices ($a=1,2$) in the Lagrangian, combined with the circling rule one has:
\begin{equation*}
    \includegraphics{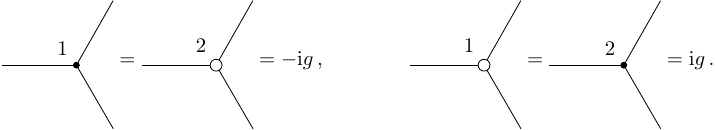}
\end{equation*}
For the propagators we have 16 different options.
Disregarding the particle nature and denoting the general propagator with $\Delta$, the cutting rules for the propagator are depicted as follows (arrows indicate momentum flow):

\begin{equation*}
    \includegraphics{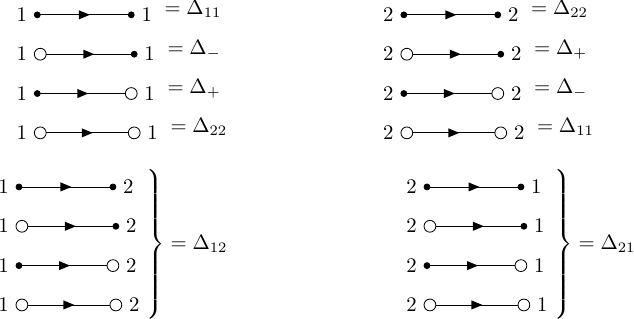}
\end{equation*}

Let $G_{\{a_i\}}\big(\{P_i\}\big)$ denote an $n$-point function ($i=1,2,\dots,n$) with external momenta $P_i$ and external vertex CTP indices $a_i = 1$ or 2.
The internal indices are labeled as $v_j$, but note that at one loop level all vertices are external vertices.
Additionally, let $\{a_i\}_{\mathbf{S}}$ denote a configuration of vertices where a set of vertices $\mathbf{S} \subset \{a_i\}$ is circled and the rest is left uncircled, with $\mathbf{S} \neq \{a_i\}$, $\varnothing$.
The imaginary part of the $n$-point function is given as \cite{Gelis:1997zv}:
\begin{equation}
    \label{eq:nPointFuncImagPartDefinition}
    \imag \Big[\ri G_{\{a_i\}}\big(\{P_i\}\big)\Big] = -\frac{1}{2}\sum_{\{v_j=1,2\}}\sum_{\mathbf{S},\mathbf{Z}}G_{\{a_i\}_{\mathbf{S}},\{v_j\}_{\mathbf{Z}}}\big(\{P_i\}\big)\,.
\end{equation}
The first sum adds all possible assignments of the CTP indices of internal vertices and the second sum denotes all possible circling of vertices with the exception of when all or none of the vertices are circled.

In the special case of one-loop diagrams all vertices are external vertices and the right-hand side of eq.~\eqref{eq:nPointFuncImagPartDefinition} reduces to a single sum over circlings of external vertices,
\begin{equation}
    \label{eq:nPoint1loopImag}
    \imag \Big[\ri G^{\rm (1)}_{\{a_i\}}\big(\{P_i\}\big)\Big] = -\frac{1}{2}\sum_{\mathbf{S}}G^{\rm (1)}_{\{a_i\}_{\mathbf{S}}}\big(\{P_i\}\big)\,.
\end{equation}
Assuming 3-particle interaction vertices, for a one loop $n$-point function there are $n$ total vertices and $2^{n}-2$ distinct circled diagrams contribute on the right-hand side of eq.~\eqref{eq:nPoint1loopImag}.
For the self-energy diagram ($n=2$) and the vertex correction ($n=3$) we have 2 and 6 diagrams respectively.
For these cases specifically it is possible to assign cuts to the circled diagrams, in particular, two diagrams with complementary circlings constitute a cut diagram, i.e., we have 1 cut for the self-energy and 3 cuts for the vertex correction.
Note that at higher loop order one generally cannot assign a cutting picture.

In summary, in this appendix we introduced the cutting rules and the definition of the imaginary part of any $n$-point function.
These $n$-point functions have specific external CTP indices.
Initially it was thought that external indices must be of type 1 (i.e., real particles) for physical processes. 
However, this is not true and a combination of diagrams with various CTP indices must be taken to find the physical amplitude, which is detailed in section~\ref{sec:RA_vertex} of the main text.

\blue{
\section{On the inclusion of the thermal mass}
\label{sec:app:thermal-mass}

In order to introduce the thermal mass, one has to consider one-loop self-energy corrections to bosonic and fermionic tree-level propagators with vacuum mass $m$, which are usually calculated in the high-temperature limit $T\gg m$.
In this limit the scalar mass squared receives a $\Delta m_{\rm scalar}^2=\mathcal{O}(T^2)$ additive correction from the various one-loop diagrams: (i) the scalar tadpole, (ii) the fermion bubble, and (iii) the gauge boson tadpole (see App.~C of ref.~\cite{Seller:2023xkn}).

For fermions one has to consider a self-energy with a scalar-fermion loop (bubble diagram).  
Compared to scalars, the situation is more complicated because the self-energy corrected propagator, called the hard thermal loop (HTL) resummed propagator in the approximation mentioned above, has two poles corresponding to two thermal excitations (quasiparticles) with different dispersion relations. 
In order to see this within the real-time formalism, one uses the self-energy $\Sigma_\textrm{HTL}$ calculated first by Weldon in the form
\begin{equation}
\Sigma_\textrm{HTL}(P;T)=-a(P;\Delta m_f)\slashed{P}-b(P;\Delta m_f)\slashed{u},
\end{equation}
where $u_\mu$ is the four velocity of the heat bath satisfying $u_\mu u^\mu=1$, $a$ and $b$ are two scalar functions given in eq.~(2.5) of ref.~\cite{Weldon:1982bn} and $\Delta m_f(T)$ is the
thermal correction to the tree-level fermion mass $m$.

As shown in ref.~\cite{Kobes:1985kc}, the self-energy enters the propagator $\mathcal{D}(P;m)=\ri/(\slashed{P}-m-\Sigma_\textrm{HTL}(P;T) + \ri\varepsilon)$, in terms of which the $2\times2$ HTL resummed propagator matrix $\mathcal{S}$ is defined:
\begin{equation}
\label{eq:resummed_propagator_matrix}
\mathcal{S}(P) = \mathcal{V}(P;T)\left(
\begin{matrix} \mathcal{D}(P;m) & 0 \\ 0 & \mathcal{D}^*(P;m)
\end{matrix}\right) 
\mathcal{V}(P;T)\,.     
\end{equation} 
The orthogonal matrix $\mathcal{V}$
depends on the Fermi-Dirac statistical factor, as given below eq.~(2.7b) of ref.~\cite{Kobes:1985kc}. 
Its form is such that for $\Sigma_\textrm{HTL}\to 0$ one recovers from $\mathcal{S}$ the tree-level matrix elements $S_{ij}$ given below eq.~\eqref {eq:boson_propagator_11_12}.

In the rest frame of the plasma, where the 4-velocity is $u^\mu=(1,{\bf 0})$, one has the propagator $\mathcal{D}(P;m)=\ri(\slashed{P}(1+a)+b\gamma^0+m)/\mathtt{D}_m(P)$, with denominator
\begin{equation}
\mathtt{D}_m(P)=\prod_{s=\pm}\big[p_0(1+a)+b-s \sqrt{\p^2(1+a)^2+m^2}\ \big] = \prod_{s=\pm}\mathtt{D}_m^s(p_0,|\p|;\Delta m_f).
\end{equation}
In the massless limit the propagator reduces to
\begin{equation}
\label{eq:propagator_massless_limit}
\mathcal{D}(P;0)=\sum_{s=\pm} \frac{\ri \gamma_0 \Lambda_{-s}(p)}{(1+a)(p_0-s|\p|)+b} = 
\sum_{s=\pm} \frac{\ri \gamma_0 \Lambda_{-s}(\p)}{\mathtt{D}_0^s(p_0,|\p|;\Delta m_f)},
\end{equation}
where $\Lambda_{\pm}(\p)=\frac12(1\pm \gamma_0 \bm{\gamma}\cdot \p/|\p|)$ are two orthogonal projectors.

The poles of the propagator for $p_0>0$, given by $\mathtt{D}_m^\pm(p_0,|\p|;\Delta m_f)=0$, result in dispersion relations of the form $p_0=\omega^\pm_m(|\p|;\Delta m_f)$, with
\begin{equation}
\label{eq:omega_pm_dispersion}
\omega^\pm_m(|\p|;\Delta m_f) = m_\pm + \frac{1}{\Delta m_f^2+m_\pm^2}\left[\frac{\Delta m_f^2}{3m_\pm}
\pm \frac{1}{2 m}\left(\frac{3 m_\pm^2 - \Delta m_f^2}{3 m_\pm}\right)^2\right] \p^2 + \mathcal{O}(\p^4),
\end{equation}
where $m_\pm =(\sqrt{m^2+4\Delta m^2} \pm m)/2$.
These were studied in ref.~\cite{Petitgirard:1991mf}, while the corresponding dispersion relations of the massless case were studied in ref.~\cite{Weldon:1982bn}.
The $\omega^+$ dispersion relation corresponds to a particle-like excitation, while the $\omega^-$ one to a hole-like excitation. 
The residue of the latter vanishes exponentially for increasing $|\p|/T$ and it is already negligible at $|\p|/T\simeq 0.06$. 
For this reason the hole-like excitation is usually neglected in integrals involving resummed HTL propagators. 
As a result, only the particle-like dispersion relation $\omega^+_m(|\p|;\Delta m_f)$ has to be considered. 

In the massless case, which applies to leptons at temperatures above the electroweak phase transition, $\omega_0^+(|\p|,\Delta m_L)$ can be well approximated by the usual relativistic dispersion relation $\omega(|\p|)=\sqrt{\p^2+\Delta m_L^2}$ (see ref.~\cite{Giudice:2003jh}, although in most of their calculations the full structure of the fermion propagator is kept), therefore one can use the replacement $m^2\to \overline{m}_L^2\approx \Delta m_L^2$ in the tree-level propagator matrix. 
However, according to eq.~\eqref{eq:propagator_massless_limit}, this thermal mass only appears in the denominator of the propagator, but not in the numerator as one would expect for massive fermions.
In the calculation of the CP-asymmetry factor the structure of the numerator only matters for the spinor trace evaluated in eq.~\eqref{eq:DiracTrace}, which in this case leads to the same expression regardless of using the tree-level fermion propagator with mass exchanged as above, or using eq.~\eqref{eq:resummed_propagator_matrix} with \eqref{eq:propagator_massless_limit}.

When the vacuum mass is much larger than the thermal corrections ($m^2\gg \Delta m^2$), which applies to our right-handed neutrinos, one has from eq.~\eqref{eq:omega_pm_dispersion} an approximate dispersion relation $\omega_m^+(|\p|;\Delta m_f)\simeq M+p^2/2M$ with an effective mass squared $M^2\simeq m^2+2\Delta m_f^2$ \cite{Petitgirard:1991mf}.
The same is obtained in the large-momentum limit $\Delta m\ll |\p|\ll T$, where the dispersion relation is $\omega_m^+(|\p|,\Delta m_f)\simeq |\p| + (m^2 + 2\Delta m^2)/2|\p|$ \cite{Chaudhuri:2023djv}. 
Therefore, for the right-handed neutrinos one can use the replacement of the form $m_N^2\to \overline{m}^2_N \simeq m^2_N + \Delta m^2_N$ in the tree-level propagator matrix.
}

\section{Retarded-advanced formalism}
\label{sec:app:RAformalism}

In this appendix we introduce the retarded-advanced (RA) propagator formalism for thermal field theory and discuss its connection to the Kobes-Semenoff (KS) formalism.
In particular, here we introduce the matrices connecting the two schemes and show how an arbitrary $n$-point function may be expressed.
As an example, we re-derive the constraint equation of eq.~\eqref{eq:MainTextVertexConstraint} from a different approach than the diagrammatic method presented in the main text.

Following ref.~\cite{Aurenche:1991hi} we can derive useful relations related to the finite temperature vertex function in the RA formalism.
The rotation matrix between the usual CTP indices (latin indices, $a=1,2$) and those of the RA formalism (greek indices, $\alpha={\rm A,B}$) in the symmetric assignment is:
\begin{equation}
    \label{eq:Vmatrix_RA_KS}
    \mathbf{V}_{\alpha a}^{(\eta)}(K) =
    \begin{pmatrix}
        1 & \eta\exp\left(-\frac{\beta k^0}{2}\right) \\
        -\eta f_\eta(k^0) & -\eta f_\eta(k^0)\exp\left(\frac{\beta k^0}{2}\right)
    \end{pmatrix}\,,
\end{equation}
Here $\eta=\pm$ corresponds to bosons and fermions respectively with
\begin{equation}
    f_\pm(k^0) \equiv f_{\rm B/F}(k^0) = \frac{1}{\exp(\beta k^0)\mp 1}\,.
\end{equation}
In calculations we will use the notation $\mathbf{V}^{(\rm B/F)}_{\alpha a}(K)$ instead of having $\eta=\pm$ in the superscript.
In the CTP index notation of the RTF one has two types of vertices on the Lagrangian level (see, e.g.~chapter 2.5 in \cite{Das:1997gg}): $g_{111}=g$ and $g_{222}=-g$ (all other components are zero).
Then the matrix given in eq.~\eqref{eq:Vmatrix_RA_KS} gives the connection between the tree-level vertex functions with all incoming kinematics as
\begin{equation}
    \label{eq:treeLevelCouplingsRA}
    -\ri \gamma_{\alpha\beta\rho}^{\eta_1\eta_2\eta_3}(P_1,P_2,P_3) = -\ri V^{(\eta_1)}_{\alpha a}(P_1) V^{(\eta_2)}_{\beta b}(P_2) V^{(\eta_3)}_{\rho c}(P_3) g_{abc}\,.
\end{equation}
As we see, in the RA formalism the thermal dependence appears in the vertex functions.
Using momentum conservation $\sum_iP_i=0$ and denoting $P_1\equiv P$ and $P_3\equiv Q$, the final formula for the tree level vertex becomes
\begin{equation}
\label{eq:general_stripped_tree_vertex_RA}
\begin{aligned}
    &\Tilde{\gamma}_{\alpha\beta\rho}^{\eta_1\eta_2\eta_3}(P,Q) = \\
    &\,\left[-f_{\eta_1}(p^0)\right]^{\delta_{\alpha \rA}}\left[-f_{\eta_2}(-p^0-q^0)\right]^{\delta_{\beta \rA}}\left[-f_{\eta_3}(q^0)\right]^{\delta_{\rho \rA}}\left[\eta_1^{\delta_{\alpha \rA}}\eta_2^{\delta_{\beta \rA}}\eta_3^{\delta_{\rho \rA}}-\eta_1\eta_2\eta_3e^{-\beta X^0}\right]\,,
\end{aligned}
\end{equation}
where $X^0=(\delta_{\alpha \rR}-\delta_{\beta\rR})p^0+(\delta_{\rho\rR}-\delta_{\beta\rR})q^0$ and $f_{\pm}(p^0)\equiv f_{\rm B/F}(p^0)$, and the tilde denotes the stripped vertex function, where the coupling and any possible spinor structure of the vertex function is decoupled.

In physical calculations one generally has incoming as well as outgoing particles in a vertex.
In ref.~\cite{Aurenche:1991hi} the authors defined a notation where incoming and outgoing particles are separated with a semicolon, e.g.,
\begin{equation}
    \gamma_{\alpha\beta;\rho}(P,Q;R) \equiv \gamma_{\alpha\beta\Bar{\rho}}(P,Q,-R)\,,\text{ where }P+Q=R\,.
\end{equation}
Here $\Bar{\rho}$ indicates the ``conjugation" of the index, i.e., the exchange of R$\leftrightarrow$A\,.
We can generalize the matrix defined in eq.~\eqref{eq:Vmatrix_RA_KS} to be applicable to outgoing particles instead of only incoming ones.
For an $n$-point function with $n_{\rm in}$ incoming and $n_{\rm out}=n-n_{\rm in}$ outgoing momenta we write
\begin{equation}
    \label{eq:RA_KS_correspondence_general}
    F_{\{\alpha_i\};\{\sigma_j\}}\big(\{P_i\};\{Q_j\}\big) = \left[\prod_{i=1}^{n_{\rm in}}\mathbf{V}^{(\eta_i)}_{\alpha_i a_i}(P_i)\right] F_{\{a_i\};\{c_j\}}\big(\{P_i\};\{Q_j\}\big)\left[\prod_{j=1}^{n-n_{\rm in}}\mathbf{U}^{(\eta_j)}_{c_j\sigma_j}(Q_j)\right]\,,
\end{equation}
where the matrix acting on the outgoing states is given by
\begin{equation}
    \label{eq:Umatrix_RA_KS}
    \mathbf{U}^{(\eta)}_{a\alpha}(K) = 
    \begin{pmatrix}
        f_\eta(k^0) \exp(\beta k^0) & 1 \\
        f_\eta(k^0) \exp\left(\frac{\beta k^0}{2}\right) & \eta\exp\left(\frac{\beta k^0}{2}\right)
    \end{pmatrix}\,.
\end{equation}
As for the $\mathbf{V}$-matrix, in calculations we will use the notation $\mathbf{U}^{\rm (B/F)}_{\alpha a}(K)$ instead of having $\eta=\pm$ in the superscript.
Physically, if all indices are of the retarded type ($\alpha_i={\rm R}$ and $\sigma_j={\rm R}$), then eq.~\eqref{eq:RA_KS_correspondence_general} gives the amplitude of a process with incoming particles with momentum $P_i$ and outgoing particles with momentum $Q_j$ \cite{Aurenche:1991hi}.

In this paper the central physical process is that involving the $N\phi L$ vertex and we continue with presenting an example calculation with the RA formalism using this process.
We consider the vanishing of the RRR component of the thermal vertex function, $\Gamma_{\rm RR;A}(P,Q;R)\equiv\Gamma_{\rm RRR}(P,Q,-R)=0$ (see eq.~\eqref{eq:FBF_relations} and below).
By definition this is related to the vertex functions of the CTP index notation via eq.~\eqref{eq:RA_KS_correspondence_general},
\begin{equation}
    \label{eq:GammaRRpA}
    \Gamma_{\rm RR;A}^{L\phi N}(P,Q;R) = \sum_{a,b,c=1,2}\mathbf{V}^{(\rm F)}_{{\rm R}a}(P)\mathbf{V}^{\rm (B)}_{{\rm R}b}(Q)\Gamma^{L\phi N}_{abc}(P,Q,R)\mathbf{U}^{\rm (F)}_{c{\rm A}}(R)\,. 
\end{equation}
Since we are only interested in the imaginary parts of the vertex function we make use of the following relations (see eqs.~\eqref{eq:ABB_imaginary_part_relations} and \eqref{eq:AAA_imaginary_part_relations} in the main text):
\begin{align}
    \imag\Gamma_{111}^{L\phi N} = \imag\Gamma_{222}^{ L\phi N}&\,,\quad \imag\Gamma_{112}^{L\phi N} = -\imag\Gamma_{221}^{L\phi N}\,, \\
    \imag\Gamma_{121}^{L\phi N} = \imag\Gamma_{212}^{L\phi N}&\,,\quad \imag\Gamma_{211}^{\rm L\phi N} = -\imag\Gamma_{122}^{L\phi N}\,.
\end{align}
Using $P+Q=R$ and the above imaginary part relations, eq.~\eqref{eq:GammaRRpA} simplifies to a constraint equation between 4 components of the thermal vertex function: 
\begin{equation}
    \label{eq:GammaRRRconstraint}
    \imag\Gamma_{111}^{L\phi N} + \imag\Gamma_{121}^{L\phi N}\cosh\left(\frac{\beta q^0}{2}\right) + \imag\Gamma_{211}^{L\phi N}\sinh\left(\frac{\beta p^0}{2}\right)-\imag\Gamma_{112}^{L\phi N}\sinh\left(\frac{\beta r^0}{2}\right)=0\,.
\end{equation}
This equation may be used to eliminate an additional degree of freedom leading to a thermal vertex function with a total of three independent components.
We found the same relation in eq.~\eqref{eq:MainTextVertexConstraint} with the proper assignment of momenta ($P=-P_L$, $Q=-P_\phi$, $R=-P_N$) and noting that $\Gamma^{L\phi N}_{abc}=\Gamma^{N\phi L}_{cba}$, latter being the convention used in the main text.

\section{Gaudin method applied to the thermal scalar vertex function}
\label{sec:app:Gaudin}

Gaudin's method~\cite{gaudin65} is a systematic way to perform Matsubara sums algebraically. 
The method is especially efficient in case of multiloop Feynman graphs, where for $L$ loops there are $L$ Matsubara sums.
The method uses the spectral representation of the propagators, hence it is applicable also when full propagators are used instead of the Feynman ones.
For details, subtleties, and applications of the method the reader is referred to refs.~\cite{Espinosa:2005gq, Blaizot:2004bg, Mottola:2009mi}. 

We write the $L$-loop Feynman graph containing all $I$ internal propagators in the spectral representation (see~eq.~\eqref{Eq:spectral_rep}).
Gaudin's method then consists of two steps: 
(i) first the product of $I$ propagators results in a product of fractions whose denominators are linear in the Matsubara frequencies, which is then decomposed into a sum of terms, each of them containing the product of $L$ original fractions and $I-L$ fractions for which the Matsubara frequencies have been expressed as functions of the external Matsubara frequencies and the real energies of the $L$ unchanged fractions; 
(ii) a regulator is assigned to each of the remaining $L$ original fractions and the corresponding Matsubara sum is performed.

The result of (i) is a decomposition formula of the form
\begin{equation}
\label{app:Gaudin_decomp}
    \prod_{i=1}^{I}\frac{1}{k_i^0-\ri\omega_{n_i}} = \sum_{g \in \mathcal{G}}
    \bigg[
    \prod_{j\in \mathcal{T}_g} \frac{1}{k_j^0-\ri\Omega_j\big(\{k_i^0\}; \{\ri \omega_{e_i}\}\big)}
    \,\prod_{l\in \bar{\mathcal{T}}_g} \frac{1}{k_l^0-\ri\omega_{n_l}}
    \bigg]\,.
\end{equation}
In this formula $I$ is the total number of internal lines and $\omega_{n_i}$ denotes a generic Matsubara frequency corresponding to the $i$th propagator, with $n_i\in\mathbb{Z}$.
Additionally, $\mathcal{G}$ is the set of Gaudin graphs, $\mathcal{T}_g$ denotes those lines of a particular graph $g\in\mathcal{G}$ that form a \emph{Gaudin tree} (lines that connect all vertices without forming a closed loop), while $\bar{\mathcal{T}}_g$ is the complement of $\mathcal{T}_g$ within the graph $g$.
It follows that $|\bar{\mathcal{T}}_g|=L$ and $|\mathcal{T}_g|+|\bar{\mathcal{T}}_g|=I$.
For each $g$, the frequencies associated to the lines of $\bar{\mathcal{T}}_g$ are considered independent Matsubara frequencies, while those associated to the lines of $\mathcal{T}_g$ are denoted by $\Omega_j$ and expressed in terms of real energies $k_l^0$ ($l\in\Bar{\mathcal{T}}_g$) and external Matsubara frequencies $\omega_{e_i}$, with $e_i\in\mathbb{Z}$.  

In step (ii) of the method, a regulator of the form $\exp(\ri \omega_{n_i}\tau_i)$ is assigned to all internal propagators ($i=1,2\dots I$), where $\tau_i = \kappa_i\theta$ with $\kappa_i\in\mathbb{Z}^+$ and $\theta>0$.
The regulator is removed upon taking the limit $\theta\to 0^+$.
Then the following summation formula is applied to the independent Matsubara frequencies $\omega_{n_l}$, $l\in\Bar{\mathcal{T}}_g$:
\begin{equation}
    \label{app:Gaudin_summation}
    \lim_{\theta\to 0^+} T\sum_{n_l\in\mathbb{Z}}\frac{e^{\ri \omega_{n_l} \theta\,\mathscr{T}_l}}{k_l^0-\ri\omega_{n_l}} = 
    \begin{cases}
        \sgn(\mathscr{T}_l) \BE\big(\sgn(\mathscr{T}_l) k_l^0\big)\,, & \quad\text{for}\quad \omega_{n_l} = 2 n_l \pi T\,, \\
        -\sgn(\mathscr{T}_l) \FD\big(\sgn(\mathscr{T}_l) k_l^0\big)\,, & \quad\text{for}\quad \omega_{n_l} = (2n_l+1)\pi T\,.
    \end{cases}
\end{equation}
This formula associates a statistical factor to each line of $\bar{\mathcal{T}}_g$, resulting in each Gaudin graph being proportional to the product of $L$ statistical factors for an $L$-loop Feynman diagram.
To obtain $\mathscr{T}_l$, one first specifies the orientation of each line of the Feynman diagram (usually chosen to be the momentum flow), and considers a closed loop $\mathscr{C}_l$ formed by the line $l\in\bar{\mathcal{T}}_g$ and the lines of $\mathcal{T}_g$.
The orientation of $\mathscr{C}_l$ is specified by the orientation of the line $l$.
One then sums over $\tau_i$ along all lines in $\mathscr{C}_l$ and defines $\mathscr{T}_l = {\displaystyle \sum_{i\in\mathscr{C}\!_l}\mathscr{O}_i\kappa_i}$\,, where $\mathscr{O}_i=\pm$ if the orientation of the line $i$ coincides/is opposite to the orientation of $\mathscr{C}_l$.\footnote{This value of $\mathscr{T}_l$ is obtained also from eq.~\eqref{app:Gaudin_decomp} if one expresses all $\omega_{n_j}$ ($j\in\mathcal{T}_g$) in terms of $\omega_{n_l}$ ($l\in \bar{\mathcal{T}}_g$) and the external frequencies $\omega_{e_i}$ by using energy conservation at the vertices when assigning the regulators.}
The integers $\kappa_i$ are arbitrary as far as $\mathscr{T}_l$ is non-vanishing for any of the Gaudin graphs.

\begin{figure}[t]
    \begin{center}
    \includegraphics[width=0.75\textwidth]{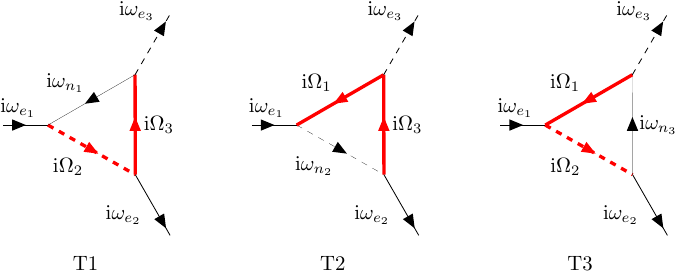}
    \caption{Diagrammatic representation of the three Gaudin graphs associated to the scalar triangle integral shown in figure~\ref{fig:scalartriangle}.
    Thick red lines form the tree $\mathcal{T}_g$ where $g={\rm T1},\,{\rm T2},\,{\rm T3}$, while thin black lines represent the lines of $\bar{\mathcal{T}}_g$, the complement of a tree graph within a Gaudin graph $g$. 
    In each Gaudin graph $\ri \omega_{e_i}$ are the external Matsubara frequencies, $\ri \omega_{n_i}$ are the Matsubara frequencies of the internal lines in $\Bar{\mathcal{T}}_g$, and $\ri\Omega_{j}$ are frequencies of internal lines in $\mathcal{T}_g$.
    \label{fig:ST_Gaudin}
    }
    \end{center}
\end{figure}

In case of the scalar triangle diagram there are three Gaudin graphs, as shown in figure~\ref{fig:ST_Gaudin}. 
To illustrate the method we give explicitly the contribution of the first Gaudin graph denoted by T1 in the figure.
Energy conservation gives $\ri\Omega_{2}(k_1^0;\ri \omega_{e_1})=k_1^0+\ri \omega_{e_1}$ and $\ri \Omega_{3}(k_1^0;\ri \omega_{e_3}) = k_1^0+\ri \omega_{e_3}$.
We associate the regulators to the internal lines as e.g. $\kappa_1=1$, $\kappa_2=2$, and $\kappa_3 = 4$, then one has $\mathscr{T}_1=\kappa_1+\kappa_2+\kappa_3=7$, hence $\sgn(\mathscr{T}_1)=1$.
Therefore, from eqs.~\eqref{app:Gaudin_decomp} and \eqref{app:Gaudin_summation} the contribution of the first Gaudin graph of figure~\ref{fig:ST_Gaudin} is
\begin{equation}
    \textrm{T1} \quad \longrightarrow\quad \frac{-\FD(k_1^0)}{(k_2^0-k_1^0-\ri \omega_{e_1})(k_3^0 - k_1^0 - \ri \omega_{e_3})}, 
\end{equation}
as appears in the first term of eq.~\eqref{Eq:Matsu_sum_res} with the identification $\omega_{e_1}=\nu_n$ and $\omega_{e_3}=\omega_h$.

\section{Parametrization of the integrals in the expressions for the thermal vertex cuts \label{sec:app:ThermalVertexCuts}}

In this appendix we detail the evaluation strategy of the integrals that appear in the expressions for the second and third thermal vertex cuts as introduced in section \ref{sec:vertex}, see eqs.~\eqref{eq:ImGammaCut2}--\eqref{eq:ImGammaCut3}.
First, we indicate the differences as compared to the first cut.
Second, we discuss the specific coordinate system and integration variable that is most convenient for the problem.

The amplitudes corresponding to the cuts in question are proportional to the integral of products of $\delta$ distributions,
\begin{gather}
    \label{eq:exampleM23}
    \imag\Gamma^{\sf (cut\,2,3)}\propto \int\frac{\rd^4 K}{(2\pi)^4}\delta\Big(K^2-m_{L,\phi}^2\Big)\delta\Big((K-P_{\phi,L})^2-m_{N_j}^2\Big)(...)\,.
\end{gather}
The four-momentum $K$ is the momentum of the internal lepton in $\imag\Gamma^{\sf (cut\,2)}$, and of the internal scalar in $\imag\Gamma^{\sf (cut\,3)}$.
Expanding the $\delta$ distributions of the cut neutrino propagators we find scalar products of the loop momentum and the outgoing momenta ($K\cdot P_L$ and $K\cdot P_\phi$). 
These scalar products depend non-linearly on the cosine of the loop azimuth angle $\theta'$ and they also depend on the loop polar angle $\varphi'$, whereas $K\cdot P_N$ was linear in $\cos\theta'$ and was independent of $\varphi'$ allowing for a simple evaluation of the $\cos\theta'$-integral in the case of cut 1.
It is then convenient to choose a coordinate system such that $\p_{L,\phi}\parallel \hat{x}$ (instead of $\p_N\parallel \hat{x}$ as done for cut 1), where the integrals are evaluated identically to those in the first cut.
In addition to the spatial rotation of the coordinate system we also find that the natural choice for the independent energy variable is $E_{L,\phi}$ instead of $E_N$.

\begin{figure}[t]
    \centering
    \includegraphics[width=\linewidth]{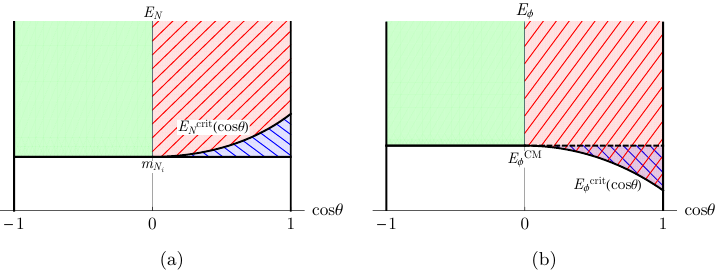}
    \caption{Integration regions for the $N_i\to\phi+L$ decay: (a) with the original decaying particle energy $E_N$ and (b) with the final state scalar energy $E_\phi$.
    The similarly hatched or shaded regions are identified with each other.
    The red and blue hatched regions are distinct in $E_N$ but are overlapping in $E_\phi$, showing the breakdown of the one-to-one relation between $E_N$ and $E_\phi$.
    In the left panel we introduced the shorthand notation $E_N^{\rm crit}(\cos\theta)\equiv E_N\big(E_\phi^{\rm crit}(\cos\theta),\cos\theta\big)$.}
    \label{fig:IntegrationRegions}
\end{figure}

As a starting point we first describe the change of variables $E_N\to E_{L,\phi}$.
We define an example integral with a well-behaved function $\mathcal{F}$ over the physical domains of the incoming neutrino energy $E_N$ and center of mass scattering angle $\theta$:
\begin{equation}
    \label{eq:example_integral}
    \mathcal{I}=\int_{m_{N_i}}^\infty\rd E_N\int_{-1}^1\rd\cos\theta~\mathcal{F}(E_N,\cos\theta)\,.
\end{equation}
In the changing of the energy variable $E_N$ there is a great deal of symmetry between $E_{L,\phi}$ and thus we present them together.
To find the function $E_N(E_{L,\phi},\cos\theta)$ we need to invert the original relations introduced for the Lorentz-boosted decay kinematics, with $x=\cos\theta$ and ${\rm p}=L,\phi$ we have:
\begin{equation}
    \label{eq:ELH_EN}
    E_{\rm p}(E_N,x) = \frac{E_N}{m_{N_i}}\left[E_{\rm p}^\CM + (-1)^{\delta_{{\rm p}\phi}} \frac{\sqrt{E_N^2-m_{N_i}^2}}{E_N}p^\CM\, x\right]\,.
\end{equation}
With the exception of $x$, all involved quantities are positive and we find a relationship between the various energies and the sign of $x$:
\begin{subequations}
\begin{align}
    \label{eq:cosalpha_positive_relation}
    \frac{E_N}{m_{N_i}}>\frac{E_{\rm p}(E_N,x)}{E_{\rm p}^\CM} \quad&\text{when}\quad 
    \begin{cases}
    x<0 & {\rm for}~{\rm p}=L\,, \\
    x>0 & {\rm for}~{\rm p}=\phi\,,
    \end{cases} \\
    \label{eq:cosalpha_negative_relation}
    \frac{E_N}{m_{N_i}}<\frac{E_{\rm p}(E_N,x)}{E_{\rm p}^\CM}\quad&\text{when}\quad \begin{cases}
    x>0 & {\rm for}~{\rm p}=L\,, \\
    x<0 & {\rm for}~{\rm p}=\phi\,.
    \end{cases}
\end{align}
\end{subequations}
These relations are considered as consistency conditions when taking the square of eq.~\eqref{eq:ELH_EN} and solving the resulting quadratic equation for $E_{N}$. 
For any sign of $x$ the two solutions (labeled with $\pm$) to the quadratic equation are given by:
\begin{equation}
    \label{eq:ENplusminus_solution}
    E_N^{\pm}(E_{\rm p},x)=m_{N_i}\cdot \frac{\displaystyle \frac{E_{\rm p}}{E_{\rm p}^\CM} \pm |x|\cdot\frac{p^\CM}{E_{\rm p}^\CM} \sqrt{\left(\frac{E_{\rm p}}{E_{\rm p}^\CM}\right)^2-1+x^2\left(\frac{p^\CM}{E_{\rm p}^\CM}\right)^2}}{\displaystyle 1-x^2\left(\frac{p^\CM}{E_{\rm p}^\CM}\right)^2}\,.
\end{equation}
As it is seen from eq.~\eqref{eq:ENplusminus_solution}, the relationship between $E_N$ and $E_{\rm p}$ is not always one-to-one.
When boosts have an anti-parallel component to the momentum vector of the outgoing particle, we get the same outgoing particle energy for two distinct incoming particle energies.
In this case, the boosted energy is initially smaller than the CM energy down to a critical energy where the boosted decay angle is exactly orthogonal to the boost vector itself.
For larger boosts the energy increases indefinitely, providing a double cover between the critical energy and the CM energy, as shown in figure~\ref{fig:IntegrationRegions}.
The critical energy is defined as the smallest possible value of the boosted energy of one outgoing particle,
\begin{equation}
    \label{eq:critical_energy}
    E_{\rm p}^{\rm \, crit}(x) =
    \sqrt{\left(E_{\rm p}^\CM\right)^2 - \frac{1}{2}\big(1 + (-1)^{\delta_{{\rm p}L}} {\rm sgn}(x)\big)\left(p^\CM\right)^2x^2} \leq E_{\rm p}^\CM\,.
\end{equation}
As expected from the above discussion, depending on the sign of $x$, the minimum energy is seen to be either equal to, or smaller than the CM energy.

To remove unphysical solutions from eq.~\eqref{eq:ENplusminus_solution} we must subject the results to the consistency relations of eqs.~\eqref{eq:cosalpha_positive_relation}--\eqref{eq:cosalpha_negative_relation}.
First, the requirement of $E_N^+$ remaining real and the consistency relations provide constraints on the energy interval and on the sign of $\cos\theta$ respectively.
We find that $E_N^+(E_{\rm p},\cos\theta)$ is physical if
\begin{equation}
    \label{eq:ENplus_regions}
    E_N^+(E_{\rm p},\cos\theta):\qquad 
    \begin{cases}
        E_L\geq E_L^{\rm crit}(\cos\theta) &\text{ for }~\cos\theta<0\,, \\
        E_\phi\geq E_\phi^{\rm crit}(\cos\theta) &\text{ for }~\cos\theta>0\,.
    \end{cases}
\end{equation}
The other solution is more complicated as there are two regions in $E_{\rm p}$ that satisfy different consistency relations.
It turns out that $E_N^-(E_{\rm p},\cos\theta)$ is physical if
\begin{equation}
    \label{eq:ENminus_regions}
    E_N^-(E_{\rm p},\cos\theta):\qquad
    \begin{cases}
        E_L\geq E_L^\CM & \text{ for }~\cos\theta > 0\,, \\
        E_L^\CM\geq E_L \geq E_L^{\rm crit}(\cos\theta) & \text{ for }~\cos\theta < 0\,, \\
        E_\phi\geq E_\phi^\CM & \text{ for }~\cos\theta < 0\,, \\
        E_\phi^\CM \geq E_\phi \geq E_\phi^{\rm crit}(\cos\theta) & \text{ for }~\cos\theta>0\,.
    \end{cases}
\end{equation}
Comparing eqs.~\eqref{eq:ENplus_regions} and \eqref{eq:ENminus_regions} we find that the energy intervals $E_{\rm p}^\CM\geq E_{\rm p}\geq E_{\rm p}^{\rm crit}(\cos\theta)$ are doubly covered by having solutions with both $E_N^{\pm}$.
All other regions in the energy range are single valued.

In summary, the physical energy domain in $E_N$ can be separated into three regions as defined by eqs.~\eqref{eq:ENplus_regions}--\eqref{eq:ENminus_regions}.
We write the example integral in eq.~\eqref{eq:example_integral} as a sum of these three contributions, $\mathcal{I}_{\rm p}=\mathcal{I}_{\rm p}^{(1)}+\mathcal{I}_{\rm p}^{(2)}+\mathcal{I}_{\rm p}^{(3)}$.
The various integration regions are depicted in figure~\ref{fig:IntegrationRegions} for $E_\phi$ as an example.
In the figure, the green shaded region corresponds to $\mathcal{I}_{\phi}^{(1)}$, the red hatched region is $\mathcal{I}_{\phi}^{(2)}$, and the blue hatched region is $\mathcal{I}_{\phi}^{(3)}$.
The integrals themselves can then be written as
\begin{subequations}
\label{eq:HiggsRegionIntegrals}
\begin{align}
    \mathcal{I}_\phi^{(1)} &= \int_{0}^1\rd \cos\theta \int_{E_{\phi}^{\rm crit}(\cos\theta)}^\infty\rd E_{\phi}\,\left[\frac{\partial E_N^{+}(E_{\phi},\cos\theta)}{\partial E_{\phi}}\right]\mathcal{F}\big[E_N^{+}(E_{\phi},\cos\theta),\cos\theta\big]\,, \\
    \mathcal{I}_\phi^{(2)} &= \int_{-1}^0\rd\cos\theta\int_{E_\phi^\CM}^\infty\rd E_\phi\,\left[\frac{\partial E_N^{-}(E_{\phi},\cos\theta)}{\partial E_{\phi}}\right]\mathcal{F}\big[E_N^{-}(E_{\phi},\cos\theta),\cos\theta\big]\,, \\
    \mathcal{I}_\phi^{(3)} &= \int_{0}^1\rd\cos\theta\int_{E_\phi^\CM}^{E_\phi^{\rm crit}(\cos\theta)}\rd E_\phi\,\left[\frac{\partial E_N^{-}(E_{\phi},\cos\theta)}{\partial E_{\phi}}\right]\mathcal{F}\big[E_N^{-}(E_{\phi},\cos\theta),\cos\theta\big]\,.
\end{align}
\end{subequations}
Similarly, for the lepton energy one has:
\begin{subequations}
\label{eq:LeptonRegionIntegrals}
\begin{align}
    \mathcal{I}_L^{(1)} &= \int_{-1}^0\rd \cos\theta \int_{E_{L}^{\rm crit}(\cos\theta)}^\infty\rd E_{L}\,\left[\frac{\partial E_N^{+}(E_{L},\cos\theta)}{\partial E_{L}}\right]\mathcal{F}\big[E_N^{+}(E_{L},\cos\theta),\cos\theta\big]\,, \\
    \mathcal{I}_L^{(2)} &= \int_{0}^1\rd\cos\theta\int_{E_L^\CM}^\infty\rd E_L\,\left[\frac{\partial E_N^{-}(E_{L},\cos\theta)}{\partial E_{L}}\right]\mathcal{F}\big[E_N^{-}(E_{L},\cos\theta),\cos\theta\big]\,, \\
    \mathcal{I}_L^{(3)} &= \int_{-1}^0\rd\cos\theta\int^{E_L^{\rm crit}(\cos\theta)}_{E_L^\CM}\rd E_L\,\left[\frac{\partial E_N^{-}(E_{L},\cos\theta)}{\partial E_{L}}\right]\mathcal{F}\big[E_N^{-}(E_{L},\cos\theta),\cos\theta\big]\,.
\end{align}
\end{subequations}
The expressions for leptons and the scalar are similar, the difference lies in the boundaries of the angular integral.

The last step in setting up the convenient coordinate system for the calculation is the rotation of the spatial axis such that $\p_{L,\phi}\parallel \hat{x}$.
As the outgoing momenta in the original frame were defined such that they lay in the $x$--$y$ plane (see eq.~\eqref{eq:PLphiDef}), the required rotation will be constrained to this plane as well with a single angle $\Delta$ to be determined.
By definition, the rotation angle is obtained through solving the following matrix equation:
\begin{equation}
\label{eq:DeltaRotationMatrix}
    \begin{pmatrix}
        \sqrt{E_{\rm p}^2 - m_{\rm p}^2} \\ 0 \\ 0
    \end{pmatrix}
    = 
    \begin{pmatrix}
        \cos\Delta_{\rm p} & -\sin\Delta_{\rm p} & 0 \\
        \sin\Delta_{\rm p} & \cos\Delta_{\rm p} & 0 \\
        0 & 0 & 1
    \end{pmatrix}
    \begin{pmatrix}
        \gamma \big(v E_{\rm p}^\CM + (-1)^{\delta_{{\rm p}\phi}} p^\CM\cos\theta\big) \\
        (-1)^{\delta_{{\rm p}\phi}} p^\CM \sin\theta \\
        0
    \end{pmatrix}\,.
\end{equation}
The system of equations is easily solved for $\sin\Delta_{\rm p}$ and $\cos\Delta_{\rm p}$:
\begin{subequations}
\label{eq:sinDeltacosDeltaDef}
\begin{align}
    \label{eq:sinDelta}
    \sin\Delta_{\rm p}(E_{\rm p},\cos\theta) &= (-1)^{\delta_{{\rm p}L}}\frac{p^\CM}{\sqrt{E_{\rm p}^2-m_{\rm p}^2}}\sqrt{1-\cos^2\theta}\,, \\
    \label{eq:cosDelta}
    \cos\Delta_{\rm p}(E_{\rm p},\cos\theta) &= \frac{\gamma}{\sqrt{E_{\rm p}^2-m_{\rm p}^2}}\left(v E_{\rm p}^\CM + (-1)^{\delta_{{\rm p}\phi}} p^\CM\cos\theta\right)\,.
\end{align}
\end{subequations}
For $\cos\Delta_{L,\phi}$ the Lorentz $\gamma$-factor depends on the regions defined in eqs.~\eqref{eq:LeptonRegionIntegrals}--\eqref{eq:HiggsRegionIntegrals} through the neutrino energy as $\gamma \equiv E_N^\pm(E_{\rm p},\cos\theta)/m_{N_i}$.
This makes the expression for $\cos\Delta$ quite cumbersome, however, since $\sin^2\Delta+\cos^2\Delta=1$ should hold in each region, the only new information in eq.~\eqref{eq:cosDelta} can only be its overall sign.
We find the absolute value of the cosine to be given by
\begin{equation}
    |\cos\Delta_{\rm p}|(E_{\rm p},\cos\theta) = \sqrt{\frac{E_{\rm p}^2-\left(E_{\rm p}^\CM\right)^2+\left(p^\CM\right)^2\cos^2\theta}{E_{\rm p}^2-m_{\rm p}^2}}\,.
\end{equation}
Comparing with eq.~\eqref{eq:cosDelta} we have the simple relations:
\begin{equation}
\begin{aligned}
    \label{eq:cosDeltaFinal}
    \cos\Delta^+_{\rm p}(E_{\rm p},\cos\theta)&\equiv 
    |\cos\Delta_{\rm p}|(E_{\rm p},\cos\theta)\,, \\
    \cos\Delta^-_{\rm p}(E_{\rm p},\cos\theta)&\equiv
    (-1)^{\delta_{{\rm p}\phi}}\mathrm{sgn}(\cos\theta)|\cos\Delta_{\rm p}|(E_{\rm p},\cos\theta)\,.
\end{aligned}
\end{equation}
We use $\cos\Delta^+_{\rm p}(E_{\rm p},\cos\theta)$ in region 1 and $\cos\Delta^-_{\rm p}(E_{\rm p},\cos\theta)$ in regions 2 and 3, i.e., the $\pm$ indicates the regions where $E_N^\pm$ applies (see in eqs.~\eqref{eq:HiggsRegionIntegrals}--\eqref{eq:LeptonRegionIntegrals})\,.
With these definitions the rotation of the spatial coordinate system to that where either $\p_{L}\parallel \hat{x}$ or $\p_{\phi}\parallel \hat{x}$ is completely determined.

In this appendix we have introduced a convenient reference frame for the calculation of the integrals that appear in the thermal vertex cuts.
First, we shifted the integration variable from $E_N$ to $E_{L,\phi}$ in eq.~\eqref{eq:ENplusminus_solution} and found that different expressions hold in various energy- and decay angle domains, see eqs.~\eqref{eq:ENplus_regions}--\eqref{eq:ENminus_regions}.
In addition, to simplify the integration of the $\delta$ distributions (see eq.~\eqref{eq:exampleM23}), we also introduced a spatial rotation such that either $\p_{L}\parallel \hat{x}$ or $\p_{\phi}\parallel \hat{x}$. 
These rotation angles depend non-trivially on the energy and decay angle and are given in eqs.~\eqref{eq:sinDeltacosDeltaDef}--\eqref{eq:cosDeltaFinal}.

\section{Evaluation of the integrals in the CP-asymmetry factor}
\label{sec:EvaluationOfIntegrals}

In this appendix we calculate the loop-integrals appearing in the contributions to the CP-asymmetry factor, in eqs.~\eqref{eq:EpsilonSigmaFinalFormula} and \eqref{eq:epsilonVi}.
We assume that the thermal masses satisfy $\overline{m}_{N_i}>\overline{m}_\phi+\overline{m}_L$ while for the sterile neutrino in the loop we have $\overline{m}_{N_j}>\overline{m}_{N_i}$.
We work in the CR frame where the 4-momenta of the decay products can be expressed using the CM frame kinematics given in  eq.~\eqref{eq:CRenergyDefinitions} as
\begin{equation}
    \label{eq:PLphiDef}
    P_{L/\phi} \equiv P_{L/\phi}^\CR = 
    \begin{pmatrix}
        \overline{m}_N^{-1}\big(E_{L/\phi}^\CM E_N \pm p^\CM p_N\cos\theta\big) \\
        \overline{m}_N^{-1}\big(E_{L/\phi}^\CM p_N \pm p^\CM E_N\cos\theta\big) \\
        \pm p^\CM \sin\theta \\
        0
    \end{pmatrix}\,.
\end{equation}
The loop-momentum is a general 4-vector that can be parameterized by the energy $\omega$, the length of the 3-momentum $k>0$, and two polar angles, $\theta'\in[0,\pi]$ and $\varphi'\in[0,2\pi]$ as
\begin{equation}
    \label{eq:Kdef}
    K =
    \begin{pmatrix}
        \omega \\ k\cos\theta' \\ k\sin\theta'\cos\varphi' \\ k\sin\theta'\sin\varphi'
    \end{pmatrix}
    \,.
\end{equation}
With this parametrization the loop integration measure is rewritten as
\begin{equation}
    \int_K = \frac{1}{(2\pi)^4}\int_{-\infty}^{\infty}\!\rd\omega\int_0^\infty\!\rd k~k^2\int_0^{2\pi}\!\rd\varphi'\int_{-1}^{1}\!\rd\cos\theta'\,.
\end{equation}
In general, the entire loop integral cannot be evaluated analytically for the thermal self-energy and vertex functions.
However, we shall see in the following subsections that the formulae for the CP-asymmetry factor can be reduced to numerically more easily manageable double- or triple-integrals.

\subsection{Self-energy and the first cut of the vertex function}
\label{app:self-energy_and_first_cut_vertex}

As the loop-integral in the self-energy contribution in eq.~\eqref{eq:ImagSigmaTildeIntegralRepresentation} is rather similar to the first cut of the vertex function in eq.~\eqref{eq:ImGammaCut1}, we shall consider them together in this subsection.

As a first step, we carry out the $k$ and $\cos\theta'$ integrals with the help of the two Dirac-deltas corresponding to the two cut propagators of the lepton and scalar fields.
For some function $\mathcal{F}(K)$ one has
\begin{equation} 
    \label{eq:kintegral}
    \int_0^\infty\rd k~\delta(K^2-\overline{m}_L^2)\mathcal{F}(\omega,k,\cos\theta',\varphi') = \frac{\mathcal{F}\big(\omega,\sqrt{\omega^2-\overline{m}_L^2},\cos\theta',\varphi'\big)}{2\sqrt{\omega^2-\overline{m}_L^2}}\,,
\end{equation}
where we used that $k>0$ so only one physical root of the Dirac-delta survives integration.
Assuming that the $k$-integral had been carried out, the $\cos\theta'$ integral is rewritten as
\begin{equation}
\begin{gathered}
    \label{eq:costhetaintegral}
    \int_{-1}^{1}\rd\cos\theta'~\delta\big((P_N-K)^2-\overline{m}_\phi^2\big) = \int_{-1}^{1}\rd\cos\theta'~\frac{\delta\big(\cos\theta' - f_{\theta'}^{(1)}(E_N,\omega)\big)}{2\sqrt{E_N^2-\overline{m}_{N_i}^2}\sqrt{\omega^2-\overline{m}_L^2}}\,, \\
    f_{\theta'}^{(1)}(E_N,\omega) = \frac{\overline{m}_\phi^2-\overline{m}_{N_i}^2-\overline{m}_L^2+2E_N\omega}{2\sqrt{E_N^2-\overline{m}_{N_i}^2}\sqrt{\omega^2-\overline{m}_L^2}}\,,
\end{gathered}
\end{equation}
where we exploited our choice $\p_N\parallel \hat{x}$, implying $P_N\cdot K=E_N\omega - p_N \sqrt{\omega^2-\overline{m}_L^2} \cos\theta'$ and there is no $\varphi'$-dependence at all.
Since $\cos\theta'\in[-1,1]$ the integration constrains the value of $\omega$ to $\omega\in[\omega_-^{(1)},\omega_+^{(1)}]$, where using eq.~\eqref{eq:PLphiDef} one has
\begin{equation}
    \label{eq:omegapm1}
    \omega_\pm^{(1)} = \frac{E_N}{m_{N_i}}\left[E_L^\CM\pm \frac{p_N}{E_N} p^\CM \right]\,.
\end{equation}
This is simply the Lorentz transform of the CM frame lepton energy, which is necessarily positive.
In total, we could reduce the 4-momentum integration to the following double integral:
\begin{equation}
\label{eq:deltadeltaIntegral}
\begin{split}
    \int_K\delta(K^2&-\overline{m}_L^2)\delta\big((P_N-K)^2-\overline{m}_\phi^2\big)\mathcal{F}(\omega,k,\cos\theta',\varphi') = 
    \\
    &\frac{1}{64\pi^4\sqrt{E_N^2-\overline{m}_{N_i}^2}}\int_{\omega_-^{(1)}}^{\omega_+^{(1)}}\rd\omega\int_0^{2\pi}\rd\varphi'\,\mathcal{F}\Big(\omega,\sqrt{\omega^2-\overline{m}_L^2},f_{\theta'}^{(1)}(E_N,\omega),\varphi'\Big)\,.
\end{split}
\end{equation}
Having performed the $k$ and $\cos\theta'$ integrals, we turn now to the remaining angular integral.

We begin with the self-energy contribution where the function $\mathcal{F}$ in eq.~\eqref{eq:deltadeltaIntegral} could be defined through comparison with eq.~\eqref{eq:EpsilonSigmaFinalFormula} as
\begin{equation}
    \label{eq:FSigma}
    \mathcal{F}_{\Sigma}(K)={\rm sgn}(\omega){\rm sgn}(E_N-\omega)(K\cdot P_L)\big(1+\BE(E_N-\omega)-\FD(\omega)\big)\,.
\end{equation}
As $\omega_\pm^{(1)}>0$ and kinematics require that $E_N>\omega$ one has $\mathrm{sgn}(\omega)=\mathrm{sgn}(E_N-\omega)=1$.
The scalar product depends on $E_N,\theta,\omega,$ and $\varphi'$ and it can be expanded using eqs.~\eqref{eq:PLphiDef}--\eqref{eq:Kdef} as
\begin{equation}
\label{eq:KdotPLfull}
\begin{split}
    \big[K\cdot P_L\big](E_N,\theta,&\,\omega,\varphi') = 
    \frac{\omega}{\overline{m}_{N_i}}\left(E_N E_L^\CM+p^\CM\sqrt{E_N^2-\overline{m}_{N_i}^2}\cos\theta\right) \\ 
    &- \frac{\sqrt{\omega^2-\overline{m}_L^2}}{\overline{m}_{N_i}}f_{\theta'}^{(1)}(E_N,\omega)\left(E_L^\CM\sqrt{E_N^2-\overline{m}_{N_i}^2} + E_N p^\CM\cos\theta\right) \\
    &-p^{\rm \,(CM)}\sqrt{\omega^2-\overline{m}_L^2}\sqrt{1-\big[f_{\theta'}^{(1)}(E_N,\omega)\big]^2}~\sin\theta\cos\varphi'\,.
\end{split}
\end{equation}
The integral of the last term over $\varphi'$ vanishes so
\begin{equation}
    \int_{0}^{2\pi}\rd\varphi'\,\big[K\cdot P_L\big](E_N,\theta,\omega,\varphi') = 2\pi\big[K\cdot P_L\big]\Big(E_N,\theta,\,\omega,\frac{\pi}{2}\Big)\,.
\end{equation}
The remaining $\omega$--integral in eq.~\eqref{eq:deltadeltaIntegral} has to be done numerically.

Using that $P_N\cdot P_L=\overline{m}_{N_i}E_L^\CM$ one concludes that the amplitude-level CP-asymmetry factor at finite temperature due to the self-energy contribution defined in eq.~\eqref{eq:EpsilonSigmaFinalFormula} is given as
\begin{equation}
\begin{split}
    \label{eq:epsilon_Sigma_final}
    \epsilon_{\Sigma_i}(E_N,\theta) = \frac{G}{4\pi}&\sum_{j\neq i}\frac{\overline{m}_{N_j}}{\overline{m}_{N_i}^2-\overline{m}_{N_j}^2}\frac{1}{E_L^\CM\sqrt{E_N^2-\overline{m}_{N_i}^2}} \\ 
    \times\int_{\omega_-^{(1)}}^{\omega_+^{(1)}}&\rd\omega\,\big[K\cdot P_L\big]\Big(E_N,\theta,\omega,\frac{\pi}{2}\Big)\big(1+\BE(E_N-\omega)-\FD(\omega)\big).
\end{split}
\end{equation}
To find the thermally averaged CP-asymmetry we integrate $\epsilon_{\Sigma_i}$ over the phase space of the initial state with statistical weights as explained in section \ref{sec:CPviolatingasymmetry_FiniteTemperatureCase}, in particular in eq.~\eqref{eq:thermalAvgEpsilonNumerator}.
We note that the CM decay angle integral for $\cos\theta\in[-1,1]$ is still analytic and the final result is a double numerical integral over the neutrino and loop energies $E_N$ and $\omega$.
This double--integral is easily handled by standard numerical integrators.

For the first cut of the thermal vertex function given in eq.~\eqref{eq:ImGammaCut1} one defines the $\mathcal{F}$ function of eq.~\eqref{eq:deltadeltaIntegral} as being proportional to the contribution to the CP-asymmetry factor in eq.~\eqref{eq:epsilonVi}:
\begin{equation}
    \mathcal{F}_{V}^{\sf (cut\,1)}(K)=\mathcal{F}_{\Sigma}(K) \mathcal{P}\frac{1}{(K-P_\phi)^2-\overline{m}_{N_j}^2}\,.
\end{equation}
While the integrand is similar to that of the self-energy, here we also have non-trivial angular- and energy-dependence in the denominator through the $K\cdot P_\phi$ scalar product.
Using momentum conservation we find
\begin{equation}
\label{eq:KdotPphi}
\begin{split}
    \big[K\cdot P_\phi\big](&E_N,\theta,\omega,\varphi')
    = K\cdot (P_N-P_L) \\
    &= \omega E_N - f_{\theta'}^{(1)}(E_N,\omega)\sqrt{E_N^2-\overline{m}_{N_i}^2}\sqrt{\omega^2-\overline{m}_L^2}-\big[K\cdot P_L\big](E_N,\theta,\omega,\varphi')
\end{split}
\end{equation}
This scalar product is also linear in $\cos\varphi'$ thus in the integrand one has a fraction of two linear functions in $\cos\varphi'$.
We define the result of the $\varphi'$-integral as the function
\begin{equation}
    \label{eq:Ivarphi1}
    \mathcal{I}_{\varphi'}^{\sf (cut\,1)}(E_N,\theta,\omega) = \int_0^{2\pi}\frac{\rd\varphi'}{2\pi}\frac{\big[K\cdot P_L\big](E_N,\theta,\omega,\varphi')}{\overline{m}_L^2+\overline{m}_\phi^2-\overline{m}_{N_j}^2-2\big[K\cdot P_\phi\big](E_N,\theta,\omega,\varphi')}\,.
\end{equation}
Note that $\mathcal{I}^{\sf (cut\,1)}_{\varphi'}(E_N,\theta,\omega)$ is analytic, see eq.~\eqref{eq:integrals_varphi}, but it has a lengthy expression that we do not give here.

The amplitude-level CP-asymmetry factor at finite temperature due to the first cut of the vertex function is given as
\begin{equation}
\begin{split}
    \label{eq:epsilon_V_cut1_final}
    \epsilon_{V_i}^{\sf (cut\,1)}(E_N,\theta) = \frac{G}{8\pi}&\sum_{j\neq i}\frac{\overline{m}_{N_j}}{E_L^\CM\sqrt{E_N^2-\overline{m}_{N_i}^2}}
    \\ 
    \times&\int_{\omega_-^{(1)}}^{\omega_+^{(1)}}\rd\omega\,\mathcal{I}_{\varphi'}^{\sf (cut\,1)}(E_N,\theta,\omega)\big(1+\BE(E_N-\omega)-\FD(\omega)\big) \,.
\end{split}
\end{equation}
The integration over the initial state phase space required for the thermally averaged CP-violation rate has to be done numerically, meaning that for the vertex function the final formula is only analytically reducible to a triple integral.
Nevertheless, this triple integral can still be handled with standard numerical integrators.

\subsection{Second cut of the vertex function}

The formula for the imaginary part of the second cut of the vertex function was given in eq.~\eqref{eq:ImGammaCut2}.
As explained in appendix~\ref{sec:app:ThermalVertexCuts}, it is convenient for the calculation of the second cut to use a coordinate system where we take $E_\phi$ as the independent external energy (instead of $E_N$), and take $\p_\phi \parallel \hat{x}$ (instead of $\p_N \parallel \hat{x}$).
In this subsection we use this reference frame.

The lepton and the sterile neutrino propagators are on-shell within the loop, consequently the $k$-integral formally remains the same as in eq.~\eqref{eq:kintegral} while the $\cos\theta'$ integral becomes:
\begin{equation}
\label{eq:ftheta2}
\begin{gathered}
    \int_{-1}^1\rd\cos\theta'\,\delta\big((K-P_\phi)^2-\overline{m}_{N_j}^2\big) = \int_{-1}^1\rd\cos\theta'\,\frac{\delta\big(\cos\theta'-f_{\theta'}^{(2)}(E_\phi,\omega)\big)}{2\sqrt{E_\phi^2-\overline{m}_\phi^2}\sqrt{\omega^2-\overline{m}_L^2}}\,, \\
    f_{\theta'}^{(2)}(E_\phi,\omega) = \frac{\overline{m}_{N_j}^2-\overline{m}_L^2-\overline{m}_\phi^2+2 E_\phi\omega}{2\sqrt{E_\phi^2-\overline{m}_\phi^2}\sqrt{\omega^2-\overline{m}_L^2}}\,.
\end{gathered}
\end{equation}
This result is connected to the previous $f_{\theta'}^{(1)}$ by changing $\phi\to N_j$ (cut propagator in the loop) and $N_i\to \phi$ (external line opposite to the cut).
As before, the constraint of $\cos\theta'\in[-1,1]$ restricts the loop energy to $\omega\in[\omega_-^{(2)},\omega_+^{(2)}]$.
The limits are given by
\begin{equation}
    \label{eq:omegapm2}
    \omega_{\pm}^{(2)} = \frac{E_\phi}{\overline{m}_\phi}\left[E^{(2)}\pm \frac{p_\phi}{E_\phi}\sqrt{\big[E^{(2)}\big]^2-\overline{m}_L^2}\right],\quad\text{where}\quad E^{(2)}=\frac{\overline{m}_\phi^2+\overline{m}_L^2-\overline{m}_{N_j}^2}{2\overline{m}_\phi}\,.
\end{equation}
If $\overline{m}_{N_j}>\overline{m}_{\phi,L}$ then the loop energy is strictly negative as $\omega_{\pm}^{(2)}<0$.
At zero temperature, where the cut propagators have fixed direction of energy flow, i.e., they are proportional to $\theta(\omega)$, the negativity of $\omega$ makes this cut contribution vanish.
The evaluation of the two Dirac-delta integrals lead to
\begin{equation}
\begin{split}
    \int_K\delta(K^2&-\overline{m}_L^2)\delta\big((K-P_\phi)^2-\overline{m}_{N_j}^2\big)\mathcal{F}_V^{\sf (cut\,2)}(\omega,k,\cos\theta',\varphi') =  \\
    &\frac{1}{64\pi^4\sqrt{E_\phi^2-\overline{m}_\phi^2}}\int_{\omega_-^{(2)}}^{\omega_+^{(2)}}\rd\omega\,\int_0^{2\pi}\rd\varphi'\,\mathcal{F}_V^{\sf (cut\,2)}\Big(\omega,\sqrt{\omega^2-\overline{m}_L^2},f_{\theta'}^{(2)}(E_\phi,\omega),\varphi'\Big)\,,
\end{split}
\end{equation}
where for the second cut of the vertex function one has 
\begin{equation}
    \mathcal{F}_V^{\sf (cut\,2)}(K) = \mathrm{sgn}(\omega)\mathrm{sgn}(\omega-E_\phi)\frac{K\cdot P_L}{(P_N-K)^2-\overline{m}_\phi^2}\big(\FD(\omega)-\FD(\omega-E_\phi)\big)\,.
\end{equation}
Due to $\omega<0$ and $E_\phi>0$ we have $\mathrm{sgn}(\omega)\mathrm{sgn}(\omega-E_\phi)=1$.
With positive arguments the statistical factors become $\FD(\omega)-\FD(\omega-E_\phi)=\FD(E_\phi-\omega)-\FD(-\omega)$.
Contrary to the contributions due to the self-energy and the first cut of the vertex function, where the statistical factor term in the limit $T\to 0$ was $1+\BE(E_N-\omega)-\FD(\omega)\to 1$, in the case of the second cut the same limit of the appearing statistical factors vanishes, $\FD(E_\phi-\omega)-\FD(-\omega)\to 0$.

In appendix~\ref{sec:app:ThermalVertexCuts} we show the inversion of the energy relation $E_\phi(E_N,\cos\theta)$ in order to express $E_N(E_\phi,\cos\theta)$.
The inverse relations are not one-to-one in the full integration domain, however, one can define overlapping regions where one-to-one relations may be defined separately, see figure~\ref{fig:IntegrationRegions}.
The physical energy expressions for the incoming sterile neutrino is denoted as $E_N^\pm(E_\phi,\theta)$, see eq.~\eqref{eq:ENplusminus_solution}.
Depending on which region we are in, the inverse relations use either $E_N^\pm$ as defined in eqs.~\eqref{eq:ENplus_regions}--\eqref{eq:ENminus_regions}.
Following these expressions, we define the $\varphi'$ angular integral as
\begin{equation}
    \label{eq:phi_integral_cut2}
    \mathcal{I}_{\varphi'}^{{\sf (cut\,2)}\pm}(E_\phi,\theta,\omega) = \int_0^{2\pi}\frac{\rd\varphi'}{2\pi}\frac{[K\cdot P_L^\pm](E_\phi,\theta,\omega,\varphi')}{\overline{m}_{N_i}^2+\overline{m}_L^2-\overline{m}_\phi^2-2\big[K\cdot P^\pm_N\big](E_\phi,\theta,\omega,\varphi')}\,.
\end{equation}
The scalar products necessarily depend on the inverted energy relations, as indicated with the $\pm$ superscripts.
One then has
\begin{equation}
\begin{split}
    \big[K\cdot P_L^\pm\big]&(E_\phi,\theta,\omega,\varphi') \equiv K\cdot (P_N^\pm-P_\phi)
    \\
    &= \big[K\cdot P_N^\pm\big](E_\phi,\theta,\omega,\varphi') - \omega E_\phi + f_{\theta'}^{(2)}(E_\phi,\omega)\sqrt{\omega^2-\overline{m}_L^2}\sqrt{E_\phi^2-\overline{m}_\phi^2}\,.
\end{split}
\end{equation}
Since the spatial coordinate system is rotated with angle $\Delta_\phi$ such that $\p_\phi\parallel\hat{x}$, see eq.~\eqref{eq:DeltaRotationMatrix}, thus $\p_N$ is also transformed resulting in
\begin{equation}
\label{eq:KcdotPNpm2}
\begin{split}
    \big[K\cdot &P_N^\pm\big](E_\phi,\theta,\omega,\varphi') = \omega E_N^\pm(E_\phi,\theta)
    \\
    &-\sqrt{\omega^2-\overline{m}_L^2}f_{\theta'}^{(2)}(E_\phi,\omega)\cos\Delta_\phi^\pm(E_\phi,\theta)\sqrt{\big[E_N^\pm(E_\phi,\theta)\big]^2-\overline{m}_{N_i}^2}
    \\
    &-\sqrt{\omega^2-\overline{m}_L^2}\sqrt{1-\big[f_{\theta'}^{(2)}(E_\phi,\omega)\big]^2}\sin\Delta_\phi(E_\phi,\theta)\sqrt{\big[E_N^\pm(E_\phi,\theta)\big]^2-\overline{m}_{N_i}^2}\cos\varphi'\,.
\end{split}
\end{equation}
Here the appearing rotation angles in the various domains mentioned above are indicated with the $\pm$ sign and they are defined in eqs.~\eqref{eq:sinDeltacosDeltaDef} and \eqref{eq:cosDeltaFinal}.
The $\varphi'$-integral in eq.~\eqref{eq:phi_integral_cut2} is again analytic, see eq.~\eqref{eq:integrals_varphi}.

In summary, for the second cut of the vertex function we define the amplitude-level CP-asymmetry factors as
\begin{equation}
\begin{split}
    \label{eq:epsilon_V_cut2_final}
    \epsilon_{V_i}^{{\sf (cut\,2)}\pm}(E_\phi,\theta) = \frac{G}{8\pi}&\sum_{j\neq i}\frac{\overline{m}_{N_j}}{E_L^\CM\sqrt{E_\phi^2-\overline{m}_\phi^2}}
    \\
    \times&\int_{\omega_-^{(2)}}^{\omega_+^{(2)}}\rd\omega\,\mathcal{I}_{\varphi'}^{{\sf (cut\,2)}\pm}(E_\phi,\theta,\omega)\big(\FD(E_\phi-\omega)-\FD(-\omega)\big)\,.
\end{split}
\end{equation}
Thermal averaging over the initial state phase space has to be done according to eq.~\eqref{eq:HiggsRegionIntegrals}, i.e., we have to pay attention to the energy and decay angle regions due to the loss of the one-to-one relation between the energies $E_N(E_\phi,\cos\theta)\leftrightarrow E_\phi(E_N,\cos\theta)$.

\subsection{Third cut of the vertex function}

The third, and final cut of the thermal vertex function originally given in eq.~\eqref{eq:ImGammaCut3} is evaluated similarly to the second cut presented in the previous subsection.
In fact, the correspondence between the two calculations can be made even more pronounced if one performs a momentum shift as $K\to P_N-K$.
This shift results in the scalar propagator in the loop having momentum $K$ (instead of the lepton as before) and in eq.~\eqref{eq:ImGammaCut3} we find
\begin{equation}
\label{eq:Ivarphi_cut2}
\begin{split}
    \mathrm{sgn}(\omega-E_N)\mathrm{sgn}(\omega&-E_\phi)\delta\big((K-P_N)^2-\overline{m}_\phi^2\big)\delta\big((K-P_\phi)^2-\overline{m}_{N_j}^2\big) 
    \\ &\to 
    \mathrm{sgn}(-\omega) \mathrm{sgn}(E_L-\omega) \delta(K^2-\overline{m}_\phi^2) \delta\big((P_L-K)^2-\overline{m}_{N_j}^2\big)\,.
\end{split}
\end{equation}
As before, the two $\delta$ distributions are evaluated using the integrals for $k$ and $\cos\theta'$.
The first one is given as in eq.~\eqref{eq:kintegral} but with changing $\overline{m}_L^2\to\overline{m}_\phi^2$.
The second one is performed as in eq.~\eqref{eq:ftheta2} but with interchanging $\overline{m}_L^2\leftrightarrow\overline{m}_\phi^2$ and the exchange $E_\phi\to E_L$.
The function appearing next to the $\cos\theta'$ in the Dirac-delta is then
\begin{equation}
    f_{\theta'}^{(3)}(E_L,\omega) = \frac{\overline{m}_{N_j}^2-\overline{m}_\phi^2-\overline{m}_L^2+2\omega E_L}{2\sqrt{E_L^2-\overline{m}_L^2}\sqrt{\omega^2-\overline{m}_\phi^2}}\,.
\end{equation}
The limits for the loop energy $\omega$ resulting from the $\cos\theta'$ integral follows from eq.~\eqref{eq:omegapm2} via similar replacements of $L\leftrightarrow\phi$ as before and we find
\begin{equation}
    \label{eq:omegapm3}
    \omega_\pm^{(3)}=\frac{E_L}{\overline{m}_L}\left[E^{(3)}\pm \frac{p_L}{E_L}\sqrt{\big[E^{(3)}\big]^2-\overline{m}_\phi^2}\right]\,,\quad\text{where}\quad E^{(3)}=\frac{\overline{m}_L^2+\overline{m}_\phi^2-\overline{m}_{N_j}^2}{2\overline{m}_L}\,.
\end{equation}
As for the second cut, here one also finds that if $\overline{m}_{N_j}>\overline{m}_{L,\phi}$ then necessarily $\omega_\pm^{(3)}<0$ and the loop-energy is always negative.
In conclusion the phase space integral for the loop-momentum over the two Dirac-deltas can be summarized as
\begin{equation}
\begin{split}
    \int_K&\delta(K^2-\overline{m}_\phi^2)\delta\big((P_L-K)^2-\overline{m}_{N_j}^2\big)\mathcal{F}_V^{\sf (cut\,3)}(\omega,k,\cos\theta',\varphi')
    \\
    &=\frac{1}{64\pi^4\sqrt{E_L^2-\overline{m}_L^2}}\int_{\omega_-^{(3)}}^{\omega_+^{(3)}}\rd\omega\int_0^{2\pi}\rd\varphi'\mathcal{F}_V^{\sf (cut\,3)}\Big(\omega,\sqrt{\omega^2-\overline{m}_L^2},f_{\theta'}^{(3)}(E_L,\omega),\varphi'\Big)\,,
\end{split}
\end{equation}
where for the third cut of the vertex function the integrand is given by
\begin{equation}
    \label{eq:FV3}
    \mathcal{F}_V^{\sf (cut\,3)}(K) = \mathrm{sgn}(-\omega)\mathrm{sgn}(E_L-\omega)\frac{(P_N-K)\cdot P_L}{(P_N-K)^2-\overline{m}_L^2}\big(\BE(-\omega)+\FD(E_L-\omega)\big)\,.
\end{equation}
Since $\omega<0$, the sign functions evaluate to $+1$, and the statistical factors already have positive arguments.
Note that due to the shift we performed in the loop momentum, the scalar product in the numerator originating from the Dirac-trace $\mathcal{T}(K)$ (see, eq.~\eqref{eq:DiracTrace}) is also modified. 

In the reference frame convenient for the third cut of the vertex function we have $\p_L\parallel \hat{x}$ (see, appendix.~\ref{sec:app:ThermalVertexCuts}) so the scalar product in the numerator of $\mathcal{F}_V^{(3)}$ is independent of $\varphi'$:
\begin{equation}
    \big[(P_N-K)\cdot P_L\big](E_L,\omega) = \overline{m}_{N_i}E_L^\CM - E_L\omega + f_{\theta'}^{(3)}(E_L,\omega)\sqrt{E_L^2-\overline{m}_L^2}\sqrt{\omega^2-\overline{m}_\phi^2}\,.
\end{equation}
In contrast, the scalar product $K\cdot P_N^\pm$ appearing in the denominator of eq.~\eqref{eq:FV3} is non-trivial in $\varphi'$, and it can be found from eq.~\eqref{eq:KcdotPNpm2} with the replacements $E_\phi\to E_L$, $\overline{m}_L^2\to \overline{m}_\phi^2$, $\Delta_\phi\to\Delta_L$, and $f_\theta^{(2)}\to f_\theta^{(3)}$.
As the $\varphi'$-dependence only appears in the denominator of $\mathcal{F}_V^{(3)}$ we define the $\varphi'$-integral as
\begin{equation}
    \label{eq:Ivarphi_cut3}
    \mathcal{I}_{\varphi'}^{{\sf (cut\,3)}\pm}(E_L,\theta,\omega) = \int_0^{2\pi}\frac{\rd\varphi'}{2\pi}\,\frac{1}{\overline{m}_{N_i}^2+\overline{m}_\phi^2-\overline{m}_L^2-2\big[K\cdot P_N^\pm\big](E_L,\theta,\omega,\varphi')}\,.
\end{equation}
As before, the result of this integral is analytic and finite for physical kinematics, see eq.~\eqref{eq:integrals_varphi}.

In summary, the contribution of the third cut of the thermal vertex function to the amplitude-level CP-asymmetry factor is:
\begin{equation}
\begin{split}
    \label{eq:epsilon_V_cut3_final}
    \epsilon_{V_i}^{{\sf (cut\,3)}\pm}&(E_L,\theta) = \frac{G}{8\pi}\sum_{j\neq i}\frac{\overline{m}_{N_j}}{E_L^\CM\sqrt{E_L^2-\overline{m}_L^2}} \\
    \times\int_{\omega_-^{(3)}}^{\omega_+^{(3)}}&\rd\omega\,\big[(P_N-K)\cdot P_L\big](E_L,\omega)\mathcal{I}_{\varphi'}^{{\sf (cut\,3)}\pm}(E_L,\theta,\omega)\big(\BE(-\omega)+\FD(E_L-\omega)\big)\,.
\end{split}
\end{equation}
The integration over the initial state phase space is done as explained in appendix~\ref{sec:app:ThermalVertexCuts} and in particular given in eq.~\eqref{eq:LeptonRegionIntegrals}.

\section{Useful formulae \label{sec:app3}}

\begin{enumerate}
\item {\it Decomposition of products of statistical factors:}
\begin{equation}
\label{eq:statistical_factor_products}
\begin{aligned}
    \BE(x)\BE(y) &= \BE(x+y)\big(1+\BE(x)+\BE(y)\big)\,, \\
    \FD(x)\FD(y) &= \BE(x+y)\big(1-\FD(x)-\FD(y)\big)\,, \\
    \BE(x)\FD(y) &= \FD(x+y)\big(1+\BE(x)-\FD(y)\big)\,.
\end{aligned}
\end{equation}
\item {\it Energy inversion in statistical factors:}
\begin{equation}
\label{eq:1pmf}
    f_{\rm B/F}(x) = \mp 1 - f_{\rm B/F}(-x)\,,
\end{equation}
\begin{equation}
\label{eq:BEFDperFD__momentum_inversion}
    f_{\rm B/F}(x)\exp(\beta x) = \mp f_{\rm B/F}(-x) ~ \rightarrow ~ \frac{\BE(x-y)\FD(y)}{\FD(x)}=-\frac{\BE(y-x)\FD(-y)}{\FD(-x)}\,,
\end{equation}
\begin{equation}
\label{eq:useful_f_exp_relations}
\begin{aligned}
    \FD(|p^0|)\exp\left(\frac{\beta |p^0|}{2}\right) &= \frac{\sinh(\beta p^0/2)}{\sinh(\beta p^0)}\,,
    \\ 
    \BE(|p^0|)\exp\left(\frac{\beta |p^0|}{2}\right) &= \frac{\sgn(p^0)}{2\sinh\big(\beta p^0/2\big)}\,.
\end{aligned}
\end{equation}
\item {\it Hyperbolic functions:}
\begin{equation}
\label{eq:hyperbolic_magic}
\begin{aligned}
    \coth(x)\left(1-\frac{\cosh(y)}{\cosh(x-y)\cosh(x)}\right)&=\tanh(x-y)\,, \\ \coth(x)\left(1+\frac{\sinh(x-y)}{\sinh(y)\cosh(x)}\right)&=\coth(y)\,.
\end{aligned}
\end{equation}
\item {\it Integrals with} $f_{\pm}(x)=({\rm e}^x \pm 1)^{-1}$:
\begin{enumerate}
    \item with the assumptions $a_{\pm}>b>0$:
    \begin{equation}
    \label{eq:integral_fBfF_costheta}
    \begin{aligned}
        \int_{-1}^1\rd x &f_-(a_{-}-b\,x)f_+(a_{+}+b\,x) \\
        &= -2f_+(a_{-}+a_{+}) + \frac{f_+(a_{-}+a_{+})}{b}\log\left(\frac{f_+(a_{+}-b)f_-(a_{-}-b)}{f_+(a_{+}+b)f_-(a_{-}+b)}\right)\,;
    \end{aligned}
    \end{equation}
    \item  with the assumption $|a|>1$:
    \begin{subequations}
    \label{eq:integrals_varphi}
    \begin{align}
        \int_0^{2\pi}\frac{\rd\varphi'}{2\pi}\frac{1}{a+\cos\varphi'} &= \frac{\sgn(a)}{\sqrt{a^2-1}} \\
        \int_0^{2\pi}\frac{\rd\varphi'}{2\pi}\frac{\cos\varphi'}{a+\cos\varphi'} &= 1-\frac{|a|}{\sqrt{a^2-1}}
    \end{align}
    \end{subequations}
    \item Loop energy integrals (Li$_2(x)$ is the
    dilogarithm):
    \begin{equation}
        \int \rd x \, x f_{\mp}(x) = \pm x \log\big(1\mp \exp(-x)\big)\mp\mathrm{Li}_2\big(\pm \exp(-x)\big) + \mathrm{const.}
        \label{eq:polylog}
    \end{equation}
\end{enumerate}
\end{enumerate}


\providecommand{\href}[2]{#2}\begingroup\raggedright\endgroup

\end{document}